\documentclass{article}
\usepackage[english]{babel}
\usepackage{graphicx,url,algorithm,algpseudocode,fullpage}
\newcommand{\vir}[1]{“#1”}

\makeatletter
\def\BState{\State\hskip-\ALG@thistlm}
\makeatother

\begin{document}

\title{Standard Vs Uniform Binary Search and Their Variants in Learned  Static Indexing: The Case of the Searching on Sorted Data Benchmarking Software Platform \protect\thanks{This research is funded in part by MIUR Project of National Relevance 2017WR7SHH “Multicriteria Data Structures and Algorithms: from compressed to learned indexes, and beyond”. We also acknowledge an NVIDIA Higher Education and Research Grant (donation of a Titan V GPU). Additional support to RG has been granted by Project INdAM - GNCS  “Modellizazzione ed analisi di big knowledge graphs per la risoluzione di problemi in ambito medico e web”}} 

\author{Domenico Amato$^1$\\
	\and
	Giosu\'e Lo Bosco$^1$\footnote{corresponding author, email: giosue.lobosco@unipa.it}\\
	\and
	Raffaele Giancarlo$^1$}

\date{
	$^1$Dipartimento di Matematica e Informatica\\ 
	Universit\'a degli Studi di Palermo, ITALY\\
	\today
}


\maketitle
\begin{abstract}
	Learned Indexes are a novel approach to search in a sorted table. A model is used to predict an interval in which to search into and a Binary Search routine is used to finalize the search. They are quite effective. For the final stage, usually, the {\bf lower\_bound} routine of the Standard C++ library is used, although this is more of a natural choice rather than a requirement. However, recent studies, that do not use Machine Learning predictions,  indicate that other implementations of Binary Search or variants, namely k-ary Search, are better suited to take advantage of the features offered by modern computer architectures. With the use of the Searching on Sorted Sets {\bf SOSD} Learned Indexing benchmarking software, we investigate how to choose a Search routine for the final stage of searching in a Learned Index. Our results provide indications that better choices than the {\bf lower\_bound} routine can be made. We also highlight how such a choice may be dependent on the computer architecture that is to be used. Overall, our findings provide new and much-needed guidelines for the selection of the Search routine within the Learned Indexing framework.
\end{abstract}

\section{Introduction}\label{M-sec1}
\begin{figure}[tbh]
    	\centering
    	\includegraphics[scale=0.8]{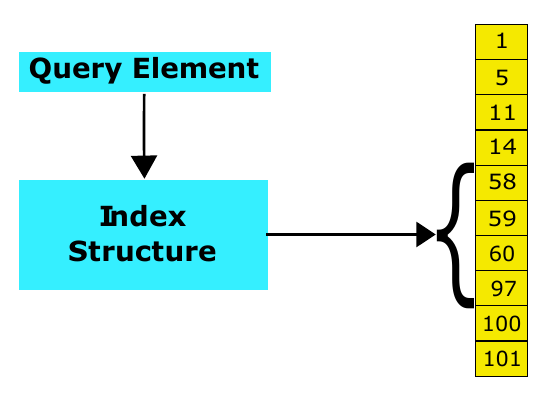}
    	\caption{{\bf  A General Paradigm of  Learned Searching in a Sorted Set}\cite{Marcus20}. The model is trained on the data in the table. Then, given a query element, it is used  to predict the interval in the table where to search (included in brackets in the figure).}
    	\label{M-fig:Par}
    \end{figure}

Learned Static Indexes, introduced by Kraska et al. \cite{kraska18case} (but see also  \cite{Ao11}),  with follow-up in 	\cite{kraska18case,amato2021learned, amato2021lncs, Ferragina:2020pgm,Ferragina:2020book,FERRAGINA21, MaltryVLDB, Kipf20},  are a novel approach to search in a sorted table, quite effective with respect to existing Procedures and Data Structures, e.g., B-trees \cite{comer1979ubiquitous},  used in important application domains such as  Data Bases \cite{rao1999cache} and Search Engines \cite{Morin17}.  
With reference to Figure \ref{M-fig:Par}, a  generic paradigm for Learned Searching in sorted sets consists of a model, trained over the data in a sorted table. 
As described in Section \ref{M-sec:PS},  such a model may be as simple as a straight line or more complex, with a tree-like structure, as the ones mentioned in Section \ref{M-sec:models}. It is used to make a prediction regarding where a  query element may be in the sorted table.  Then, the search is limited to the interval so identified and performed via Standard Binary Search. The use of this latter routine is more of a natural choice rather than a requirement. In fact, the {\bf lower\_bound}  routine from the standard C++ library is almost exclusively used. 

In order to place our contributions on the proper ground, it is useful to recall that two major studies \cite{Morin17,Schulz18}  have recently investigated which Binary Search routines or variants are better suited to take advantage of modern computer architectures.  Those experimental findings hold in the \emph{stand alone} scenario, i.e. when no prediction to reduce the search interval is performed, and they provide useful indications on which routine to use in which circumstances.  However, to what extent the recommendations coming out of those studies actually hold also for Learned Indexes has not been investigated. As a matter of fact, which version of Binary Search to use,  and when,  is unresolved for Learned Indexes, relying on the natural choice mentioned earlier. 

With the use of the  {\bf SOSD} \cite{Kipf19} benchmarking software platform, we address such a question by experimenting with various Binary Search routines on both synthetic and real datasets, with executions on two different architectures, i.e., Intel I7/9 and Apple M1. On the one hand, our results further validate and extend the ones provided in \cite{Morin17,Schulz18} for the \emph{stand alone} scenario and, on the other, they provide novel indications on when to use {\bf SOSD} for Learned Indexing executions and with which Binary Search routine. Our findings are a significant advance with respect to the research performed on Learned indexes outlined above. For completeness, we mention that our results hold for the static case of Learned Indexes, i.e., when no insertions or deletions are allowed. For the dynamic case, Learned Indexing solutions exist \cite{Ferragina:2020pgm, Ding20}. However, how to phrase a research analogous to ours in that setting is open, to the best of our knowledge.

In order to make our experiments replicable, the software we developed or modified is available at \cite{gitbvb, gitbvbStand}, while the datasets are available at \cite{gitanonym}. 

\section{A Simple View of Learned Searching in  Sorted Sets }\label{M-sec:PS}

Consider a sorted table $A$ of $n$ keys, taken from a universe $U$.  It is well known that Sorted Table Search can be phrased as the  Predecessor Search Problem:  for a given query element $x$, return the  $A[j]$ such that $A[j] \leq x < A[j+1]$. Kraska et al. \cite{kraska18case} have proposed an approach that transforms such a problem into a learning-prediction one. With reference to Figure \ref{M-fig:Par}, the model learned from the data is used as a predictor of where a query element may be in the table.  To fix ideas, Binary Search is then performed only on the interval returned by the model. 
    
\begin{figure}[tbh]
    	\centering
    	(a)
    	\begin{minipage}{0.25\textwidth}
    		\includegraphics[width=\linewidth]{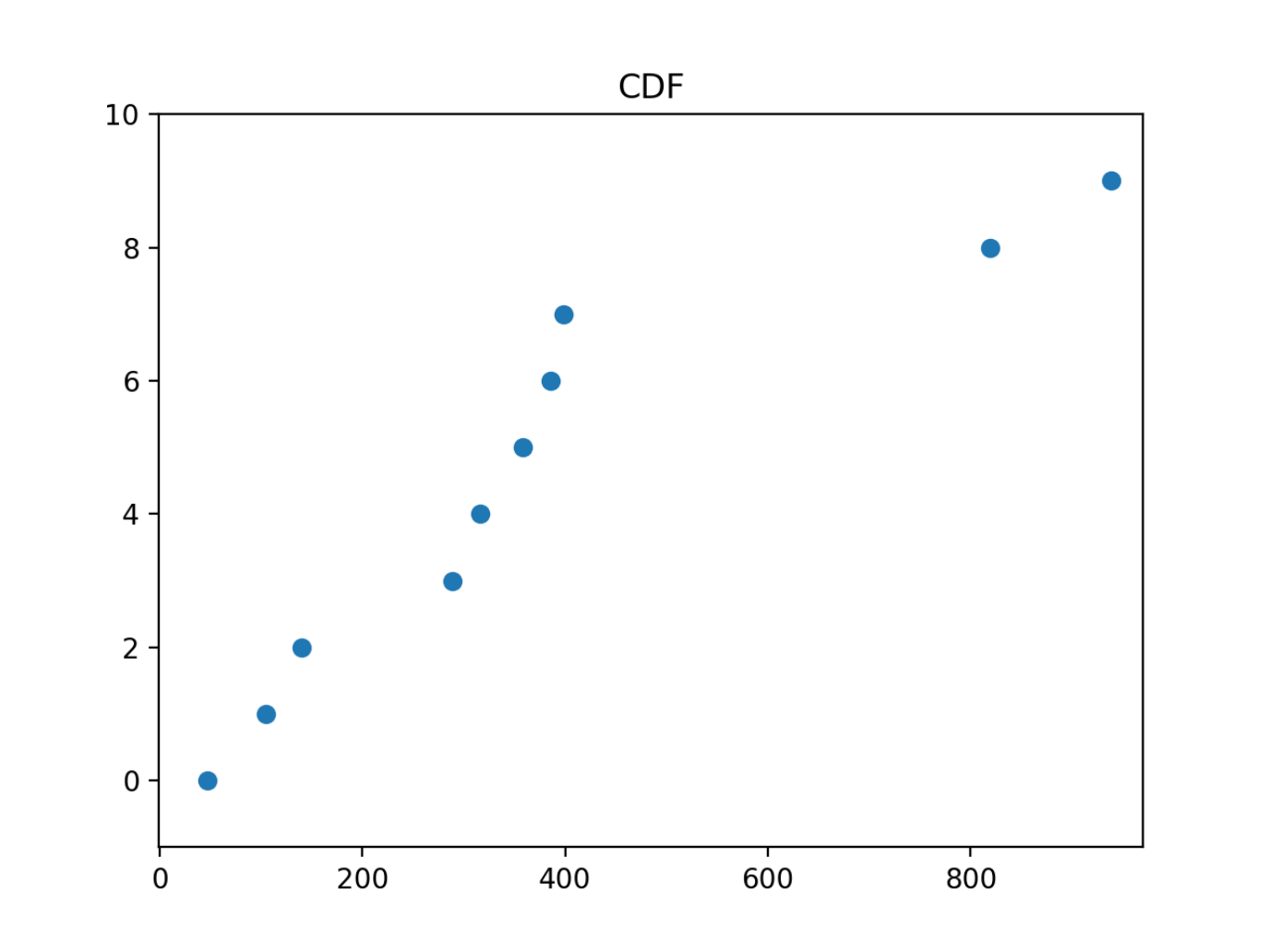}
    	\end{minipage}\hfill
    	(b)
    	\begin{minipage}{0.25\textwidth}
    		\includegraphics[width=\linewidth]{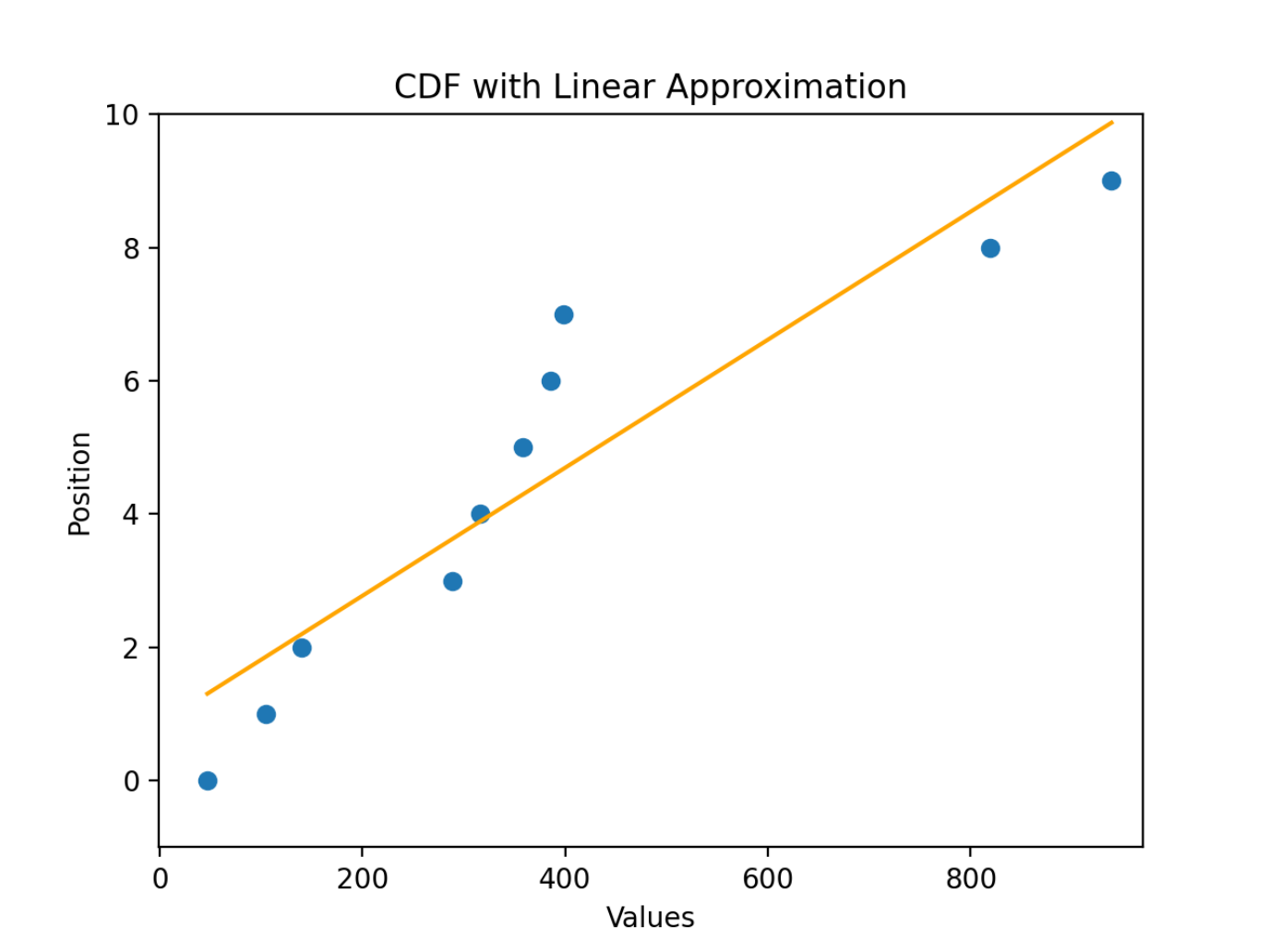}
    	\end{minipage}\hfill
    	(c)
    	\begin{minipage}{0.25\textwidth}%
    		\includegraphics[width=\linewidth]{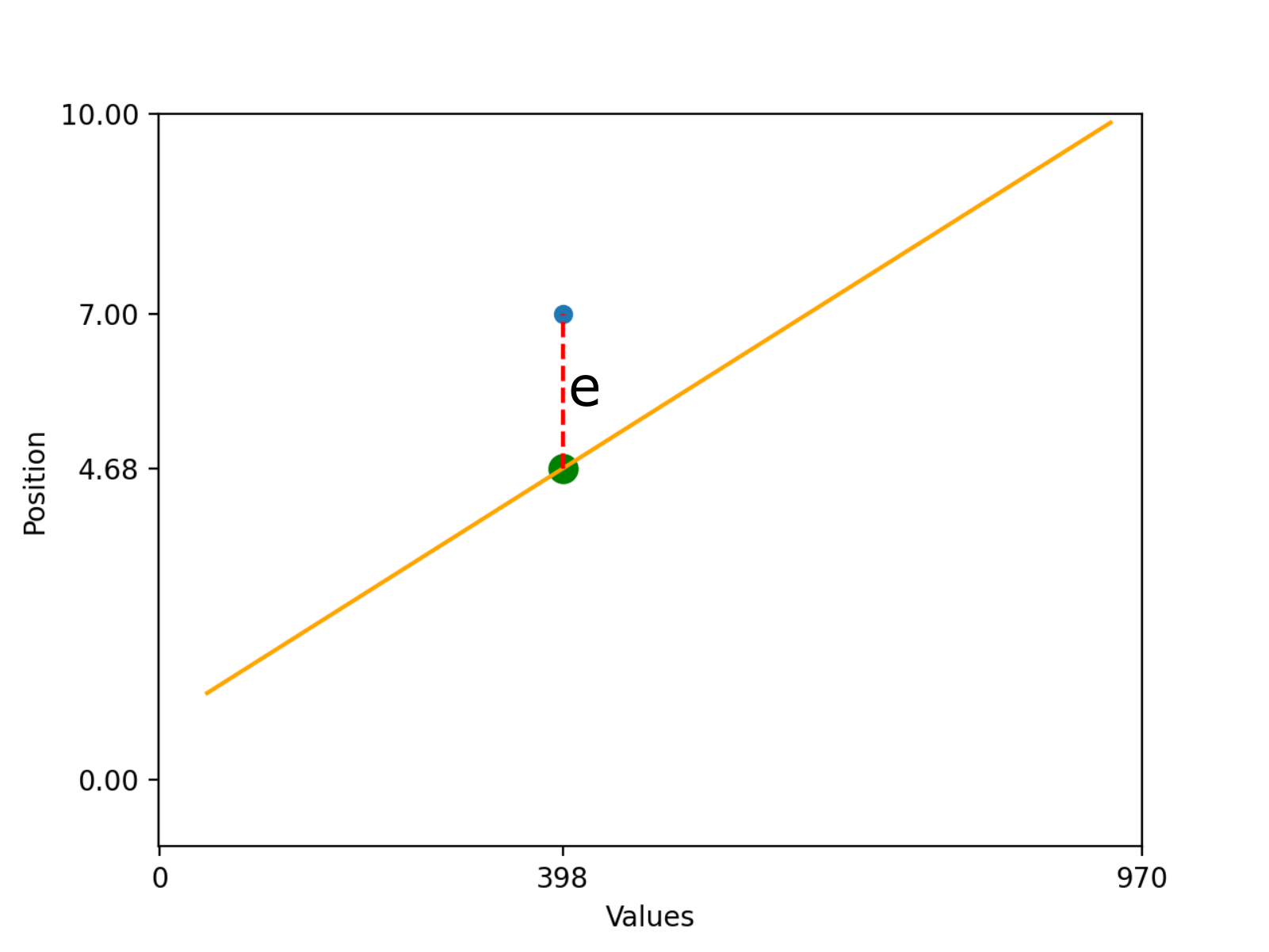}
    	\end{minipage}
    	\caption{{\bf  The Process of Learning a Simple Model via Linear Regression.} Let $A$ be $[47, 105, 140, 289, 316, 358, 386, 398, 819, 939]$. (a) The CDF of A. In  the diagram, the abscissa indicates the value of an element in the table, while the ordinate is its rank. (b) The straight line  $F(x)=ax+b$ is  obtained by determining $a$ and $b$ via Linear Regression, with Mean Square Error Minimization.  (c) The maximum error $\epsilon$ one can incur in using $F$ is also important. In this case, it is $\epsilon=3$, i.e., accounting for rounding,  it is the maximum distance between the rank of  a point in the table and its rank as predicted by $F$. In this case, the interval to search into, for a given query element $x$,  is given by $[F(x)-\epsilon, F(x)+\epsilon]$. }
    	\label{M-fig:CDF}
    \end{figure}

We now outline the basic technique that one can use to build a model for $A$. It relies on Linear Regression,  with Mean Square Error Minimization \cite{FreedmanStat}. Consider the mapping of elements in the table to their relative position within the table. Since such a function is reminiscent of the Cumulative Distribution Function over the universe $U$ of elements from which the ones in the table are drawn, as pointed out by Marcus et al. \cite{Marcus20} in their benchmarking study on Learned Indexes, we refer to it as CDF. 
With reference to the example in  Figure \ref{M-fig:CDF},  and assuming that one wants a linear model, i.e., $F(x)=ax+b$, Kraska et al. \cite{kraska18case} note that they can fit a straight line to the CDF and then use it to predict where a point $x$ may fall in terms of rank and accounting also for approximation errors. More in general, in order to perform a query, the model is consulted and an interval in which to search is returned. Then, to fix ideas,  Binary Search on that interval is performed.  
Different models may use different schemes to determine the required range, as outlined in Section  \ref{M-sec:models}. 
The reader interested in a rigorous presentation of those ideas can consult Marcus et al \cite{Marcus20}.

For this research, it is important to know how much of the table is discarded once the model makes a prediction on a query element. For instance, Binary Search, after the first test, discards $50\%$ of the table. Because of the diversity across models to determine the search interval, and in order to place all models on a par, we estimate the reduction factor of a model, i.e., the percentage of the table that is no longer considered for searching after a prediction, empirically. That is, with the use of the model and over a batch of queries, we determine the length of the interval to search into for each query. Based on it, it is immediate to compute the reduction factor for that query. Then, we take the average of those reduction factors over the entire set of queries as the reduction factor of the model for the given table.

\section{Experimental Methodology}
Our experimental set-up follows closely the one outlined in the already mentioned benchmarking study by Marcus et al. \cite{Marcus20} regarding Learned Indexes, with some variations. 
In particular, here we concentrate on the study of how different kinds of Binary and k-ary Searches can affect the performance of Learned Indexes. 
Moreover, following  Amato et al \cite{amato2021learned, amato2021lncs}, we use datasets of varying sizes in order to understand how the data structures perform on the different levels of the internal memory hierarchy.  In addition to that, we also use datasets generated as in  \cite{Morin17}, in order to establish that the Binary Search routines we use behave consistently with the findings in the mentioned paper. Details of the entire methodology are provided next. 
    
\subsection{Computer Architectures and Compilers}\label{M-sec:arch}
  
All the experiments have been performed on two different architectures, i.e,  x86 and ARM, using three different CPUs: Intel i7-8700, Intel I9-10850, and Apple M1. In the following, the specifications of the three systems used are reported. 
\begin{itemize}
 \item The i7-8700 works with a 3.2GHz clock and uses 64kb of L1 cache per core, 256kb of L2 cache per core, and 12Mb of shared L3 cache. The amount of system memory is 32 Gbyte of DDR4. The OS is Ubuntu LTS 20.04.
 \item The I9-10850 works with a 3.6 GHz clock and uses 64kb of L1 cache per core, 256kb of L2 cache per core, and 20Mb of shared L3 cache. The amount of system memory is 32 Gbyte of DDR4. The OS is Ubuntu LTS 20.04.
 \item The Apple M1 works with a 3.2Ghz clock and uses two levels of cache (L1 and L2). The amount of L1 memory cache depends on the cores, which are of two kinds. Namely, high-performance and high-efficiency cores. Moreover, in the Apple M1 architecture, the L1 memory cache is divided into a part for instructions and one for data. In our study, only one core is used, which is of the high-performance kind. As consequence, the L1 cache size is 192KB for instruction and 128 KB for data. The L2 cache is 12 Mb. The amount of system memory is 8 Gbyte DDR4. The OS is macOS Monterey 12.3.1.
\end{itemize}
         
The adopted compiler is the same for all the operating systems we have used, i.e. GCC compiler with optimization flag -O3. In order to better explain the results obtained, we use a hardware profiler on the Intel architecture, i.e. \emph{Intel Vtune}, which makes possible the extraction of several performance parameters of interest. They are detailed in Section \ref{S-sec:profiler} of the Appendix. The profiler is provided free of charge by Intel. Somewhat unfortunately, as far as the M1 architecture is concerned, we are not aware that an analogous profiler is available. More in general, for the ARM architectures such as Neon, there is a profiler that may give useful information (\emph{Arm Forge Ultimate}), but it is not free of charge. Profilers such as \emph{gperf-tool-profiler} do not give useful information. Based on such a State of the Art, for this research, no profiling for the Apple M1 architecture is performed. 
In what follows, we report results on the Intel I7 and Apple M1 architectures, while, for conciseness, results on the Intel I9 are omitted when analogous to the I7.

\subsection{Algorithms, Code and Software Platforms}
 \subsubsection{Binary Search and Its Variants}\label{M-sec:methods}
We use the standard, i.e., textbook,  Binary Search 	routine \cite{KnuthS,Cormer2009}, reported in Algorithm \ref{M-AL:S-BS} and referred to as {\bf S-BS}. We also use another version of the Binary Search strategy,  presented in \cite{KnuthS} (see also \cite{Morin17, Schulz18}) under the name Uniform. The corresponding code is provided 
in  Algorithm \ref{M-AL:U-BS} and it is referred to as {\bf U-BS}.  We point out that in all the algorithms we present unless otherwise stated, we include prefetching instructions since their use may be of advantage in terms of execution time \cite{Morin17}. However, whether or not to use them in our research is a fact that needs evaluation.  The {\bf U-BS} routine differentiates itself from the Standard one because there is no test for exit within the main loop.  Moreover,  depending on the compiler and the architecture,  instruction 8  of  Algorithm \ref{M-AL:U-BS} may be translated into a predicated instruction, which is not a branching instruction.   Indeed, the main loop of the resulting assembly code reported in {\bf Code  \ref{M-unif-i7}} has no branches. This should be contrasted with the assembly code generated on the I7 for Algorithm \ref{M-AL:S-BS}, which is reported in {\bf Code \ref{S-CD:stand-i7}} of the Appendix. Such a branch-freeness, as discussed in \cite{Morin17},  results in better use of the processor instruction pipeline. 
            
We include in this study also the {\bf lower\_bound}  routine from the standard C++ library, which is equivalent to {\bf U-BS} in terms of source code (see Algorithm \ref{S-AL:U-LB} in the Appendix), since, in the {\bf SOSD} software platform, this routine is referred to as branchy Binary Search. More precisely, it is Uniform and Branchy. 
            
We also take into consideration k-ary Search \cite{Schlegel09} routines, indicated as {\bf S-KS} and {\bf U-KS} (Algorithms \ref{S-AL:S-KS} and \ref{S-AL:U-KS} in the Appendix) in  their standard and uniform versions respectively, using $k=3$  as recommended in the work by  \cite{Schulz18}. The reason is that  k-ary Search is  superior to  {\bf U-BS} and { \bf lower\_bound} on modern computer architectures, according to results in \cite{Schulz18}.
            
Finally, we include also the Eytzinger Layout Binary Search routine ({\bf U-EL} for short, see Algorithm \ref{S-AL:U-EL} in the Appendix) because, although not directly usable within the Leaned Indexing framework,  it is a useful baseline to compare against: it is superior to {\bf U-BS} and {\bf S-BS} \cite{Morin17}. In addition to that, it has not been included by  Schulz et al. \cite{Schulz18} in their benchmarking experiments. 
            
            \begin{algorithm}
            	{
            	\caption{{\bf C++ Implementation of Standard Binary Search with Prefetching.} The version without prefetching is obtained by deleting lines 4-5.}
            	\label{M-AL:S-BS}
            	\begin{algorithmic}[1]
            		\BState int StandardBinarySearch(int *A, int x,  int left, int right)\{
            		\State  \ \ \ \ \ while (left $<$ right) \{
            		\State	\ \ \ \ \ \ \ \ \ int m = (left + right) / 2
            		\State	\ \ \ \ \ \ \ \ \  \_\_builtin\_prefetch(\&(data[lo + m / 2]), 0, 0);
            		\State	\ \ \ \ \ \ \ \ \ \_\_builtin\_prefetch(\&(data[m + m / 2]), 0, 0);
            		\State	\ \ \ \ \ \ \ \ \ if(x $<$ A[m]) rigth = m;
            		\State	\ \ \ \ \ \ \ \ \ else if( x $>$ A[m]) left = m+1;
            		\State  \ \ \ \ \ \ \ \ \ \  \ \ \ \ else return m;
            		\State \ \ \ \ \ \ \ \}
            		\State \ \ \ \ \ \ \ return right;
            		\BState \}	
            	\end{algorithmic}
            }
            \end{algorithm}

            \begin{algorithm}
            	{
            	\caption{{\bf C++ Implementation of Uniform Binary Search with Prefetching}. The code is as in   \cite{Morin17} (see also \cite{KnuthS,Schulz18}). The version without prefetching is obtained by deleteing lines 6-7.}
            	\label{M-AL:U-BS}
            	\begin{algorithmic}[1]
            		\BState int prefetchUniformBinarySearch(int *A, int x,  int left, int right)\{
            		\State \ \ \ const int *base = A;
            		\State \ \ \ int n = right;
            		\State \ \ \ while (n $>$ 1) \{
            		\State	\ \ \ \ \ const int half = n / 2;
            		\State	\ \ \ \ \ \_\_builtin\_prefetch(base + half/2, 0, 0);
            		\State	\ \ \ \ \ \_\_builtin\_prefetch(base + half + half/2, 0, 0);
            		\State	\ \ \ \ \ base = (base[half] $<$ x) ? \&base[half] : base;
            		\State	\ \ \ \ \ n -= half;
            		\State	\ \ \ \}
            		\State \ \ \ return (*base $<$ x) + base - A;
            		\BState \}	
            	\end{algorithmic}
            }
            \end{algorithm}

           \subsubsection{Index Model Classes in SOSD}\label{M-sec:models}
            From the many models available in {\bf SOSD}, we choose the ones that have been the most successful among the ones benchmarked in \cite{Marcus20}. That is, the Recursive Model Index ({\bf RMI}, for short)  \cite{kraska18case}, the Radix Spline ({\bf RS}, for short)  \cite{Kipf20} and the Piecewise Geometric Model  ({\bf PGM}, for short) \cite{Ferragina:2020pgm,FERRAGINA21}. For the convenience of the reader, a brief outline of each of those indexes  is provided in Section \ref{S-sec:LI} of the Appendix. We point out that we have modified the {\bf SOSD} library so that an implementation of a Learned Index can use one of  {\bf U-BS},  {\bf S-BS}, {\bf U-KS} or {\bf S-KS}, for the final search stage.   
            It is to be noted that each of those models can err in making a prediction. However, each of them has a mechanism to correct for such a mistake in order to return a valid interval in which to search into. The interested reader is referred to the original papers for a description of those mechanisms, which are somewhat more complex than the one we have considered in Figure  \ref{M-fig:CDF}. It is also worth recalling that those Models can only use Binary Search procedures with a sorted table layout, i.e., {\bf U-EL} cannot be used with those Models.

    \subsection{Computer Architecture and  Compilers: The Production of Branch-Free Code}\label{M-banchy}
        
        Given the Binary Search routines and their variants described in Section \ref{M-sec:methods}, it is not clear that branch-free assembly code is actually produced by the compiler. Therefore, we have inspected the assembly code generated by the compiler in each of the used architectures. The results are summarized in  Table \ref{M-branchfreeCode}. For conciseness, we report only the branch-free assembly code in Section \ref{S-sec:assembly} of the Appendix, in addition to {\bf Code \ref{S-CD:stand-i7}}. The remaining code regarding Branchy Binary Search, rather lengthy, is available upon request. Furthermore, it is to be noted that the table does not report the case for the k-ary Searches, because the program to extract the assembly code (Linux obj-dump) does not provide specific details capable to determine the presence of predicated instructions. 
        
        Interestingly, the {\bf lower\_bound} routine is translated into branchy code on the Intel architectures and in branch-free code on the Apple M1. Such a difference may be explained as follows. Although {\bf lower\_bound} is a Uniform routine, its inner loop makes explicit use of if-then-else  (lines 7-11 of Algorithm \ref{S-AL:U-LB} in the Appendix) rather than a conditional  operator (line 8 in Algorithm \ref{M-AL:U-BS}). This accounts for the difference between the two assembly codes on the Intel architectures. As for the Apple M1, being an ARM architecture, it has an extensive set of predicated instructions \cite{Morin17}, which apparently the compiler is able to use even in the presence of simple if-then-else constructs in the high-level code.  
    
        \begin{table}
        	\begin{center}
        			\caption{{\bf Branchy and Branch-free Assembly Code Production.} The first row indicates the computer architecture, while the first column the routines. For each entry, we report the kind of code produced by the compiler, i.e., Branchy or Branch-free code.}
        			\label{M-branchfreeCode}
        			\begin{tabular}{|c|c|c|c|}
        				\hline 
        				& Intel I7 & Intel I9 & M1 \\ \hline
        				S-BS & Branchy & Branchy & Branchy \\ \hline
        				U-BS & Branch-free & Branch-free & Branch-free \\ \hline
        				lower\_bound & Branchy & Branchy & Branch-free \\ \hline
        				
        			\end{tabular}
        		
        	\end{center}
        \end{table}

        \setcounter{algorithm}{0}
        \begin{algorithm}
        \floatname{algorithm}{Code}
        	\caption{{\bf Assembly Code of Uniform Binary Search on the Intel I7 (Only Main Loop)}. The predicated instruction is line 1485 (in bold). No prefetching is used.}
        	\label{M-unif-i7}
        	\begin{algorithmic}
        		\BState 0000000000001460 $<$\_Z21Uniform\_Binary\_SearchPmmmm$>$:
        		\State \ \		.
        		\State \ \		.
        		\State \ \		1473:	76 1d                	  jbe    1492 $<$\_Z21Uniform\_Binary\_SearchPmmmm+0x32$>$
        		\State \ \		1475:	0f 1f 00             	 nopl   (\%rax)
        		\State \ \		1478:	48 89 ca                mov    \%rcx,\%rdx
        		\State \ \		147b:	48 d1 ea             	 shr    \%rdx
        		\State \ \		147e:	4c 8d 04 d0            lea    (\%rax,\%rdx,8),\%r8
        		\State \ \		1482:	49 3b 30             	cmp    (\%r8),\%rsi
        		\State \ \		 {\bf 1485:	49 0f 43 c0            cmovae \%r8,\%rax}
        		\State \ \		1489:	48 29 d1             	 sub    \%rdx,\%rcx
        		\State \ \		148c:	48 83 f9 01          	cmp    \$0x1,\%rcx
        		\State \ \		1490:	77 e6                	   ja     1478 $<$\_Z21Uniform\_Binary\_SearchPmmmm+0x18$>$
        		\State \ \		.
        		\State \ \		.
        	\end{algorithmic}
        
        \end{algorithm}

    \subsection{Datasets and Index Model Training }\label{M-sec:Datasets}
      
        We have used two kinds of datasets. The first kind was generated as described in \cite{Morin17}, i.e. we generated 24 synthetic datasets with an increasing number $n$ of elements from $2^4$ to $2^{28}$, containing only odd 64-bit integers in $[1,2n+1]$. For each of these, we generated a two million query file consisting of 1 million-odd element present in the dataset of reference and 1 million even elements. 
        
        The second kind of datasets have origin  from the carefully chosen ones  in \cite{Marcus20} (and therein referred to as  {\bf amzn32}, {\bf amzn64}, {\bf face}, {\bf osm}, {\bf wiki}). They have been derived from them in \cite{amato2021learned, amato2021lncs}, in order to fit well each level of the main memory hierarchy with respect to the Intel I7 architecture. The essential point of the derivation is that, for each of the generated datasets, the CDF of the corresponding original dataset is well approximated.  
        The details are as follows, where $n$ is the number of elements in a table.
      
        \begin{enumerate}
        	\item [$\bullet$] {\bf Fitting in L1 cache: cache size 64Kb.}  Therefore, we choose $n=3.7K $. For each dataset, the table corresponding to this type is denoted with the suffix {\bf  L1}, e.g., {\bf amzn32-L1}, when needed.

        	\item [$\bullet$] {\bf Fitting in L2 cache: cache size 256Kb.} Therefore, we choose $n= 31.5K$. For each dataset, the table corresponding to this type is denoted with the suffix {\bf  L2}, when needed.

        	\item [$\bullet$] {\bf Fitting in L3 cache: cache size 8Mb.} Therefore, we choose $n=750K$. For each dataset, the table corresponding to this type is denoted with the suffix {\bf  L3}, when needed.  
        	
        	\item [$\bullet$] {\bf Fitting in PC Main Memory: memory size 32Gb.} Therefore, we choose $n=200M$, i.e., the entire dataset. For each dataset, the table corresponding to this type is denoted with the suffix  {\bf L4}.
        	
        \end{enumerate}	
        
        As for query dataset generation, for each of the tables built as described above, we extract uniformly and at random (with replacement) from the Universe $U$ a total of two million elements, 50\% of which are present and 50\% absent, in each table. 
     
        For the case of the Apple M1 architecture, we take into consideration only datasets fitting at most into the L3 cache of the Intel I7, because our system is equipped with 8 Gbyte of main memory. Indeed, when the code used in this research is executed on the Apple M1 using as input each of the full datasets, the performance degrades due to substantial swapping. 
        As for model training,  {\bf SOSD} has ten predefined Models  each for the  {\bf PGM} and {\bf RS}. For the {\bf RMI} Model family, following the Literature, we use   {\bf CDFShop} \cite{Marcus20},  which returns up to ten versions of an {\bf RMI} for a given table. The selection process is heuristic and tries to choose good models in terms of query time that use little space. The interested reader can find details in \cite{Marcus20}

        All the experiments involving the mentioned datasets are reported in full in the Appendix and in part here. The query time that we report is an average taken on a batch of two million queries executed by a search routine or a Learned Index. This is essential for Learned Indexes: a measure of a single query performance would be unreliable \cite{kipfEmail}, while the method we choose is compliant with the Literature \cite{Marcus20}. Such a limitation makes it unreliable to measure some relevant performance parameters of a Learned Index, as for instance, for each query, the amount of time spent for prediction and the amount of time spent for searching. In fact, to the best of our knowledge, none of the papers reporting on Learned Indexing provides such a breakdown. Rather they concentrate on the accuracy of a prediction. For completeness, and in terms of theoretic worst-case analysis, the prediction for the {\bf RMI}s used here takes $O(1)$ time and $O(log n)$ time for the {\bf PGM} and the {\bf RS}.

{
\section{Experiments: Searching in Constant Additional Space, With or Without SOSD}\label{M-Exp:Morin}
    The aim of this Section is to shed light on the consistency of our experimental setting with the current Literature, i.e., Khough and Morin \cite{Morin17} and Schultz et al. \cite{Schulz18}. However, the results reported here provide also useful indications regarding the use of {\bf SOSD} with Binary Search routines only. This scenario is meaningful since those routines require only constant additional space
    with respect to the table to be searched into, while the Learned Indexes may require additional space that depends on the parameters of the model. It may be a small percentage or a really large one. The interested reader can find a study in \cite{amato2021learned, amato2021lncs}. The scenario we consider is the one in which one can only use constant additional space with respect to the input table. To the best of our knowledge, it is not clear whether {\bf SOSD} is worth using and with which routine. As a further result, we get also indications on how to set up the search routines in {\bf SOSD}, when Learned Indexing is to be used.

    \subsection{Replication of The Experiments by Khough and Morin}\label{M-sec:morin}
        According to a study by Khough and Morin \cite{Morin17}, modern processor architectures are best used with branch-free Binary Search code. In order to assess to what extent those findings hold also in our experimental set-up,  we have experimented with all the routines mentioned in Section \ref{M-sec:methods}, that have been executed as \emph{stand-alone} C++ code ( as in  \cite{Morin17}) and as code included into the highly engineered {\bf SOSD} platform. We have considered all the architectures mentioned in Section \ref{M-sec:arch}, the synthetic datasets described in Section \ref{M-sec:Datasets}, with the inclusion also of the {\bf osm} dataset for completeness. 
        No prefetching is used here since the advantage of its use is discussed separately within this Section. 
        
        In regard to the {\bf lower\_bound} routine, which is the standard within {\bf SOSD} and the associated benchmarking of Learned Indexes, we have compared it with all the other routines, finding that it is inferior to all of them. The full set of experiments is available upon request and, for brevity, we only report some interesting observations regarding this important routine as compared with {\bf S-BS} and {\bf U-BS}. On the Intel processors, despite being a Uniform Binary Search, its assembly code is branchy (code omitted for brevity and available upon request). The experiments reported in Figure \ref{S-fig:I7lowerboundSOSD} of the Appendix point to the fact that, on those architectures, there is very little difference with {\bf S-BS}. On the Apple M1 architecture, its assembly code is branch-free (Code \ref{S-AL:M1-LB} in the Appendix)  but, as shown in Figure \ref{S-fig:M1lowerboundSOSD} of that  File, both {\bf S-BS} and {\bf U-BS} are better. For those reasons, we no longer consider {\bf lower\_bound} in this study. 
        
        For the remaining routines, the results are reported in Figure \ref{M-fig:Morin} on synthetic datasets and in Figure \ref{S-fig:I7M1osm} of the Appendix for the {\bf osm} dataset. We also anticipate that,  from the experiments reported and discussed in this Section, there is no substantial difference in performance between  {\bf S-KS} and {\bf U-KS}. For this reason, {\bf U-KS} is not considered in the following  Sections. 

        \paragraph{\emph{Stand-alone}}

        In this setting, on both the Intel and the ARM architectures, the results in \cite{Morin17} are confirmed (see Figures \ref{M-fig:Morin}(b) and (d) for synthetic datasets and Figures \ref{S-fig:I7M1osm}(b) and (d) of the Appendix for the {\bf osm} datasets). That is, Uniform branch-free is better than Standard branchy for datasets fitting in the cache memory. It is useful to recall that, for the ARM architecture, we use only datasets that fit in the cache memory. Moreover, there is also confirmation of the findings in  \cite{Schulz18}, stating that k-ary Search is better than both of those routines for datasets fitting in main memory. A new finding, overlooked both in \cite{Morin17} and in \cite{Schulz18}, is that {\bf U-EL} is always better than k-ary Search, on both architectures we have considered. 

        \paragraph{SOSD}

        We discuss first the Intel architectures. In this setting, we find the same results as in the \emph{stand-alone} case, with some notable differences: (a) for tables fitting in the cache memory, e.g., of size in the interval $[1,2^{22}]$, the gap between the performance of {\bf U-BS} and {\bf S-BS} is reduced and virtually unnoticeable for small tables, as shown by the two curves in Figure \ref{M-fig:Morin}(a) and (b); (b) {\bf S-KS} and {\bf U-KS} are always better than {\bf S-BS} and {\bf U-BS}. In regard to the {\bf osm} datasets, the same results are confirmed  (see   Figure \ref{S-fig:I7M1osm}(a) and (c) of the Appendix).
        Concerning the Apple M1 architecture, and in reference to Figure \ref{M-fig:Morin}(c), we find that {\bf S-BS}, {\bf U-BS} and {\bf U-EL} perform analogously, while {\bf S-KS} and {\bf U-KS} are always better.

        \paragraph{\emph{Stand-alone} Vs SOSD}

        It is also of interest to assess  whether there are differences in terms of query execution time between the \emph{stand-alone} and the {\bf SOSD} case. With reference to Figure \ref{S-fig:I7sosdnososd} of the Appendix for the Intel I7, it is to be noted that the  {\bf S-BS} and {\bf U-BS} routines perform better in their \emph{stand-alone} version with datasets fitting the memory cache. Moreover, in regard to  {\bf U-EL}, the \emph{stand-alone} version is always better than the {\bf SOSD} counterpart. Finally, the  {\bf S-KS} and {\bf U-KS} perform always better in the {\bf SOSD} case.

        \paragraph{Profiler Analysis}

        In order to gain insights into the differences outlined above, we make use of the \emph{Intel Vtune} profiler on two synthetic datasets, i.e. a small table (size $2^{10}$ elements)  and a large one (size $2^{24}$ elements). The results are reported in Tables \ref{S-T:ProfilingMorin10} and \ref{S-T:ProfilingMorin24} in the Appendix. Although the profiling does not give a definite indication regarding the superiority of one setting over the other, there are a few facts that are worth noting. 
        
        Concerning the {\bf U-EL}, all the profiler parameters get worse within {\bf SOSD}, except for the Front-end one. In particular, the Bad Speculation parameter (which indicates "wrong prediction" such as in the case of branches), gets much worse in going from  \emph{stand-alone} to {\bf SOSD}. Interestingly, the same thing happens with {\bf U-BS}. On the other hand, such a parameter decreases for {\bf S-BS}. Although no definitive conclusion can be drawn, those facts indicate that the optimizations within {\bf SOSD} seem to be oriented towards branchy code.

        \begin{figure}[tbh]
        	\begin{center}
        	(a)
        	\begin{minipage}{0.45\textwidth}
        		\includegraphics[width=\linewidth]{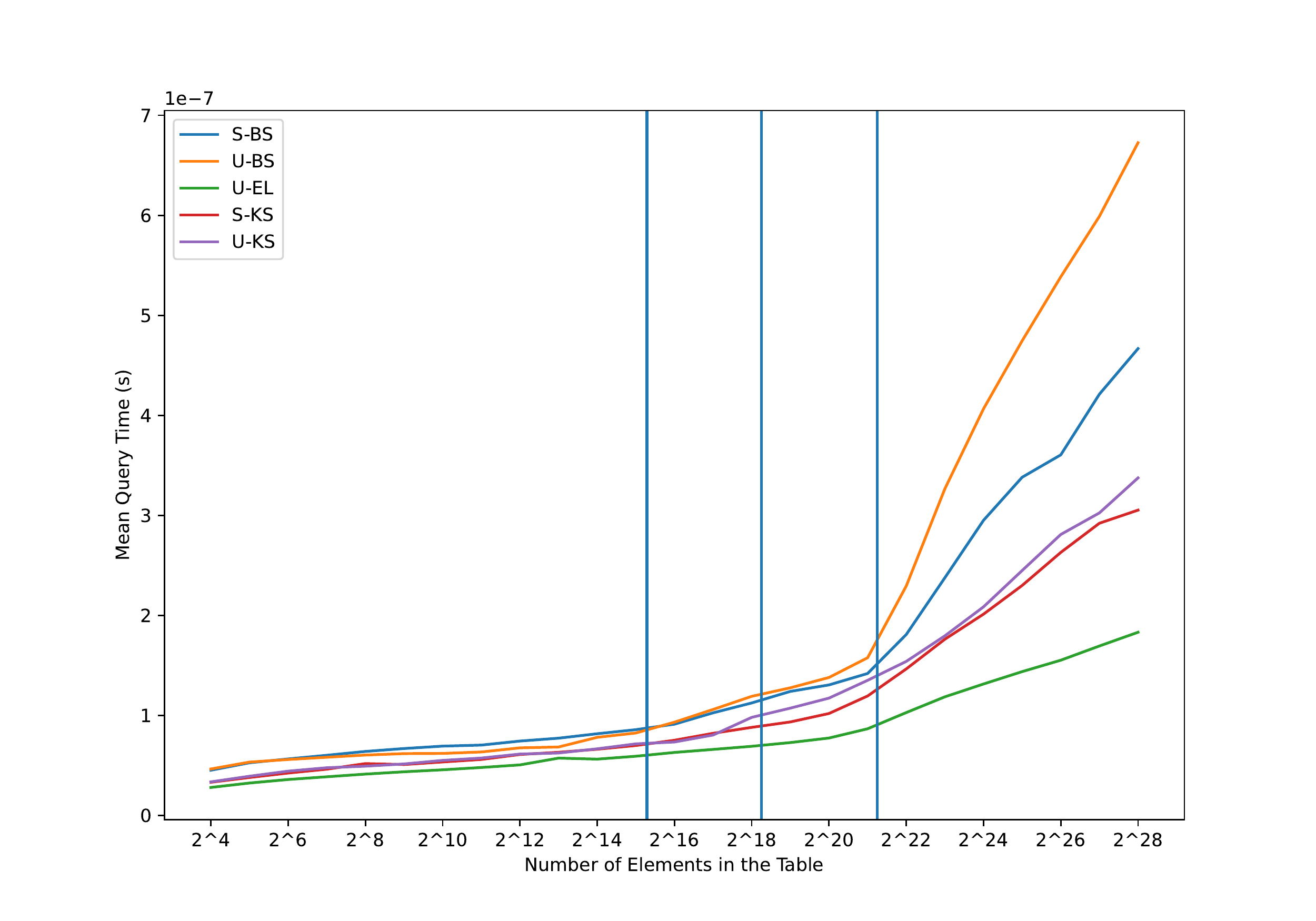}
        	\end{minipage}\hfill
        	(b)
        	\begin{minipage}{0.45\textwidth}
        		\includegraphics[width=\linewidth]{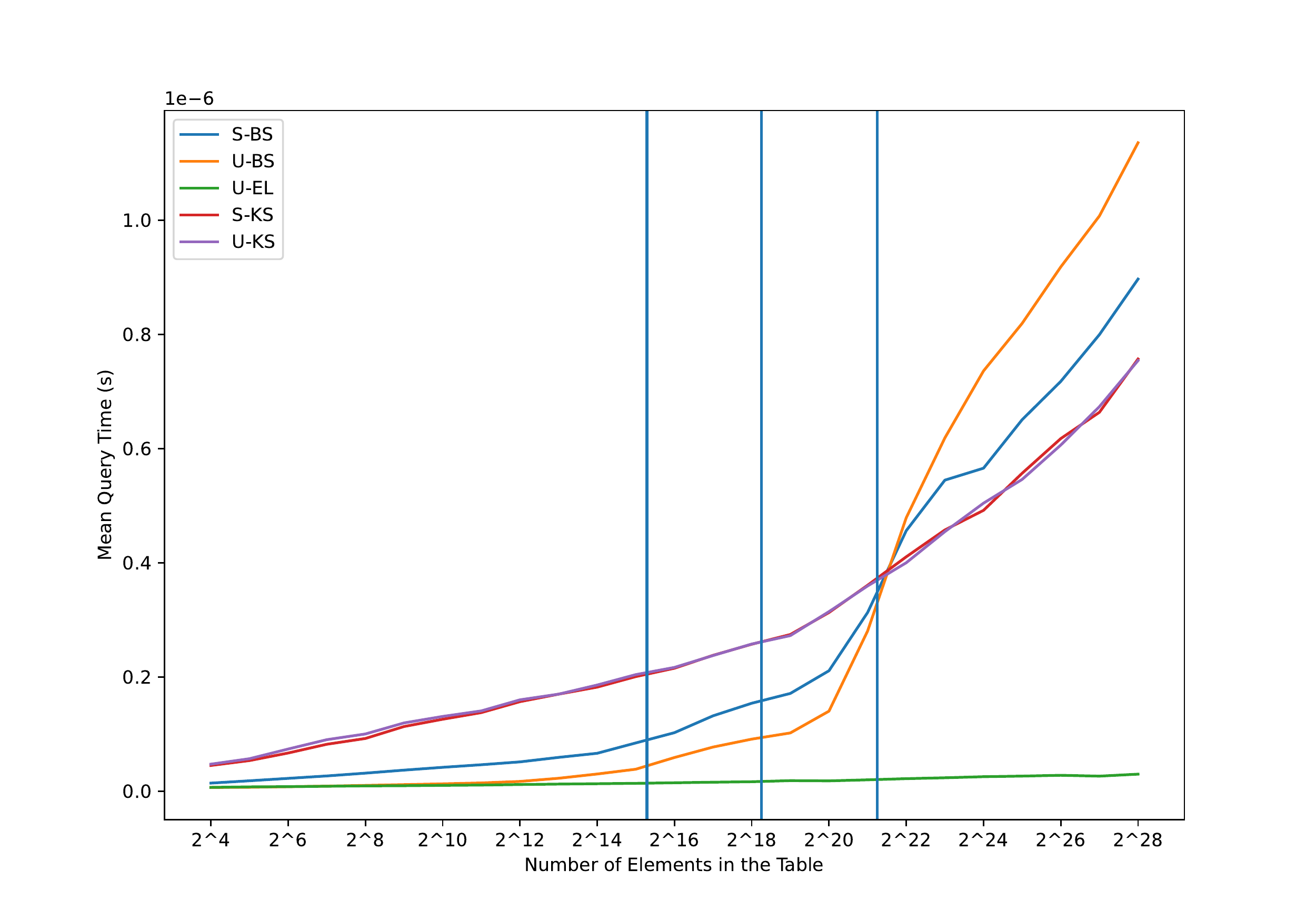}
        	\end{minipage}\hfill
        	\\
        	(c)
        	\begin{minipage}{0.45\textwidth}
        		\includegraphics[width=\linewidth]{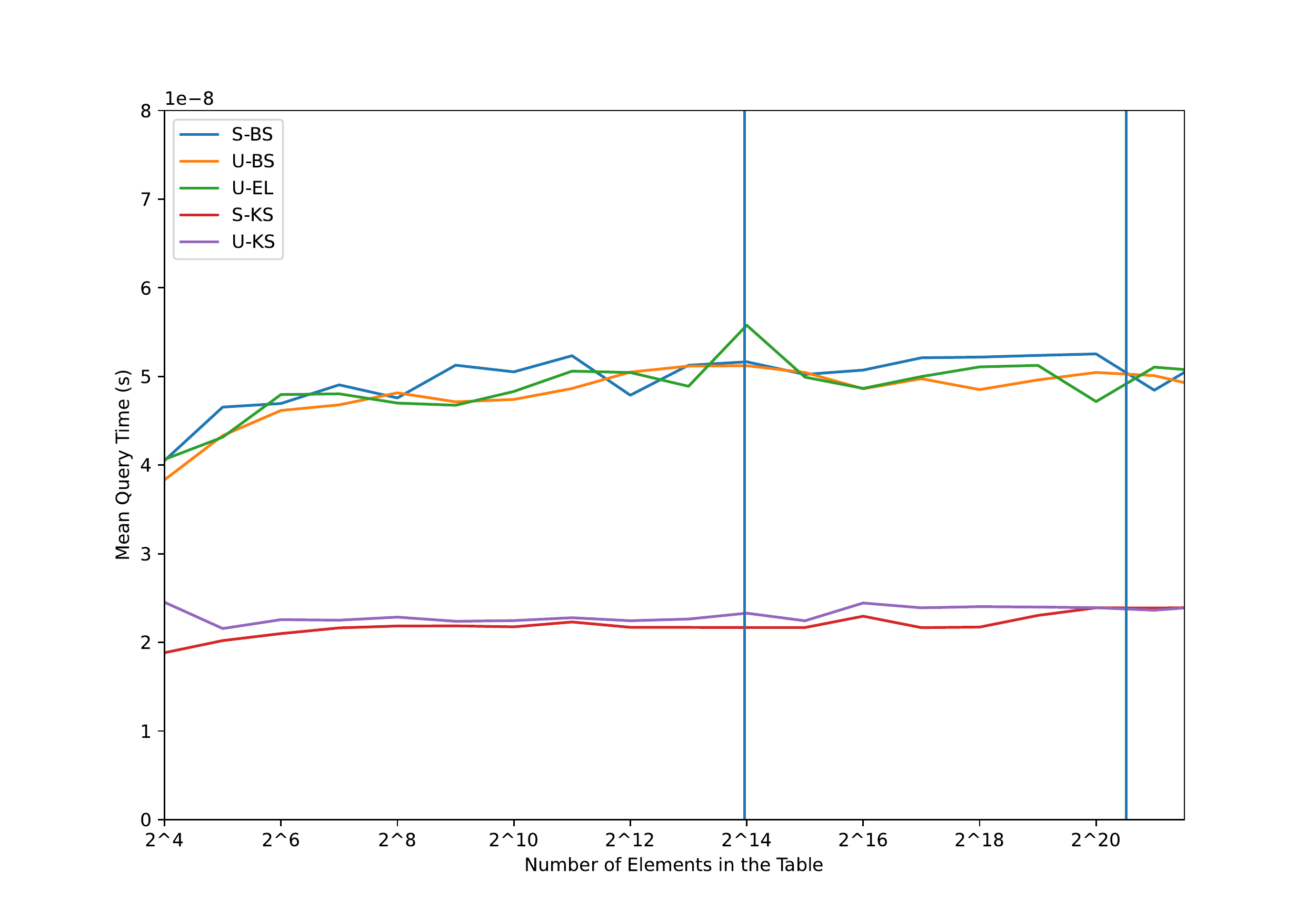}
        	\end{minipage}\hfill
        	(d)
        	\begin{minipage}{0.45\textwidth}
        		\includegraphics[width=\linewidth]{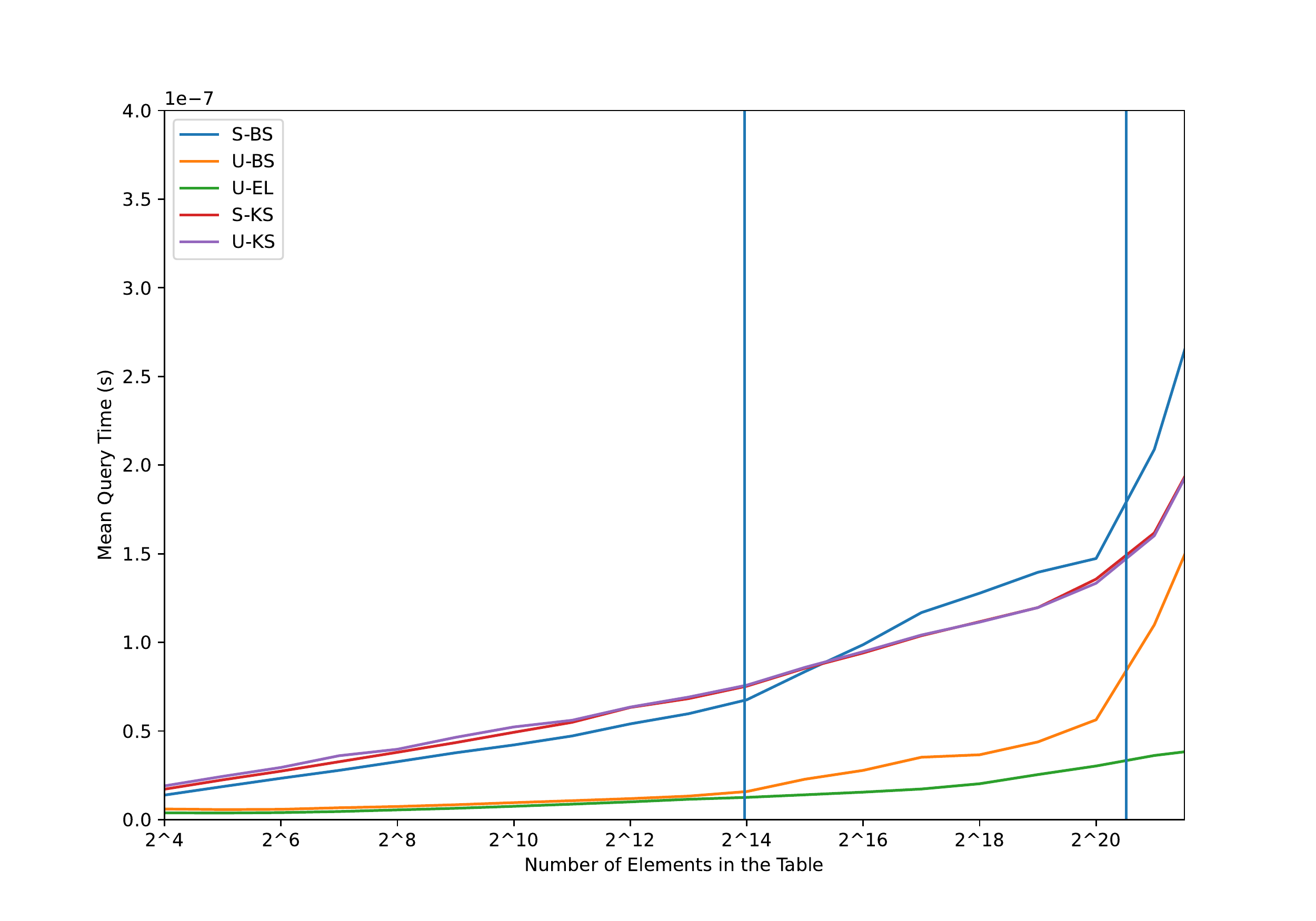}
        	\end{minipage}\hfill
            \caption{{\bf Mean Query Times of Search Methods on Synthetic Data.} Figure (a) and (c) report results using {\bf SOSD} on the Intel I7 and the Apple M1, respectively. Figures (b) and (d) reports \emph{Stand-Alone} implementation for the same architectures. The abscissa reports number of elements in the input Table, while the ordinate the mean query time in seconds. The vertical lines indicate the size of each cache memory level.}\label{M-fig:Morin}
        	\end{center}
        
        \end{figure}

    \subsection{Pros and Cons of Prefetching}\label{M-sec:pref}

        As pointed out in \cite{Morin17}, explicit prefetching can improve performance by loading blocks of data into the cache memory before they are accessed, avoiding processor stalls. However, this operation is expensive. Therefore, the advantage of using it must be carefully evaluated.
        To this end, using the same set-up of Section \ref{M-sec:morin}, we have studied whether explicit prefetching can improve the performance of the search routines included in this research. 
        
        The results are summarized in Figure \ref{M-fig:I7Prefetch} for the Intel I7 architecture and Figure \ref{M-fig:M1BBSPrefetch} for the Apple M1. We report only results for the execution within {\bf SOSD} because in the \emph{stand-alone} case they are analogous.
        
        The findings of \cite{Morin17} for the use of explicit prefetching are confirmed on the Intel I7 architectures, i.e. it is never useful for {\bf S-BS} (Figure \ref{M-fig:I7Prefetch}a), it is useful for {\bf U-BF} only for the on big-sized datasets (Figure \ref{M-fig:I7Prefetch}b), and  it is always useful for {\bf U-EL} (Figure \ref{M-fig:I7Prefetch}c). In regard to the Apple M1 architecture, prefetching is not useful across any search method. Possibly, this is due to the fact that we only use datasets fitting the cache memory.

        As for the Intel architecture, in order to get insights into those findings, we have again profiled the code, as in the previous Section. The results are reported in Section \ref{S-sec:prefetch} of the Appendix. With reference to them, it is to be noted that the profiler parameters confirm, for datasets larger than the cache memory, that the number of stalls due to data cache misses decreases with the use of explicit prefetching.

        \begin{figure}[tbh]
        	\centering
        	(a)
        	\begin{minipage}{0.45\textwidth}
        		\includegraphics[width=\linewidth]{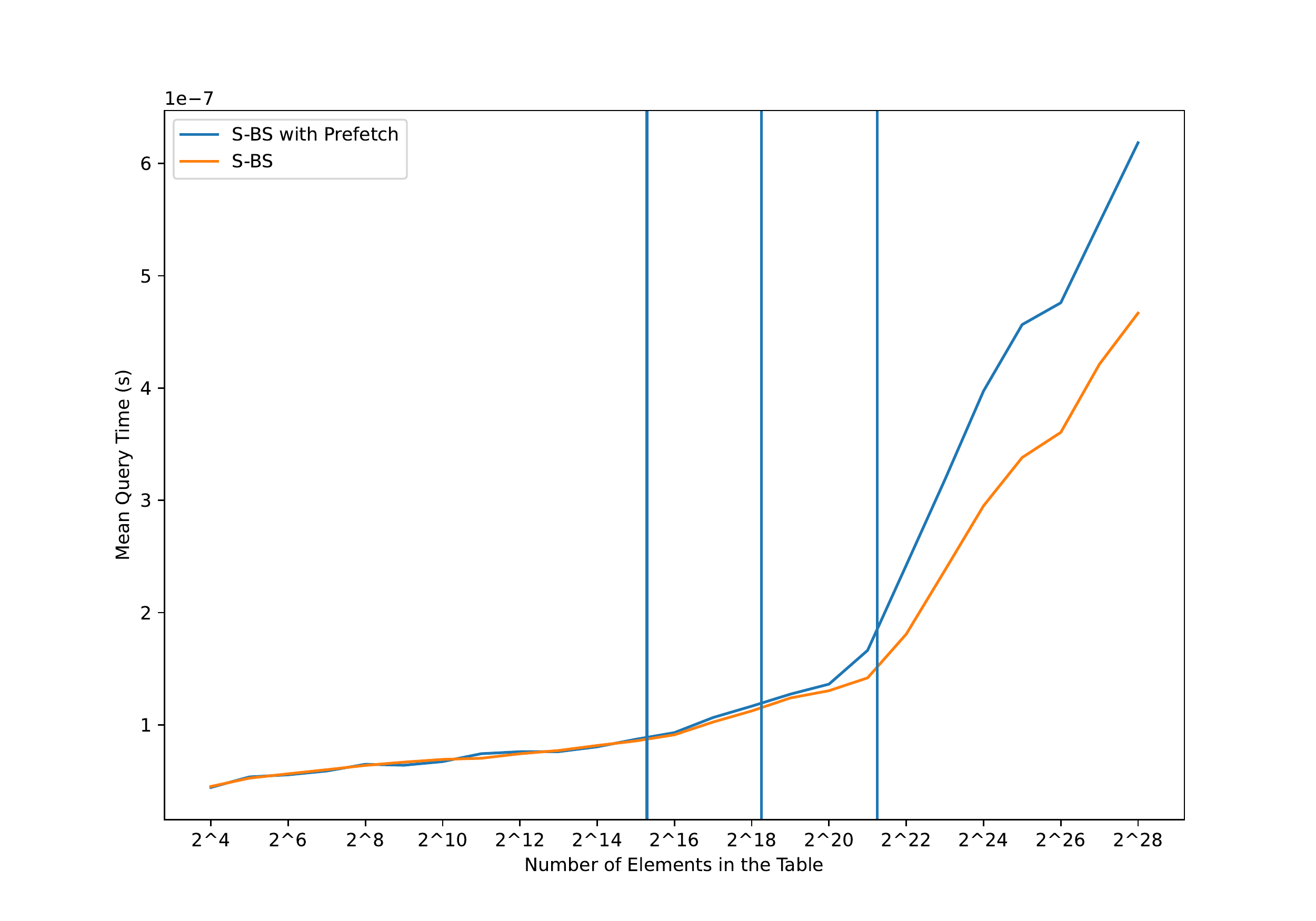}
        	\end{minipage}\hfill
        	(b)
        	\begin{minipage}{0.45\textwidth}
        		\includegraphics[width=\linewidth]{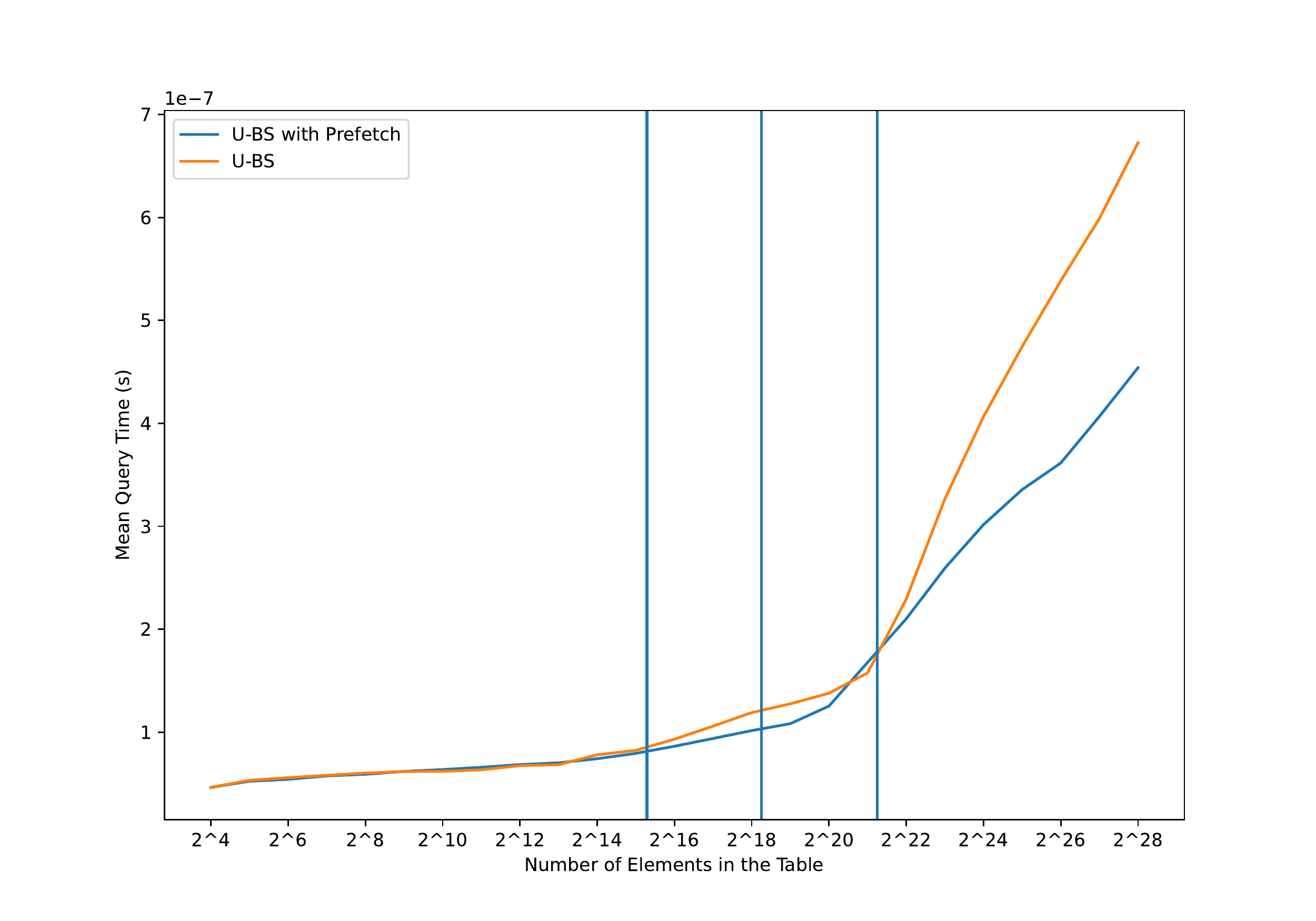}
        	\end{minipage}\hfill
        	\\(c)
        
        	\begin{minipage}{0.45\textwidth}
        		\includegraphics[width=\linewidth]{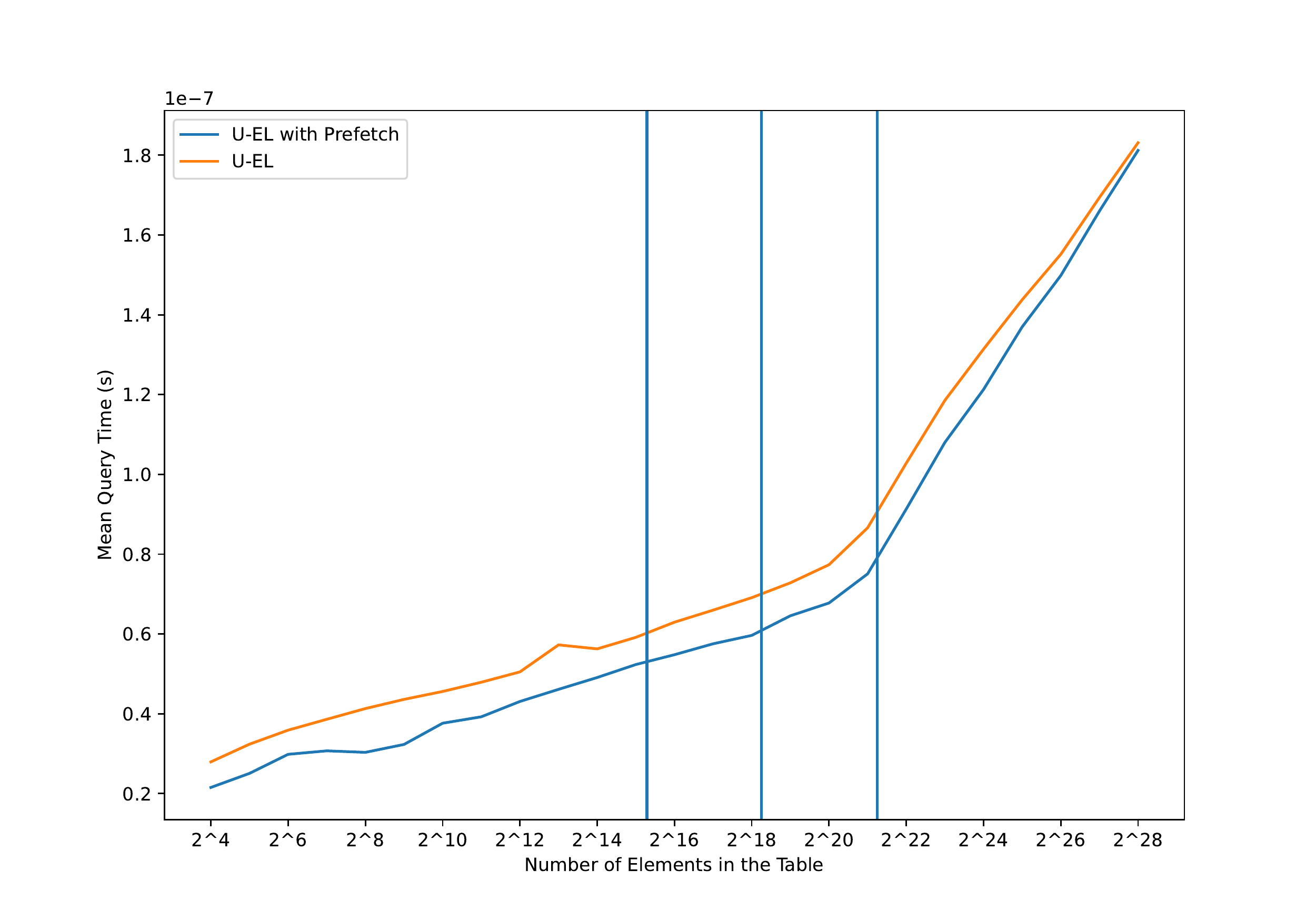}
        	\end{minipage}\hfill
        	\caption{{\bf Comparison between routines with and without explicit prefetching on SOSD using the Intel I7}. The Figures show the comparison between routines with ad without explicit prefetching. In particular, we report {\bf S-BS} in Figure (a), {\bf U-BS} in Figure (b) and {\bf U-EL} in Figure (c). On the abscissa, we report the number of elements in the table and, on the ordinates, the mean query time in seconds. The vertical lines indicate the size of each cache memory level.}
        	\label{M-fig:I7Prefetch}
        \end{figure}
        
        \begin{table}[h]
        	\begin{center}
        		\caption{{\bf Worst-case Average Predicted Search Intervals Length for L4 datasets.} For each of the largest of the benchmark datasets and each model class, we have considered the ten model instances used in {\bf SOSD}. For each, we have computed the average predicted interval length and its standard deviation over a query dataset obtained as described in Section \ref{M-sec:Datasets}. Then, we have taken the maximum average, which we report in the table for each dataset and model class, together with its standard deviation.}\label{M-T:SearchRange}
        		\begin{tabular}{|c|r|r|r|}
        			\hline 
        			& \multicolumn{1}{c|}{{\bf RMI}} & \multicolumn{1}{c|}{{\bf PGM}} &  \multicolumn{1}{c|}{{\bf RS}}\\
        			\hline
        			amzn32 & 3.44e5 $\pm$ 1.18e6 & 4.10e3 $\pm$ 2.37e3 & 2.71e2 $\pm$ 1.58e2 \\ 
        			\hline
        			amzn64 & 6.19e4 $\pm$ 5.88e5 & 2.05e3 $\pm$ 2.37e3 & 1.35e2 $\pm$ 1.57e2 \\ 
        			\hline
        			face & 8.59e4 $\pm$ 4.19e5 & 4.10e3 $\pm$ 2.37e3 & 5.31e2 $\pm$ 3.08e2 \\ 
        			\hline
        			osm & 5.26e6 $\pm$ 2.73e6 & 2.05e3 $\pm$ 2.37e3 & 3.65e2 $\pm$ 4.23e2 \\ 
        			\hline
        			wiki & 4.25e5 $\pm$ 2.61e6 & 2.05e3 $\pm$ 2.37e3 & 2.50e2 $\pm$ 2.90e2\\ 
        			\hline
        		\end{tabular}
        		
        	\end{center}
        \end{table}
        
        \begin{figure}
        	\center
        	\includegraphics[width=0.7\textwidth]{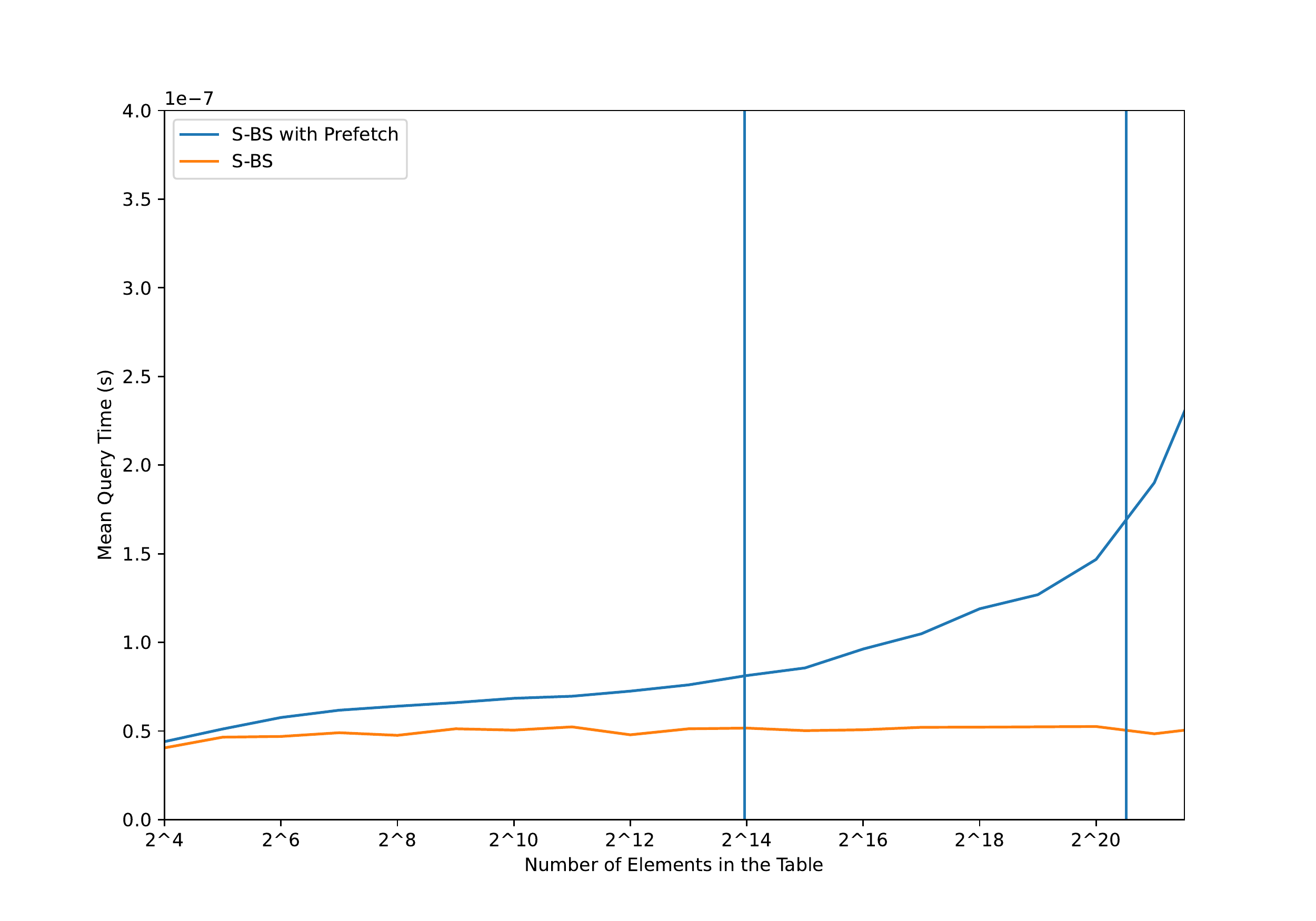}
        	\caption{{\bf Comparison between S-BS with and without explicit prefetching on SOSD using the Apple M1}. The Figure shows the comparison between {\bf S-BS} with ad without explicit prefetching. The results for {\bf U-BS} and {\bf U-EL} are the same. On the abscissa, we report the number of elements in the table and, on the ordinates, the mean query time in seconds. The vertical lines indicate the size of each cache memory level.}\label{M-fig:M1BBSPrefetch}
        \end{figure}

    \subsection{A Summary of Useful Indications}\label{M-sec:indication}

        Based on the experiments reported so far, we get some useful indications regarding the use of Binary Search routines that account also for the {\bf SOSD} platform.
        
        \begin{itemize}
            \item {\bf No Learned Indexing}. This scenario indicated the constraint of constant additional space with respect to the table to be searched into. The indication is to use {\bf U-EL} as a \emph{stand alone} routine, with prefetching on the Intel architectures and without it on the M1 architecture.  For this latter, the table must fit in the cache memory, to avoid swapping.

            \item {\bf Learned Indexing}. With the exclusion of {\bf U-EL} that cannot be used by the current models, the experiments conducted so far provide the indication that prefetching is not really needed for routines that complete the search of a Learned Index.  As far as the Apple M1 architecture is concerned, the fact that prefetching is of no advantage is evident, since we use only dataset fitting in the cache memory. As for the Intel architectures,  the scenario is more complex. Indeed, {\bf S-BS} never takes advantage of explicit prefetching, while {\bf U-BS} achieves a substantial improvement with its use for datasets larger than the cache memory, i.e. datasets with a number of elements greater than about $4.19e6$ ( see Figure \ref{M-fig:I7Prefetch}(b) again). That is, the predicted interval of a model must be of length of at least $4.19e6$, in order to consider prefetching in the final search routine of that model.  Now, it is worth recalling that, for models that are effective, large reduction factors are expected that, in turn, correspond to small tables to be searched into. Unless the original table is really large or the model particularly bad, the case of predicted interval lengths where prefetching can be useful with  {\bf U-BS} seems unlikely, as the experiment reported in Table \ref{M-T:SearchRange} seems to indicate. As evident from that table, only in the case of the worst {\bf RMI} on the {\bf osm-L4} dataset,
            there may be some marginal benefit in using prefetching with {\bf U-BS}. Therefore, in what follows, explicit prefetching is not used. 
        \end{itemize}

\section{Experiments: Searching using Learned Indexes, With or Without SOSD}\label{M-Exp:Learned_Index}

    In order to better describe the experimental work presented in this Section, it is useful to recall that {\bf SOSD} has been designed to provide an environment in which to evaluate the relative merits of existing and possibly future, Learned Index Models. 
    Given a dataset, how models are trained is briefly described in Section \ref{M-sec:Datasets}.  Although originally designed for benchmarking, {\bf SOSD} can also be used to identify among the models it has available for a given dataset, the best performing one. The list of those Models is reported in Table \ref{S-T:BestModels} of the Appendix. Here we focus on average query time, although space may be of importance also \cite{amato2021learned, amato2021lncs}. 
    For a given dataset, this amounts to up to thirty different Models to choose from. The routine used for the final search stage is, by default, the {\bf lower\_bound} routine. Here we consider three search routines for the final stage, i.e., {\bf U-BS, S-BS} and {\bf S-KS}. As mentioned earlier, the {\bf lower\_bound} routine has been excluded since it is redundant with respect to the other ones. In summary, for a given dataset, one has up to ninety possible Model configurations to choose from. Once such a Model configuration has been identified, and in view of the results reported in Section \ref{M-sec:morin}, it is not clear whether the actual deployment of the selected Model in an Application Domain, e.g., Data Bases \cite{rao1999cache} or Search Engines \cite{Morin17},   it should be executed within the {\bf SOSD} platform or as a \emph{stand-alone} software program. As a matter of fact,  such a point has not been addressed in the Literature. The main goal of this part of our experimental work is to investigate the mentioned aspects, which also provide useful indications on which search routine to use.
    
    \paragraph{Selecting a Best Model and Associated Search Routine via SOSD}

    For each dataset described in Section \ref{M-sec:Datasets},  we consider up to ninety possible Model configurations. Then, for each model class and search routine so obtained, the Model with the best mean query time is chosen, based on its execution within  {\bf SOSD}. 
    The results are reported in Figures \ref{M-fig:osmI7I9M1}(a) and (c) only for the {\bf osm} dataset, while the remaining ones are reported in Figures \ref{S-fig:amzn32I7I9M1}-\ref{S-fig:wikiI7I9M1} of the Appendix. From those Figures, we can extract the following findings regarding the search routines, for which we also provide a justification.
    
    \begin{itemize}
        \item {\bf S-KS is the best}. It is self-evident from the results reported in the mentioned Figures. Quite remarkably, they are consistent across all datasets, memory levels and architectures we have considered. This is a novel finding in this area, since only the {\bf lower\_bound} or the {\bf S-BS} routine have been used for the terminal stage of searching in a Learned Index, as well documented in experimental studies prior to this one (see \cite{kraska18case} and \cite{Ferragina:2020pgm}). It is worth noting that the results reported here are coherent with the ones of the previous Section, in which we have evaluated {\bf S-KS} as a generic Binary Search routine rather than as a terminal to a Learned Index. In order to explain such a coherence, we need to point out that the models resulted to be the best with {\bf S-KS} provide a quite small predicted interval, on average: for instance on the {\bf osm\_L4} dataset and for the {\bf RMI} model, the average predicted interval length is 4.41e+2 with a standard deviation of 2.73e+3. The full data regarding this point is not shown and is available upon request. For small tables, as far as the Apple M1 architecture is concerned and when the search routines are performed within {\bf SOSD} without a prediction phase, Figure \ref{M-fig:Morin}(b) and Figure \ref{S-fig:I7M1osm}(b) of the Appendix report a wide margin in average query time between {\bf S-KS} and the other routines, which apparently turns out to be preserved also in the Learned Indexing framework executed within {\bf SOSD}. As for the Intel architectures, to better highlight the performance of such routines on small tables, we need to \vir{zoom-in} in Figure \ref{M-fig:Morin}(a) and Figure \ref{S-fig:I7M1osm}(a) of the Appendix. The corresponding Figures are \ref{S-fig:I7MorinOsmZoom}(a) and (b) in the Appendix. It is evident that also in this case {\bf S-KS} is always better than the other routines on small tables. It is worth noting that, compared to the results on the Apple M1, the margin in average query time between {\bf S-KS} and the other routines is smaller, which is reflected in a smaller margin also in the Learned Indexing framework, executed within {\bf SOSD}.
        A more refined analysis regarding such a coherence behaviour within {\bf SOSD} would require individually measure prediction and search time for each query: as already mentioned in Section \ref{M-sec:Datasets}, this would yield unreliable results.
        
        \item {\bf U-BS and S-BS}. In contrast to the case of {\bf S-KS}, the results for {\bf S-BS} and {\bf U-BS} indicate that there is no clear winner between those two procedures. Indeed, which procedures to pick seems to depend, for both architectures, on the memory level in which the input table fits, the input table itself and the Model we are using. Those latter results do not contradict the ones described in Section \ref{M-sec:morin}. In fact, as already mentioned in the discussion regarding {\bf S-KS}, the best Models have a quite small average predicted interval length and, for small tables, the performance of {\bf S-BS} and {\bf U-BS} within {\bf SOSD} is quite close. 
    \end{itemize}

    \begin{figure}[tbh]
    	\begin{center}
    	(a)
    	\begin{minipage}{0.45\textwidth}
    		\includegraphics[width=\linewidth]{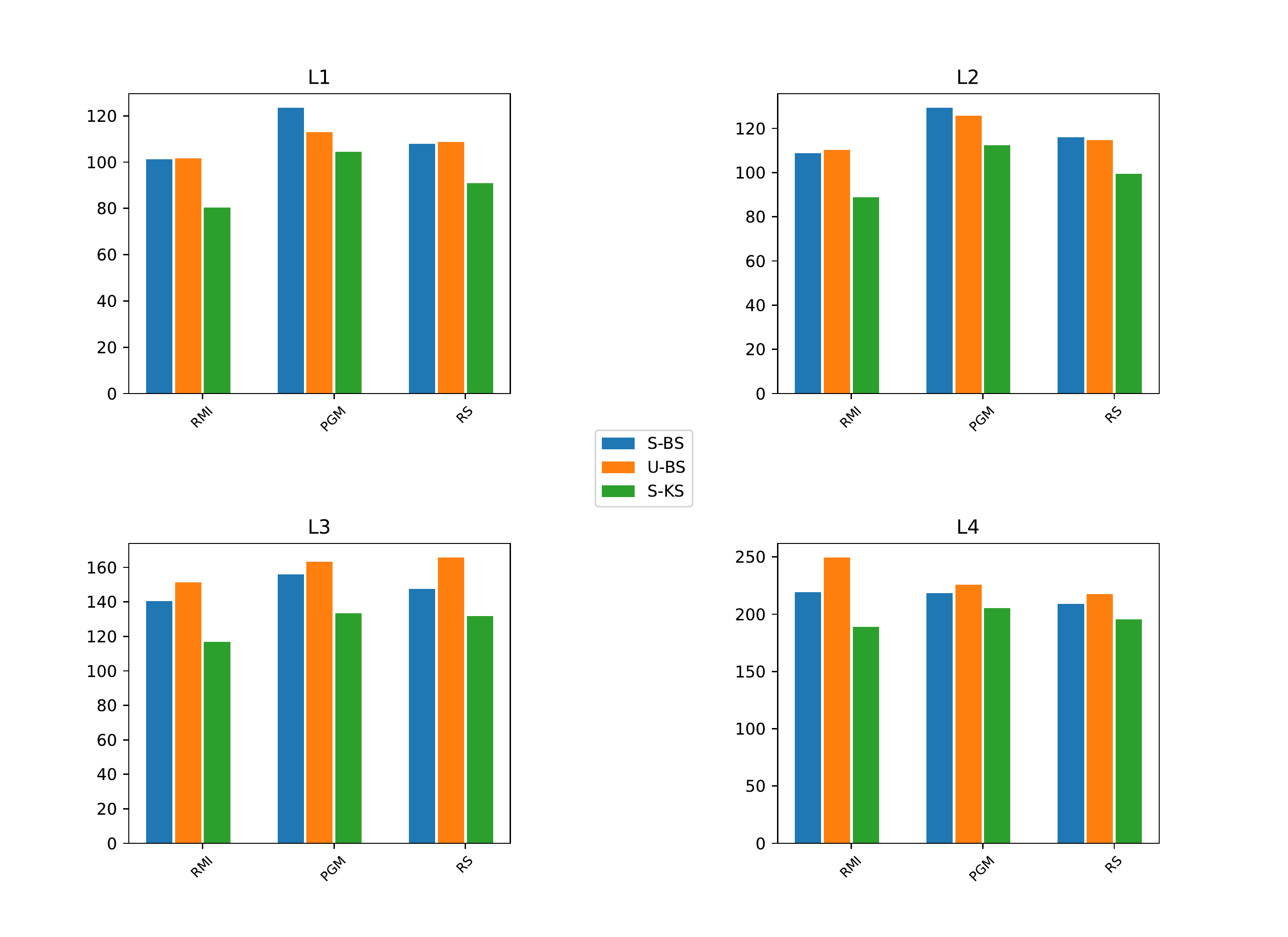}
    	\end{minipage}\hfill
    	(b)
    	\begin{minipage}{0.45\textwidth}
    		\includegraphics[width=\linewidth]{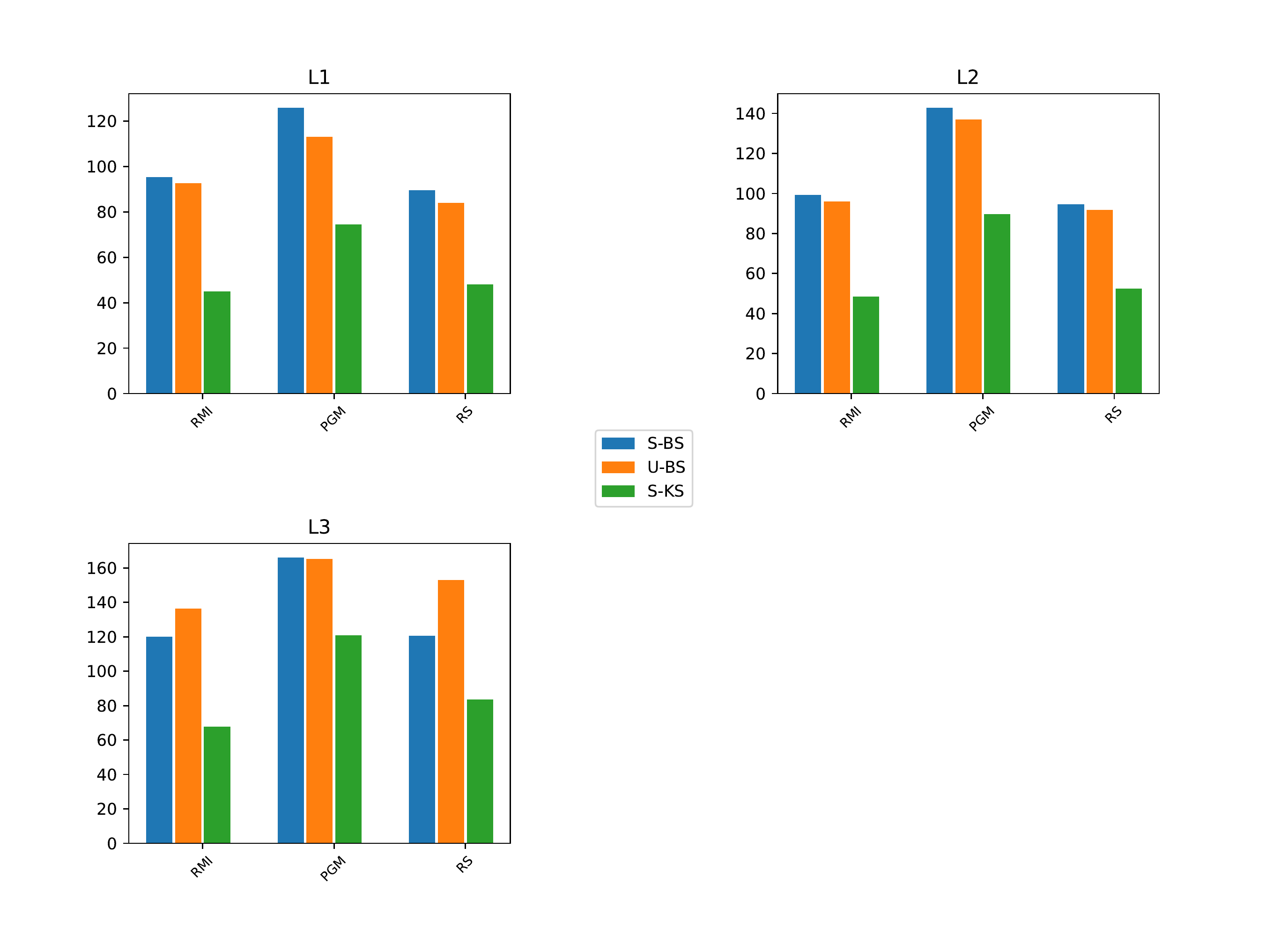}
    	\end{minipage}\hfill
        \caption{{\bf SOSD Mean Query Times of Best Learned Indexes on the osm Dataset}. Figure (a) and (b) report results using {\bf SOSD} on the Intel I7 and Apple M1, respectively. For each model class, we report the mean query time of the best Learned Indexes adopting in their last stage the routines described in Section \ref{M-sec:methods}. In particular, the blue bar is {\bf S-BS}, the orange bar is {\bf U-BS} and the green bar is {\bf S-KS}.}\label{M-fig:osmI7I9M1}
    	\end{center}
    
    \end{figure}
    
    \paragraph{Convenience of the Execution of the Best Model within SOSD or \emph{Stand-alone}}
    
    For each experiment described above, we select the best performing Model. As anticipated, we execute each of those Models in a \emph{stand-alone} setting, with the routines {\bf S-BS}, {\bf U-BS} and {\bf S-KS} as terminal for the search stage. The results are reported in Figure \ref{M-fig:allI7M1best}. It is evident that the convenience of using {\bf SOSD} with the selected Model rather than in a \emph{stand-alone} setting is architecture-dependent. In particular, the experiments indicate that it is advisable to use {\bf SOSD}, with {\bf S-KS} as a terminal for the search stage, on the Apple M1 architecture. As for the Intel I7 architecture, we find that, for the largest datasets (level L4), it is advisable to use {\bf SOSD}, again with {\bf S-KS} as a terminal for the search stage. On the other memory levels, the performance is dataset-dependent. In particular, {\bf SOSD} is better than the \emph{stand-alone} settings for {\bf face} and {\bf wiki} datasets and worse for the remaining ones. As far as {\bf S-BS} and {\bf U-BS} is concerned, in \emph{stand-alone} setting {\bf U-BS} seems to be the method of choice. This is in agreement with the result reported in Section \ref{M-sec:morin} (\emph{Stand-alone} setting).
    
    All of the above provides factual indications regarding the use of {\bf SOSD}. As for their justification, a profiler analysis would be required for the Apple M1 architecture. Unfortunately, as already stated, this is not possible at this time with free software. As for the Intel architecture, the fact that {\bf SOSD} is to be used on large datasets, may be attributed to the design and use of {\bf SOSD}: the vast majority of the experiments conducted with {\bf SOSD} are on large datasets. Moreover, as pointed out in Section \ref{M-sec:morin}(Profiler Analysis), it seems to favour branchy code.

    \begin{figure}[tbh]
    	\begin{center}
    	(a)
    	\begin{minipage}{0.45\textwidth}
    		\includegraphics[width=\linewidth]{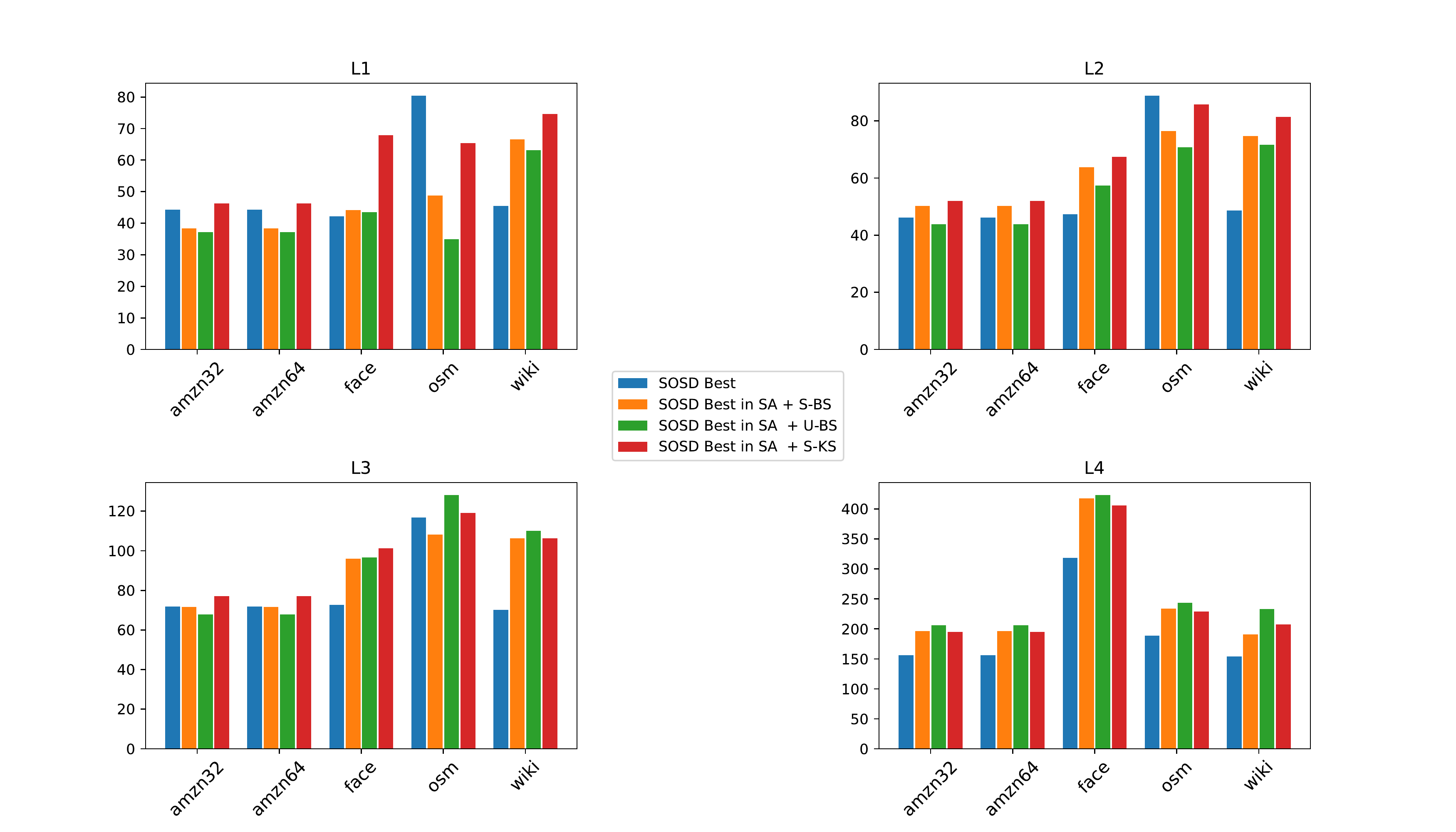}
    	\end{minipage}\hfill
    	(b)
    	\begin{minipage}{0.45\textwidth}
    		\includegraphics[width=\linewidth]{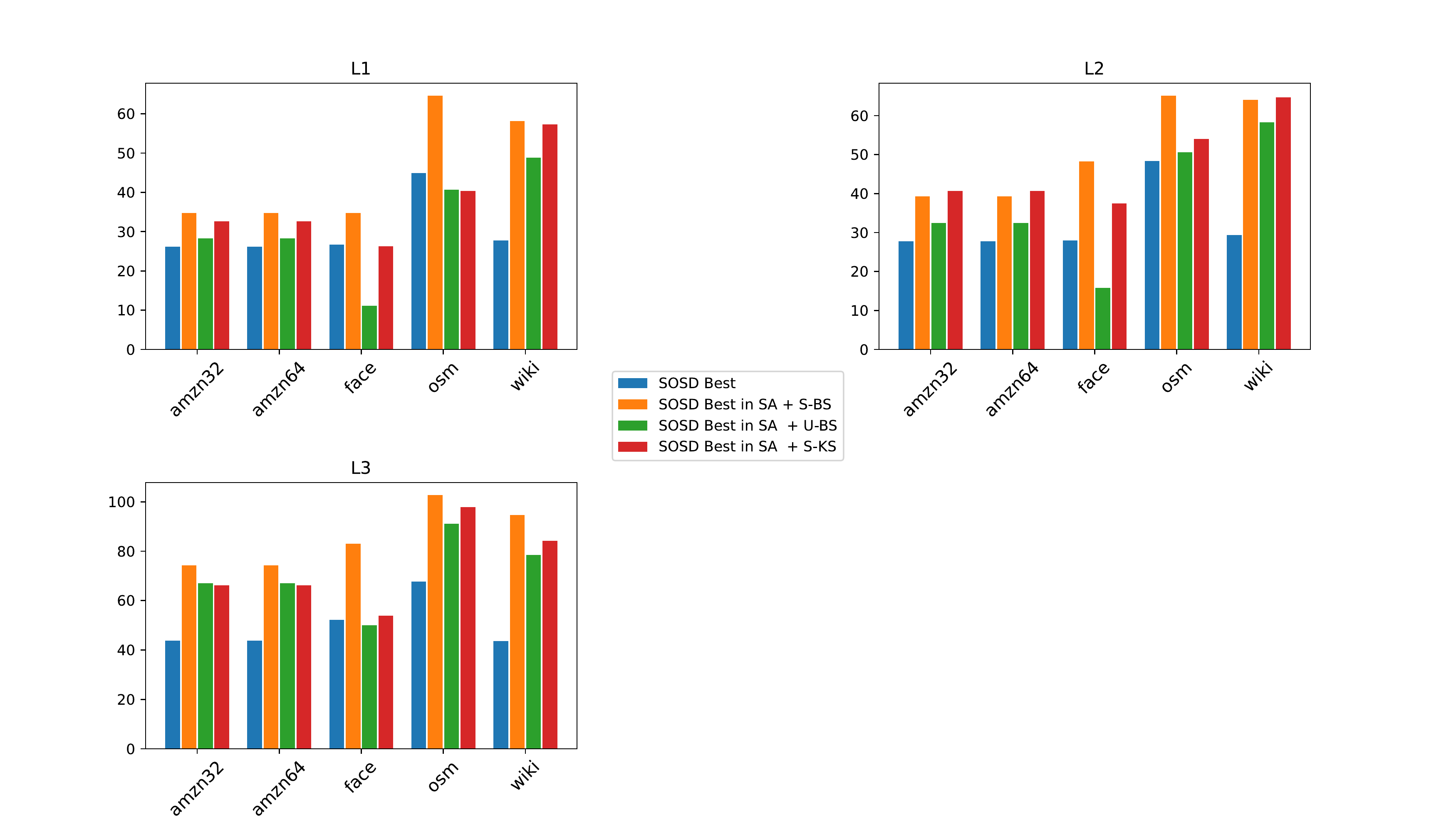}
        \end{minipage}\hfill
       \caption{{\bf Comparison Between the Best Model Indicated by SOSD and its \emph{Stand-alone} counterpart}. Figures (a) and (b) report results for the Intel I7 and the Apple M1, respectively. Each group bar is relative to a dataset, while on the ordinate we report the mean query times in nanoseconds. In each group, the blue bar indicates the best Model selected by {\bf SOSD} (also named {\bf SOSD} Best),while the next bars are the same Model when used in \emph{stand-alone} configuration (indicated as {\bf SOSD} Best in SA) together with {\bf S-BS} (orange bar), {\bf U-BS} (green bar) and {\bf S-KS} (red bar) as terminal stage routines.}\label{M-fig:allI7M1best}
       \end{center}
    
    \end{figure}

\section{Conclusions and Future Directions}

The main question we have addressed in this research is to provide indications on how the findings of Khough and Morin \cite{Morin17} and Schulz et al. \cite{Schulz18}, regarding the choice of which Binary Search routines or variants are to be used on modern computer architectures, can also be extended to the novel field of Learned Indexing, considering those routines for the final search stage. So far, for that stage, only the {\bf lower\_bound} routine has been considered. A summary of our results that can be useful both to designers and users of Learned Indexes is the following.

\begin{itemize}
    \item When no additional space with respect to the input table can be afforded, {\bf U-EL} is the best choice both in {\bf SOSD} and \emph{stand-alone} settings. This result confirms the findings by Khough and Morin \cite{Morin17} and extends them since we consider also k-ary Search.
    \item When Learned Indexing is to be used, for each model class considered in this research, {\bf SOSD} returns the best models with {\bf S-KS} as the terminal search stage. This fact holds for both the hardware architectures we have considered.
    \item When the choice between {\bf SOSD} and \emph{stand-alone} settings needs to be made to allow deployment in an Application Domain, such as  Databases \cite{rao1999cache} or Search Engines \cite{Morin17}, 
    several factors need to be considered.
    \begin{itemize}
        \item On the Apple M1 architecture, {\bf SOSD} with {\bf S-KS} as the final search stage is to be preferred.
        \item On the Intel I7 architecture, the choice depends on datasets and memory levels. Indeed, for datasets larger than cache memory, {\bf SOSD} with again as final search stage  {\bf S-KS} is to be preferred, while, for the ones fitting the cache memory, such a choice is input data-dependent. In any case, for Learned Indexes to be executed in a \emph{stand-alone} setting, {\bf U-BS} as the final search stage seems to be the more convenient choice.
    \end{itemize}

    Among the many open problems that the new area of Learned Indexing poses, we mention two that are relevant to the research we have conducted. 
    Although the Eyzinger Layout Binary Search is superior to the other routines we have considered, it cannot be used by the Index Models known so far. So, it would be very interesting to devise new models that can use such an excellent layout in the final search stage.  The second problem is in relation to the extension of our study to Dynamic Learned Indexes, i.e., the {\bf PGM} and {\bf Alex} \cite{Ding20}. Although an extension to the {\bf PGM} may be simple (such a Model uses a Binary Search routine as the final search stage, even in the Dynamic setting), it is not clear how to intervene in {\bf Alex}, given the high level of engineering that has been deployed for the realization of that Model. 
\end{itemize}

\bibliographystyle{plain}
\bibliography{references.bib}

\begin{thebibliography}{10}

\bibitem{gitbvb}
https://github.com/globosco/A-modified-SOSD-platform-for-benchmarking-Branchy-vs-Branch-free-search-algorithms.git.

\bibitem{gitbvbStand}
https://github.com/globosco/Stand-Alone-Platform-for-benchmarking-Branchy-vs-Branch-free-search-algorithms.

\bibitem{gitanonym}
https://anonymous.4open.science/r/DatasetsAndSupplementaryMaterial-E534/README.md.

\bibitem{amato2021learned}
D.~Amato, G.~{Lo Bosco}, and R.~Giancarlo.
\newblock Learned sorted table search and static indexes in small model space.
\newblock {\em CoRR}, abs/2107.09480, 2021.

\bibitem{amato2021lncs}
D.~Amato, G.~{Lo Bosco}, and R.~Giancarlo.
\newblock Learned sorted table search and static indexes in small model space
  ({E}xtended {A}bstract).
\newblock In {\em AIxIA 2021 -- Advances in Artificial Intelligence}, Cham,
  2022. AIxIA.

\bibitem{Ao11}
N.~Ao, F.~Zhang, D.~Wu, D.~S. Stones, G.~Wang, X.~Liu, J.~Liu, and S.~Lin.
\newblock Efficient parallel lists intersection and index compression
  algorithms using graphics processing units.
\newblock {\em Proc. VLDB Endow.}, 4(8):470–481, May 2011.

\bibitem{comer1979ubiquitous}
D.~Comer.
\newblock Ubiquitous {B}-{T}ree.
\newblock {\em ACM Computing Surveys (CSUR)}, 11(2):121--137, 1979.

\bibitem{Cormer2009}
T.~H. Cormen, C.~E. Leiserson, R.~L. Rivest, and C.~Stein.
\newblock {\em Introduction to Algorithms, Third Edition}.
\newblock The MIT Press, 3rd edition, 2009.

\bibitem{Ding20}
J.~Ding, U.~F. Minhas, J.~Yu, C.~Wang, J.~Do, Y.~Li, H.~Zhang, B.~Chandramouli,
  J.~Gehrke, D.~Kossmann, D.~Lomet, and T.~Kraska.
\newblock Alex: An updatable adaptive learned index.
\newblock In {\em Proceedings of the 2020 ACM SIGMOD International Conference
  on Management of Data}, SIGMOD '20, page 969–984. ACM, 2020.

\bibitem{FERRAGINA21}
P.~Ferragina, F.~Lillo, and G.~Vinciguerra.
\newblock On the performance of learned data structures.
\newblock {\em Theoretical Computer Science}, 871:107--120, 2021.

\bibitem{Ferragina:2020book}
P.~Ferragina and G.~Vinciguerra.
\newblock Learned data structures.
\newblock In {\em Recent {T}rends in {L}earning {F}rom {D}ata}, pages 5--41.
  Springer International Publishing, 2020.

\bibitem{Ferragina:2020pgm}
P.~Ferragina and G.~Vinciguerra.
\newblock The {PGM-index}: a fully-dynamic compressed learned index with
  provable worst-case bounds.
\newblock {\em {PVLDB}}, 13(8):1162--1175, 2020.

\bibitem{FreedmanStat}
D.~Freedman.
\newblock {\em Statistical Models : {T}heory and Practice}.
\newblock {Cambridge University Press}, August 2005.

\bibitem{Morin17}
Paul-Virak Khuong and Pat Morin.
\newblock Array layouts for comparison-based searching.
\newblock {\em ACM Journal of Experimental Algorithmics}, 22, may 2017.

\bibitem{Kipf20}
A.~Kipf, R.~Marcus, A.~van Renen, M.~Stoian, A.~Kemper, T.~Kraska, and
  T.~Neumann.
\newblock Radixspline: A single-pass learned index.
\newblock In {\em Proceedings of the Third International Workshop on Exploiting
  Artificial Intelligence Techniques for Data Management}, aiDM '20, pages
  1--5. ACM, 2020.

\bibitem{kipfEmail}
Andreas Kipf.
\newblock Personal communication, 2021.

\bibitem{Kipf19}
Andreas Kipf, Ryan Marcus, Alexander van Renen, Mihail Stoian, Alfons Kemper,
  Tim Kraska, and Thomas Neumann.
\newblock Sosd: A benchmark for learned indexes.
\newblock {\em NeurIPS Workshop on Machine Learning for Systems}, 2019.

\bibitem{KnuthS}
D.~E. Knuth.
\newblock {\em The {A}rt of {C}omputer {P}rogramming, Vol. 3 ({S}orting and
  {S}earching)}, volume~3.
\newblock Addison-Wesley Publishing Company, 1973.

\bibitem{kraska18case}
T.~Kraska, A.~Beutel, E.H. Chi, J.~Dean, and N.~Polyzotis.
\newblock The case for learned index structures.
\newblock In {\em Proceedings of the 2018 International Conference on
  Management of Data}, pages 489--504. ACM, 2018.

\bibitem{MaltryVLDB}
Marcel Maltry and Jens Dittrich.
\newblock A critical analysis of recursive model indexes.
\newblock {\em Proceedings of the VLDB Endowment}, 15(5):1079–1091, jan 2022.

\bibitem{Marcus20}
R.~Marcus, A.~Kipf, A.~van Renen, M.~Stoian, S.~Misra, A.~Kemper, T.~Neumann,
  and T.~Kraska.
\newblock Benchmarking learned indexes.
\newblock {\em Proc. VLDB Endow.}, 14(1):1–13, sep 2020.

\bibitem{rao1999cache}
J.~Rao and K.~A Ross.
\newblock Cache conscious indexing for decision-support in main memory.
\newblock In {\em Proceedings of the 25th International Conference on Very
  Large Data Bases}, pages 78--89. ACM, 1999.

\bibitem{Schlegel09}
B.~Schlegel, R.~Gemulla, and W.~Lehner.
\newblock K-ary search on modern processors.
\newblock In {\em Proceedings of the Fifth International Workshop on Data
  Management on New Hardware}, DaMoN '09, page 52–60, New York, NY, USA,
  2009. ACM.

\bibitem{Schulz18}
Lars-Christian Schulz, David Broneske, and Gunter Saake.
\newblock An eight-dimensional systematic evaluation of optimized search
  algorithms on modern processors.
\newblock {\em Proceedings of the VLDB Endoment}, 11(11):1550–1562, jul 2018.

\end{thebibliography}

\appendix

\section{Appendix: Algorithms, Code and Software Platforms}

\subsection{Binary Search and Its Variants - Additional Algorithms}\label{S-sec:BS}

    
    	Details about the routines mentioned in Section \ref{M-sec:methods} are given here. In particular: 
    	\begin{itemize}
    		\item {\bf Lower\_bound Function}. The routine is given in Algorithm \ref{S-AL:U-LB}.
    		\item {\bf K-ary Search}. The routines are given in Algorithms \ref{S-AL:S-KS}-\ref{S-AL:U-KS}.
    		\item {\bf Eytzinger Layout}. The sorted table is now seen as stored in a virtual complete balanced Binary Search Tree. Such a tree is laid out in Breadth-First Search order in an array. An example is provided in Fig. \ref{S-fig:EyL}. Also, in this case, we adopt a Branch-free version with prefetching of the Binary Search procedure corresponding to this layout. It is reported in Algorithm \ref{S-AL:U-EL}. 
    	\end{itemize}

    \setcounter{algorithm}{2}
    
    \begin{algorithm}
    	\caption{{\bf lower\_bound Template}. }
    	\label{S-AL:U-LB}
    	\begin{algorithmic}[1]
    		\BState ForwardIterator lower\_bound (ForwardIterator first, ForwardIterator last, const T\& val)\{
    		\State \ \	ForwardIterator it;
    		\State \ \	iterator\_traits$<$ForwardIterator$>$::difference\_type count, step;
    		\State \ \	count = distance(first,last);
    		\State \ \	while (count$>$0)\{
    		\State \ \ \ \		it = first; step=count/2; advance (it,step);
    		\State \ \ \ \		if (*it$<$val)\{
    		\State \ \ \ \ \ \			first=++it;
    		\State \ \ \ \ \ \			count-=step+1;
    		\State \ \ \ \		\}
    		\State \ \	\ \ 	else count=step;
    		\State \ \	\}
    		\State \ \	return first;
    		\BState \}
    	\end{algorithmic}
    \end{algorithm}
    
    \begin{algorithm}
    	\caption{{\bf Implementation of Standard K-ary Search}. The code is as in \cite{Schulz18}}
    	\label{S-AL:S-KS}
    	\begin{algorithmic}[1]
    		\BState int StandardKarySearch(int *arr, int x, int start, int end, int k)\{
    		\State \ \ int left = start, right = end;
    		\State \ \ while (left $<$ right)
    		\State \ \ \{
    		\State \ \ \ \ int segLeft = left;
    		\State \ \ \ \ int segRight = left + (right - left) / k;
    		\State \ \ \ \ for (int i = 2; i $<=$ k; ++i)
    		\State \ \ \ \ \{
    		\State \ \ \ \ \ \ if (x $<=$ arr[segRight]) break;
    		\State \ \ \ \ \ \ \ \ segLeft = segRight + 1;
    		\State \ \ \ \ \ \ \ \ segRight = left + (i * (right - left)) / k;
    		\State \ \ \ \ \}
    		\State \ \ \ \ \ \ left = segLeft;
    		\State \ \ \ \ \ \ right = segRight;
    		\State \ \ \}
    		\State \ \ return left;
    		\BState \}	
    	\end{algorithmic}
    \end{algorithm}
    
    \begin{algorithm}
    	\caption{{\bf Implementation of Uniform K-ary Search}. The code is as in \cite{Schulz18}}\label{S-AL:U-KS}
    	\begin{algorithmic}[1]
    		\BState int UniformKarySearch(int *arr, int x, int start, int end, int k)\{
    		\State \ \ int left = start, right = end;
    		\State \ \ while (left $<$ right)\{
    		\State \ \ \ \ int segLeft = left;
    		\State \ \ \ \ int segRight = left + (1 * (right - left)) / k;
    		\State \ \ \ \ for (int i = 2; i $<=$ k; ++i)
    		\State \ \ \ \ \{
    		\State \ \ \ \ \ \ int nextSeparatorIndex = left + (i * (right - left)) / k;
    		\State \ \ \ \ \ \ segLeft = x $>$ arr[segRight] ? segRight + 1 : segLeft;
    		\State \ \ \ \ \ \ segRight = x $>$ arr[segRight] ?  nextSeparatorIndex : segRight;
    		\State \ \ \ \ \ \ \}
    		\State \ \ \ \ left = segLeft;
    		\State \ \ \ \ right = segRight;
    		\State \ \ \}
    		\State \ \ return left;
    		\BState \}	
    	\end{algorithmic}
    \end{algorithm}
    
    \begin{figure}[h]
    	\begin{center}
    		\includegraphics[width=0.40\textwidth]{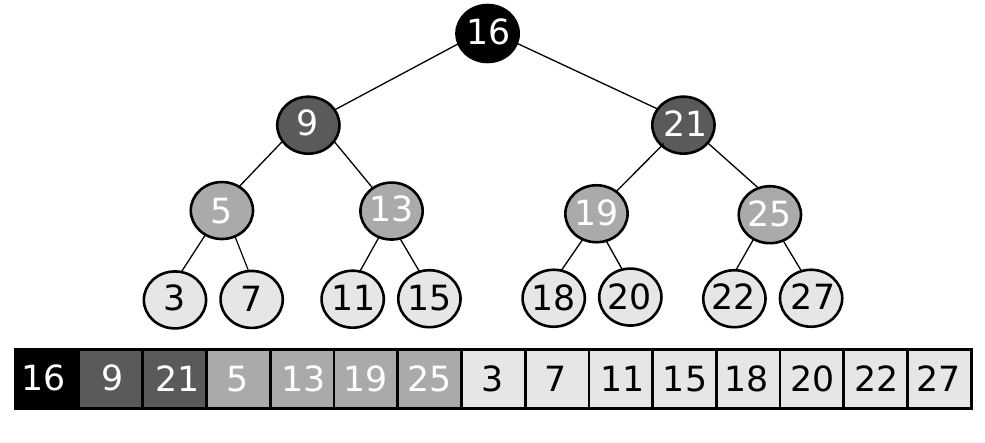}
    		\caption{{\bf An example of  Eyzinger layout of a table with 15 elements}. See also \cite{Morin17}.}
    		\label{S-fig:EyL}
    	\end{center}
    \end{figure}
    
    \begin{algorithm}
    	
    	\caption{{\bf  Implementation of Uniform Binary Search with Eytzinger Layout and Prefetching}. The code is as in   \cite{Morin17}. The version without prefetching is obtained by deleteing line 5.}
    	\label{S-AL:U-EL}
    	\begin{algorithmic}[1]
    		\BState int UniformEytzingerLayout(int *A, int x,  int left, int right)\{
    		\State \ \ \ int i = 0;
    		\State \ \ \ int n = right;
    		\State \ \ \ while (i $<$ n)\{
    		\State \ \ \ \ \ \_\_builtin\_prefetch(A+(multiplier*i + offset)); 
    		\State \ \ \ \ \ i = (x $<=$ A[i]) ? (2*i + 1) : (2*i + 2);
    		\State \ \ \ \}
    		\State \ \ \ int j = (i+1) $>>$ \_\_builtin\_ffs($\sim$(i+1));
    		\State \ \ \ return (j == 0) ? n : j-1;
    		\BState \}	
    	\end{algorithmic}
    \end{algorithm}

\subsection{Index Model Classes in SOSD - Additional Details}\label{S-sec:LI}

    In this Section details about the Learned Indexes mentioned in Section \ref{M-sec:models} are given. In particular:
    \begin{itemize}
    	\item {\bf RMI} \cite{kraska18case}. It is a multi-stage model. When searching for a given key and starting with the first stage, a prediction at each stage identifies the model of the next stage to use for the next prediction. This process continues until a final stage model is reached. This latter is used to predict the interval to search into.  An example is helpful. 
    	Figure \ref{S-fig:RMIRSPGM}(a) describes a two-stage model. The first stage model (linear) is used to pick one of the stage 2 models (cubic). Finally, the selected model provides the final interval to search into. 
    	
    	\item {\bf RS} \cite{Kipf20}. It is a two-stage model. It uses user-defined approximation parameter $\epsilon$.  With reference to Figure \ref{S-fig:RMIRSPGM}(c), a spline curve approximating the CDF of the data is built. Then, the radix table is used to identify spline points to use to refine the search interval. Indeed, for the key in the Figure, the most significant three bits are used to identify two spline points. Then, a Binary Search is performed on the table of the spline points delimited by the points identified earlier. Such a Binary Search identifies two spline points such that a simple interpolation guarantees a prediction with error   $\epsilon$. 
    	
    	\item {\bf PGM} \cite{Ferragina:2020pgm,FERRAGINA21}.It is also a multi-stage model, built bottom-up and queried top down.  It also uses a user-defined approximation parameter $\epsilon$, that controls the prediction error at each stage. With reference to Figure \ref{S-fig:RMIRSPGM}(b), the table is subdivided into three pieces. A prediction in each piece can be done via a linear model guaranteeing an error of $\epsilon$.  A new table is formed by selecting the minimum values in each of the three pieces. This new table is possibly again partitioned into pieces, in which a linear model can make a prediction within the given error. The process is iterated until only one linear model suffices, as in the case in the Figure. A query is processed via a series of predictions, starting at the root of the tree. 
    \end{itemize}
    
    \begin{figure}[tbh]
        	\begin{center}
        	(a)
        	\begin{minipage}{0.32\textwidth}
        		\includegraphics[width=\textwidth]{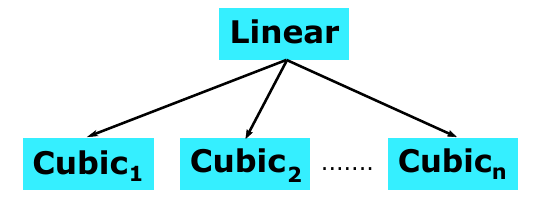}
        	\end{minipage}\hfill
        	(b)
        	\begin{minipage}{0.32\textwidth}
        		\includegraphics[width=\textwidth]{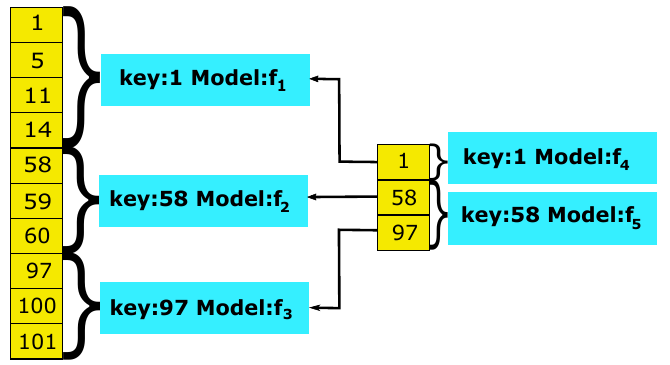}
        	\end{minipage}\hfill
        	(c)
        	\begin{minipage}{0.25\textwidth}
        		\includegraphics[width=\textwidth]{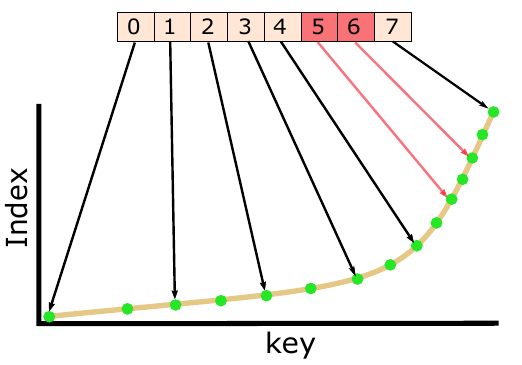}
        	\end{minipage}\hfill
            \caption{Examples of (a) RMI (b) PGM (c) RS. See also \cite{Marcus20}}\label{S-fig:RMIRSPGM}    
        	\end{center}
        
        \end{figure}

    	\subsection{Computer Architecture and  Compilers: The Production of Branch-Free Code - Additional Code}\label{S-sec:assembly}
    
    		In this Section, as indicated in Section \ref{M-banchy}, the assembly codes of the compiled routines are provided. In particular:
    		\begin{itemize}
    			\item The assembly code of Standard Binary Search on the Intel I7 is provided in Code \ref{S-CD:stand-i7}. 
    			\item The assembly codes of Uniform Binary Search on the Intel I9 and the Apple M1 are provided in Code \ref{S-AL:unif-I9} and Code \ref{S-AL:unif-M1}, respectively.
    			\item The assembly code of lower\_bound function on the Apple M1 is provided in Code \ref{S-AL:M1-LB}.
    			\item  The assembly codes of the other branchy methods are here omitted and available to the reader upon request.
    		\end{itemize}

    	\setcounter{algorithm}{1}
    	
    	\begin{algorithm}
    		\floatname{algorithm}{Code}

    			\caption{{\bf Assembly Code of Standard Binary Search on Intel I7 (Only Main Loop)}. No prefetching is used.}
    			\label{S-CD:stand-i7}
    			\begin{algorithmic}
    				\BState .0000000000001229 $<$\_ZL22Standard\_Binary\_SearchPmmii$>$:
    				\State \ \		.
    				\State \ \		.
    				\State \ \ 124d:	7f 6d                	jg     12bc $<$\_ZL22Standard\_Binary\_SearchPmmii+0x93$>$
    				\State \ \ 124f:	8b 55 dc             	mov    -0x24(\%rbp),\%edx
    				\State \ \ 1252:	8b 45 d8             	mov    -0x28(\%rbp),\%eax
    				\State \ \ 1255:	01 d0                	add    \%edx,\%eax
    				\State \ \ 1257:	89 c2                	mov    \%eax,\%edx
    				\State \ \ 1259:	c1 ea 1f             	shr    \$0x1f,\%edx
    				\State \ \ 125c:	01 d0                	add    \%edx,\%eax
    				\State \ \ 125e:	d1 f8                	sar    \%eax
    				\State \ \ 1260:	48 98                	cltq   
    				\State \ \ 1262:	48 89 45 f8          	mov    \%rax,-0x8(\%rbp)
    				\State \ \ 1266:	48 8b 45 f8          	mov    -0x8(\%rbp),\%rax
    				\State \ \ 126a:	48 8d 14 c5 00 00 00 	lea    0x0(,\%rax,8),\%rdx
    				\State \ \ 1271:	00 
    				\State \ \ 1272:	48 8b 45 e8          	mov    -0x18(\%rbp),\%rax
    				\State \ \ 1276:	48 01 d0             	add    \%rdx,\%rax
    				\State \ \ 1279:	48 8b 00             	mov    (\%rax),\%rax
    				\State \ \ 127c:	48 39 45 e0          	cmp    \%rax,-0x20(\%rbp)
    				\State \ \ 1280:	73 0c                	jae    128e $<$\_ZL22Standard\_Binary\_SearchPmmii+0x65$>$
    				\State \ \ 1282:	48 8b 45 f8          	mov    -0x8(\%rbp),\%rax
    				\State \ \ 1286:	83 e8 01             	sub    \$0x1,\%eax
    				\State \ \ 1289:	89 45 d8             	mov    \%eax,-0x28(\%rbp)
    				\State \ \ 128c:	eb b9                	jmp    1247 $<$\_ZL22Standard\_Binary\_SearchPmmii+0x1e$>$
    				\State \ \ 128e:	48 8b 45 f8          	mov    -0x8(\%rbp),\%rax
    				\State \ \ 1292:	48 8d 14 c5 00 00 00 	lea    0x0(,\%rax,8),\%rdx
    				\State \ \ 1299:	00 
    				\State \ \ 129a:	48 8b 45 e8          	mov    -0x18(\%rbp),\%rax
    				\State \ \ 129e:	48 01 d0             	add    \%rdx,\%rax
    				\State \ \ 12a1:	48 8b 00             	mov    (\%rax),\%rax
    				\State \ \ 12a4:	48 39 45 e0          	cmp    \%rax,-0x20(\%rbp)
    				\State \ \ 12a8:	76 0c                	jbe    12b6 $<$\_ZL22Standard\_Binary\_SearchPmmii+0x8d$>$
    				\State \ \ 12aa:	48 8b 45 f8          	mov    -0x8(\%rbp),\%rax
    				\State \ \ 12ae:	83 c0 01             	add    \$0x1,\%eax
    				\State \ \ 12b1:	89 45 dc             	mov    \%eax,-0x24(\%rbp)
    				\State \ \ 12b4:	eb 91                	jmp    1247 $<$\_ZL22Standard\_Binary\_SearchPmmii+0x1e$>$
    				\State \ \ 12b6:	48 8b 45 f8          	mov    -0x8(\%rbp),\%rax
    				\State \ \ 12ba:	eb 05                	jmp    12c1 $<$\_ZL22Standard\_Binary\_SearchPmmii+0x98$>$
    				\State \ \ 12bc:	8b 45 d8             	mov    -0x28(\%rbp),\%eax
    				\State \ \		.
    				\State \ \		.
    		\end{algorithmic}
    		
    	\end{algorithm}
    	
    	\begin{algorithm}
    		\floatname{algorithm}{Code}
    		
    			\caption{{\bf Assembly Code of Uniform Binary Search on Intel I9 (Only Main Loop)}. The predicated instruction is line 1475 (in bold)}
    			\label{S-AL:unif-I9}
    			\begin{algorithmic}
    				\BState 0000000000001450 $<$\_Z21Uniform\_Binary\_SearchPmmmm$>$:
    				\State \ \		.
    				\State \ \		.
    				\State \ \		1463:	76 1d                	  jbe    1482 $<$\_Z21Uniform\_Binary\_SearchPmmmm+0x32$>$
    				\State \ \		1465:	0f 1f 00             	 nopl   (\%rax)
    				\State \ \		1468:	48 89 ca             	mov    \%rcx,\%rdx
    				\State \ \		146b:	48 d1 ea             	 shr    \%rdx
    				\State \ \		146e:	4c 8d 04 d0           lea    (\%rax,\%rdx,8),\%r8
    				\State \ \		1472:	49 3b 30             	cmp    (\%r8),\%rsi
    				\State \ \		{\bf 1475:	49 0f 43 c0            cmovae \%r8,\%rax}
    				\State \ \		1479:	48 29 d1             	 sub    \%rdx,\%rcx
    				\State \ \		147c:	48 83 f9 01          	cmp    \$0x1,\%rcx
    				\State \ \		1480:	77 e6                	   ja     1468 $<$\_Z21Uniform\_Binary\_SearchPmmmm+0x18$>$
    				\State \ \		.
    				\State \ \		.
    			\end{algorithmic}
    		
    	\end{algorithm}
    	
    	\begin{algorithm}
    		\floatname{algorithm}{Code}
    		
    			\caption{{\bf Assembly Code of Uniform Binary Search on Apple Arm M1 (Only Main Loop)}. The predicated instruction is line 10000371c (in bold)}
    			\label{S-AL:unif-M1}
    			\begin{algorithmic}
    				\BState 00000001000036f8 $<$\_\_Z21Uniform\_Binary\_SearchPyyyy$>$:
    				\State \ \		.
    				\State \ \		.
    				\State \ \		100003708: 23 01 00 54 	 b.lo	0x10000372c $<$\_\_Z21Uniform\_Binary\_SearchPyyyy+0x34$>$
    				\State \ \		10000370c: 2a fd 41 d3 	  lsr	x10, x9, \#1
    				\State \ \		100003710: 0b 0d 0a 8b 	 add	x11, x8, x10, lsl \#3
    				\State \ \		100003714: 6c 01 40 f9 	   ldr	x12, [x11]
    				\State \ \		100003718: 9f 01 01 eb 	   cmp	x12, x1
    				\State \ \		{\bf 10000371c: 08 81 8b 9a 	  csel	x8, x8, x11, hi}
    				\State \ \		100003720: 29 01 0a cb 	  sub	x9, x9, x10
    				\State \ \		100003724: 3f 05 00 f1 	   cmp	x9, \#1
    				\State \ \		100003728: 28 ff ff 54 	    b.hi	0x10000370c $<$\_\_Z21Uniform\_Binary\_SearchPyyyy+0x14$>$
    				\State \ \		.
    				\State \ \		.
    			\end{algorithmic}
    		
    	\end{algorithm}

    	\begin{algorithm}
    		\floatname{algorithm}{Code}
    			\caption{{\bf Assembly Code of C++ lower\_bound Function on Apple Arm M1 (Only Main Loop)}. The predicated instruction are in lines 10000371c, 100003fa0 (in bold)}
    			\label{S-AL:M1-LB}
    			\begin{algorithmic}[1]
    				\BState 0000000100003f68 $<$\_\_Z6searchRNSt3\_\_16vectorIyNS\_9allocatorIyEEEEyPmmm$>$:
    				\State \ \		.
    				\State \ \		.
    				\State \ \		100003f7c: 60 01 00 54 	  b.eq	0x100003fa8 $<$\_\_Z6searchRNSt3\_\_16vectorIyNS\_9allocatorIyEEEEyPmmm+0x40$>$
    				\State \ \		100003f80: 29 fd 43 93 	  asr	x9, x9, \#3
    				\State \ \		100003f84: 2a fd 41 d3 	   lsr	x10, x9, \#1
    				\State \ \		100003f88: 0b 0d 0a 8b 	  add	x11, x8, x10, lsl \#3
    				\State \ \		100003f8c: 6c 85 40 f8 	   ldr	x12, [x11], \#8
    				\State \ \		100003f90: ed 03 2a aa 	  mvn	x13, x10
    				\State \ \		100003f94: 29 01 0d 8b 	  add	x9, x9, x13
    				\State \ \		100003f98: 9f 01 01 eb 	    cmp	x12, x1
    				\State \ \		{\bf 100003f9c: 29 31 8a 9a 	   csel	x9, x9, x10, lo}
    				\State \ \		{\bf 100003fa0: 68 31 88 9a 	   csel	x8, x11, x8, lo}
    				\State \ \		100003fa4: 09 ff ff b5 	     cbnz	x9, 0x100003f84 $<$\_\_Z6searchRNSt3\_\_16vectorIyNS\_9allocatorIyEEEEyPmmm+0x1c$>$
    				\State \ \		.
    				\State \ \		.
    			\end{algorithmic}
    	\end{algorithm}
	
\section{Datasets and Index Model Training - Additional Considerations}\label{S-sec:dataset}

    In this Section, we report in Figure \ref{S-fig:L3L4CDF}, the \emph{CDF} of the datasets on L3 and L4 memory levels, described in Section \ref{M-sec:Datasets}. It can be noted that the curves of {\bf face\_L3} and {\bf face\_L4} show a significant difference due to the presence of some outliers in the L4 memory level which makes the representation not very accurate.

	\begin{figure}[tbh]
		\centering
		(a)
		\begin{minipage}{0.45\textwidth}
			\includegraphics[width=\linewidth]{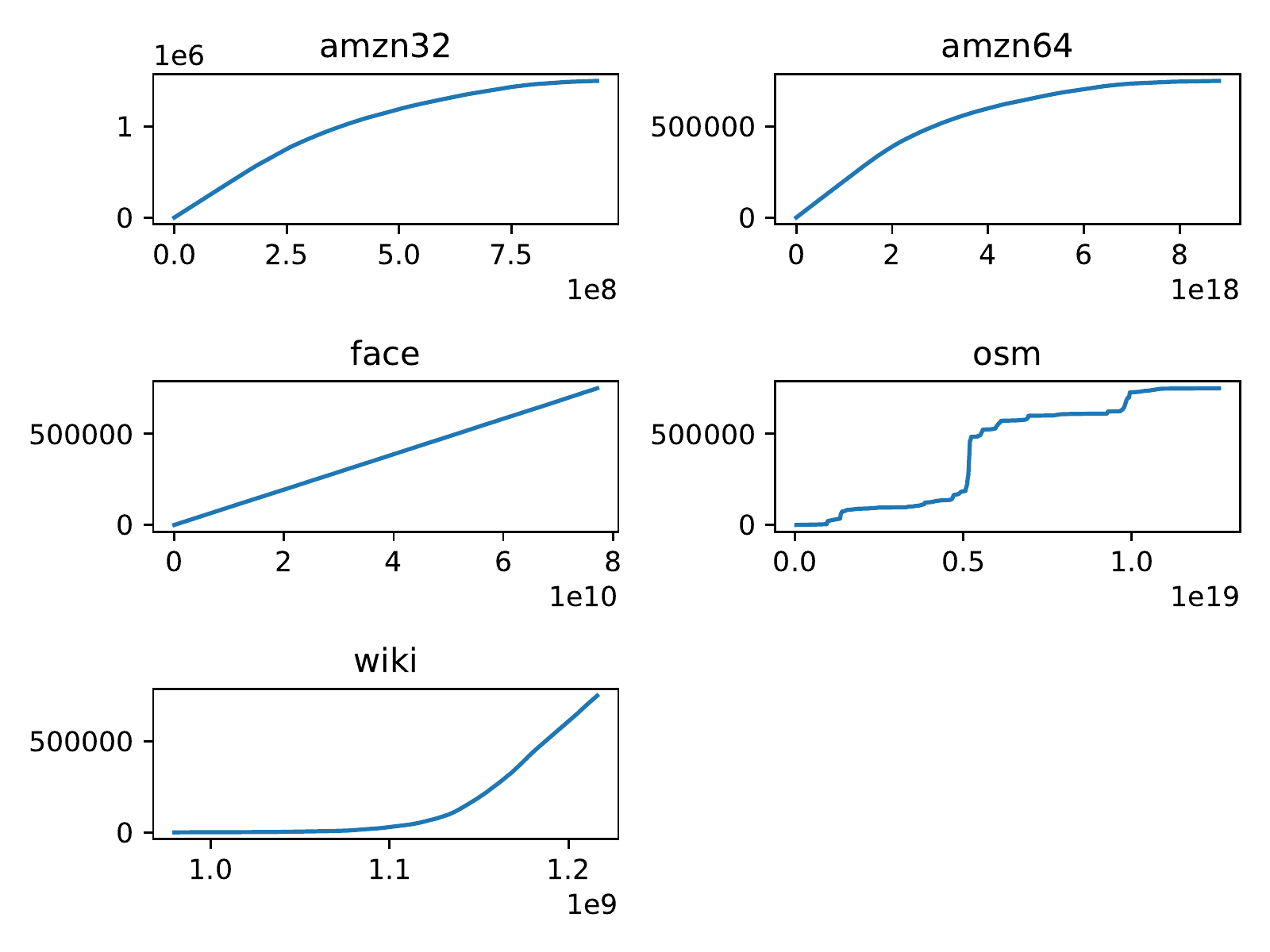}
		\end{minipage}\hfill
		(b)
		\begin{minipage}{0.45\textwidth}
			\includegraphics[width=\linewidth]{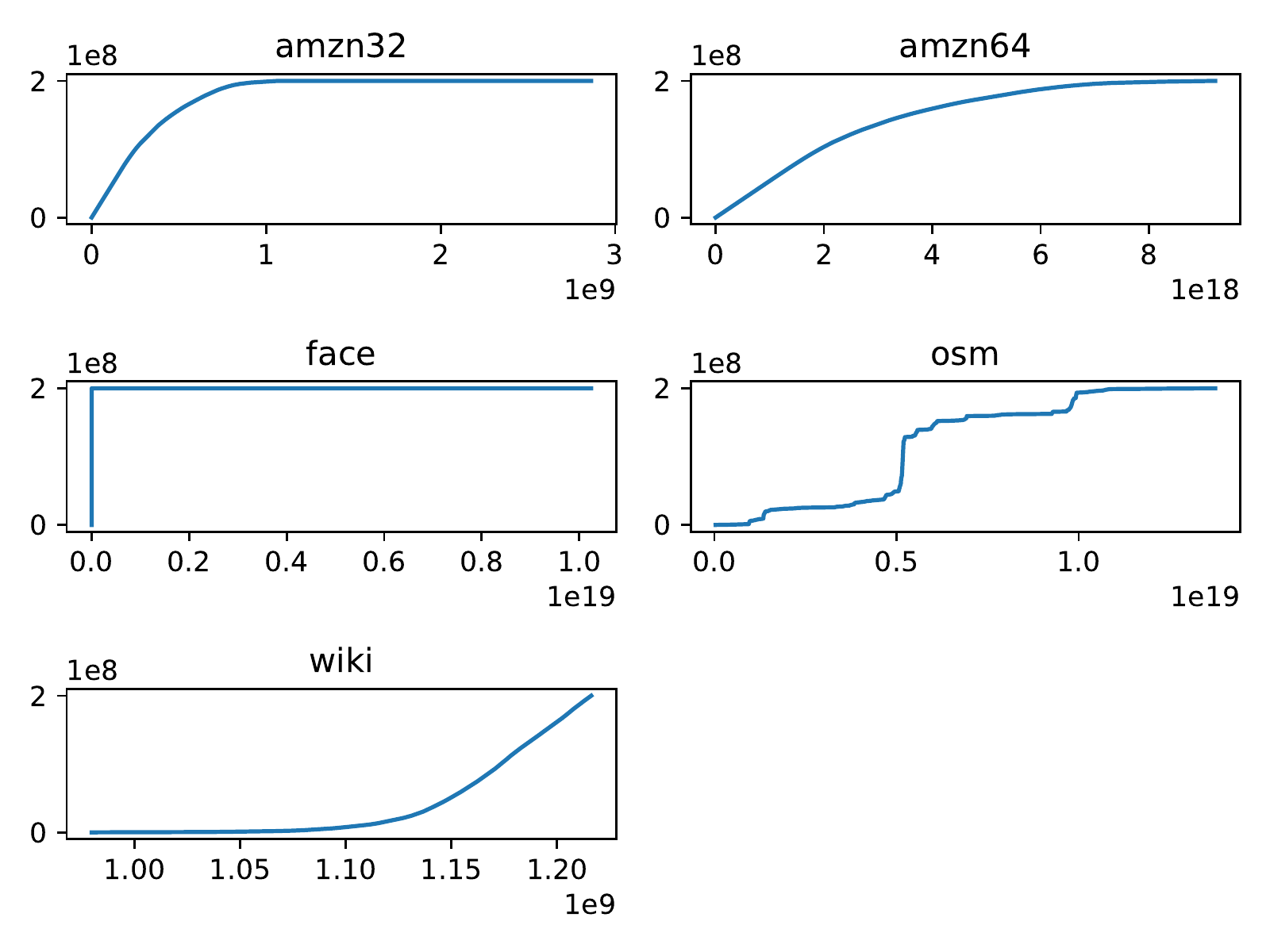}
		\end{minipage}\hfill
		
		\caption{{\bf Datesets Empirical CDF}. For each dataset, we report the value of the elements on the abscissa and their position on the ordinate. The curve so obtained is analogous to the Empirical \emph{CDF} of the Universe from which the data come. In particular, Figure (a) is referred to the L3 memory level, while (b) to L4.}
		\label{S-fig:L3L4CDF}
	\end{figure}

\section{Experiments: Searching in Constant Additional Space, With or Without SOSD -  Additional Results}

	\subsection{Replication of The Experiments by Khough and Morin}

    	{
    		We provide here the details of the experiments mentioned in Section 
    		\ref{M-sec:morin}. In particular: }

    	\begin{itemize}
    		\item The comparison between the {\bf lower\_bound} and {\bf S-BS} implementations is reported in Figure \ref{S-fig:I7lowerboundSOSD} for the case of the Intel I7 Architecture, and in Figure \ref{S-fig:M1lowerboundSOSD} for the Apple M1.
    		\item The comparison of each of the considered routines in their SOSD and \emph{stand-alone} implementations is reported in Figure \ref{S-fig:I7sosdnososd}
    		\item The comparison among the remaining routines on the {\bf osm} dataset is reported in Figure \ref{S-fig:I7M1osm}.
    		
    	\end{itemize}
    	
    	\begin{figure}[tbh]
    		\centering
    		\includegraphics[scale=0.4]{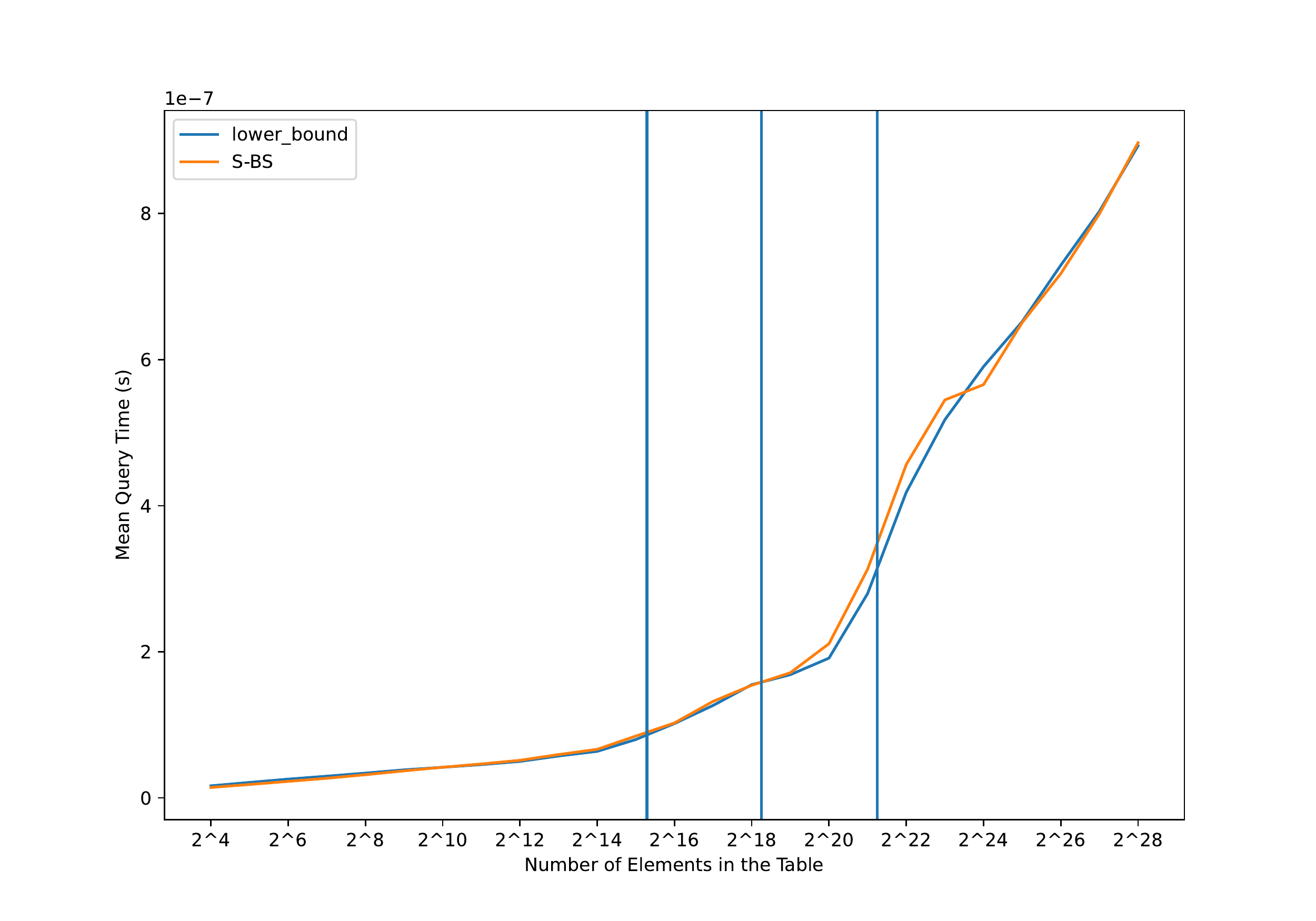}
    		\caption{{\bf Mean Query times Comparison of the lower\_bound Function and S-BS on Synthetic Datasets}. The Figure reports only the comparison of the routines executed within the {\bf SOSD} platform on Intel I7 since the results with Standard C++ implementations are the same. On the abscissa, we report the number of elements in the table and, on the ordinates, the mean query time in seconds. The vertical lines indicate the size of each cache memory level.}
    		\label{S-fig:I7lowerboundSOSD}
    	\end{figure}
    	
    	\begin{figure}[tbh]
    		\centering
    		(a)
    		\begin{minipage}{0.45\textwidth}
    			\includegraphics[width=\linewidth]{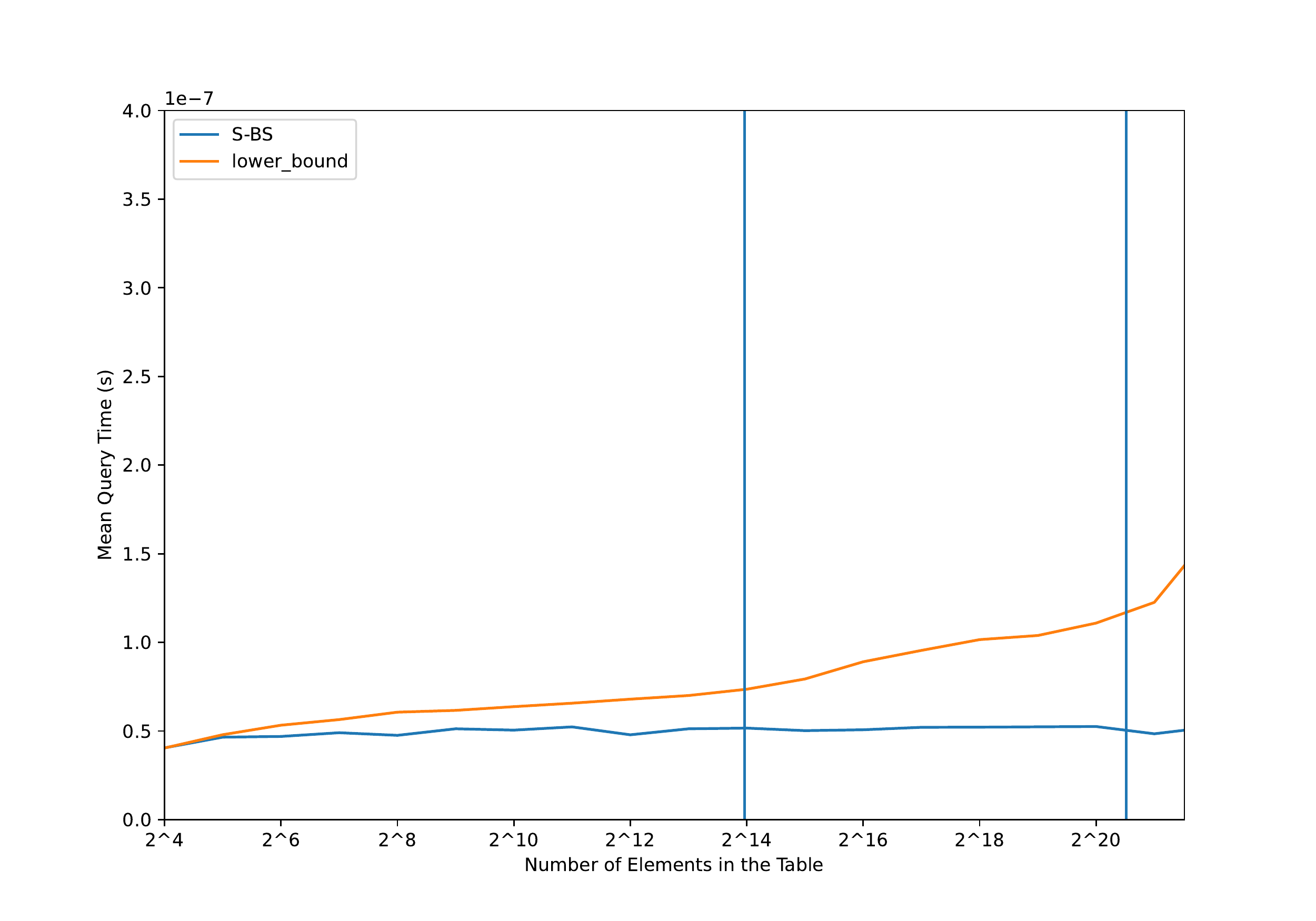}
    		\end{minipage}\hfill
    		(b)
    		\begin{minipage}{0.45\textwidth}
    			\includegraphics[width=\linewidth]{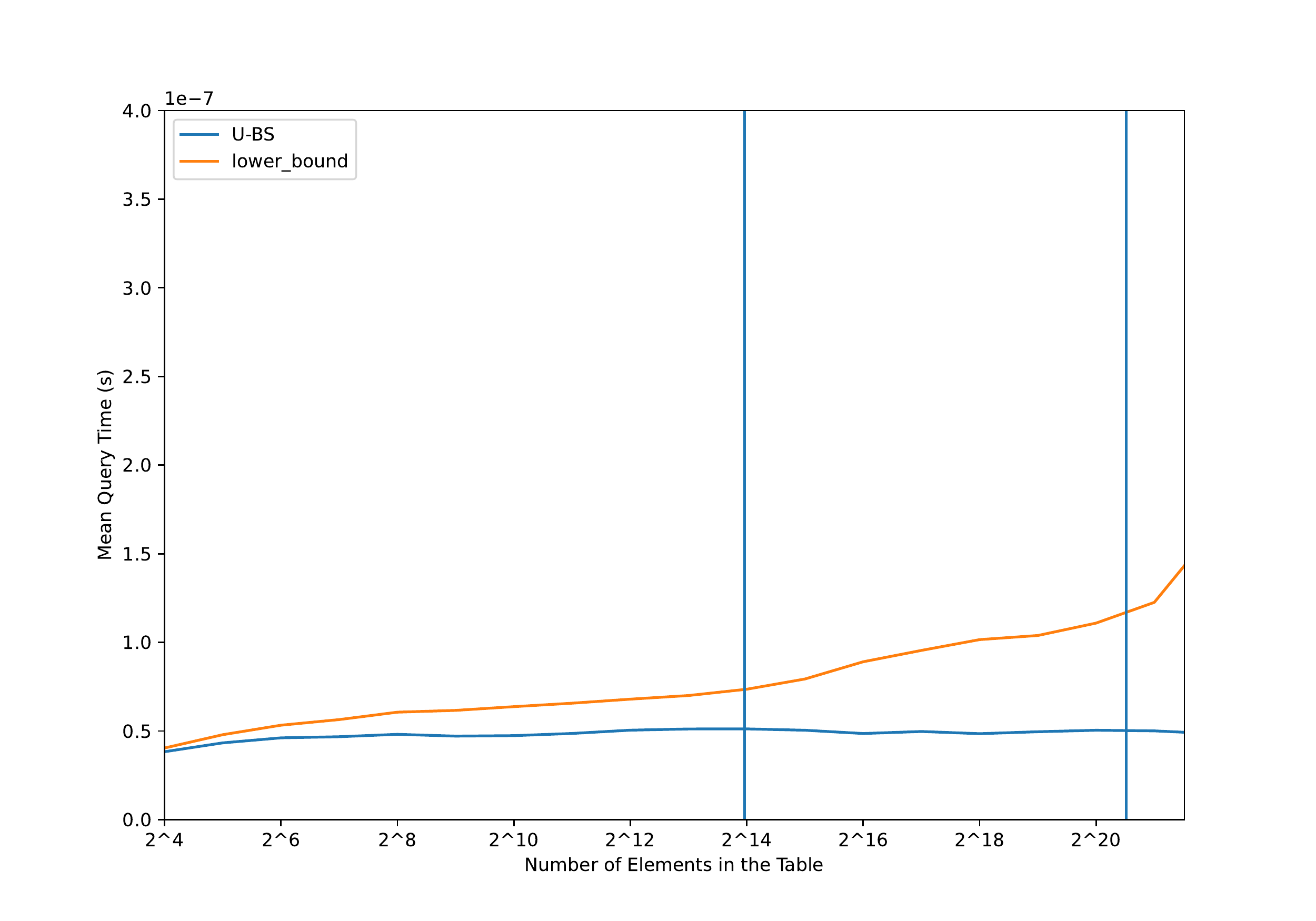}
    		\end{minipage}\hfill
    		
    		\caption{{\bf Mean Query times Comparison of  lower\_bound Function with S-BS and U-BS on Synthetic Datasets}. Figures report only the comparison of the routines executed within the {\bf SOSD} platform on Apple M1 architecture, since the results with a Standard C++ implementation are the same. We report the comparison between {\bf lower\_bound} function and {\bf S-BS} in Figure (a), while in Figure (b) the comparison between {\bf lower\_bound} function and {\bf U-BS}. On the abscissa, we report the number of elements in the table and, on the ordinates, the mean query time in seconds. The vertical lines indicate the size of each cache memory level.}
    		\label{S-fig:M1lowerboundSOSD}
    	\end{figure}

    	\begin{figure}[tbh]
    		\centering
    		(a)
    		\begin{minipage}{0.45\textwidth}
    			\includegraphics[width=\linewidth]{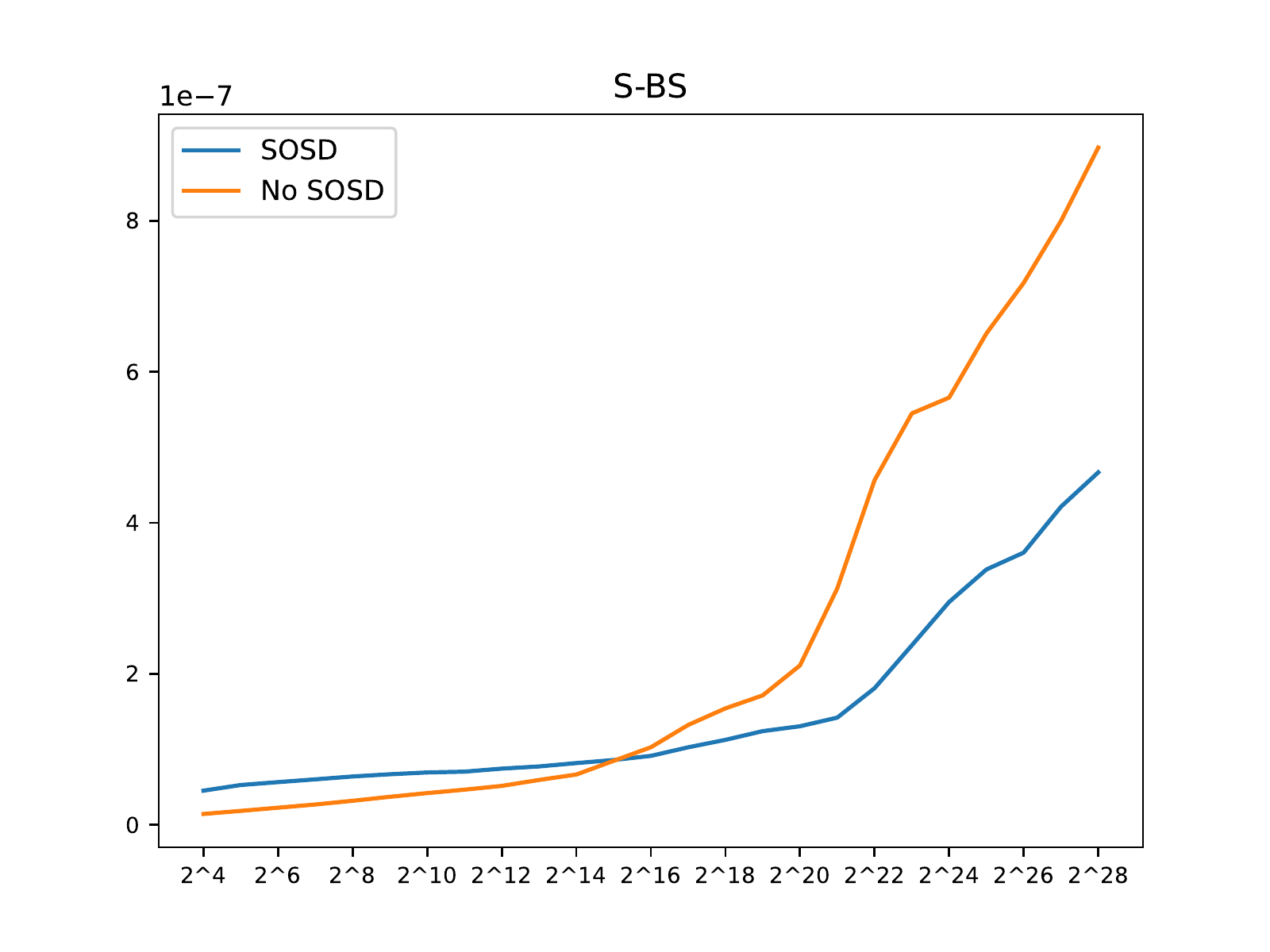}
    		\end{minipage}\hfill
    		\centering
    		(b)
    		\begin{minipage}{0.45\textwidth}
    			\includegraphics[width=\linewidth]{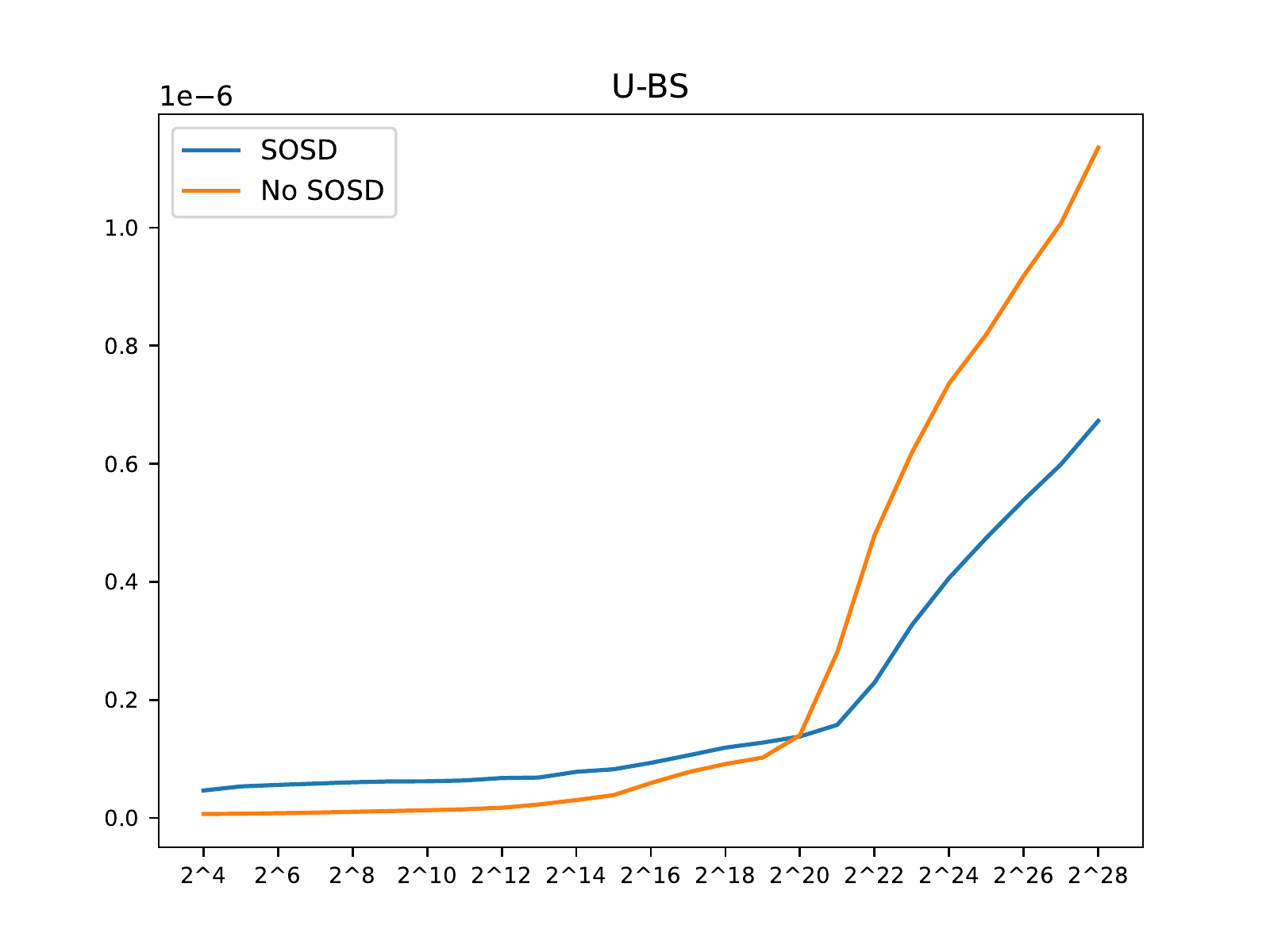}
    		\end{minipage}\hfill\\
    		\centering
    		(c)
    		\begin{minipage}{0.45\textwidth}
    			\includegraphics[width=\linewidth]{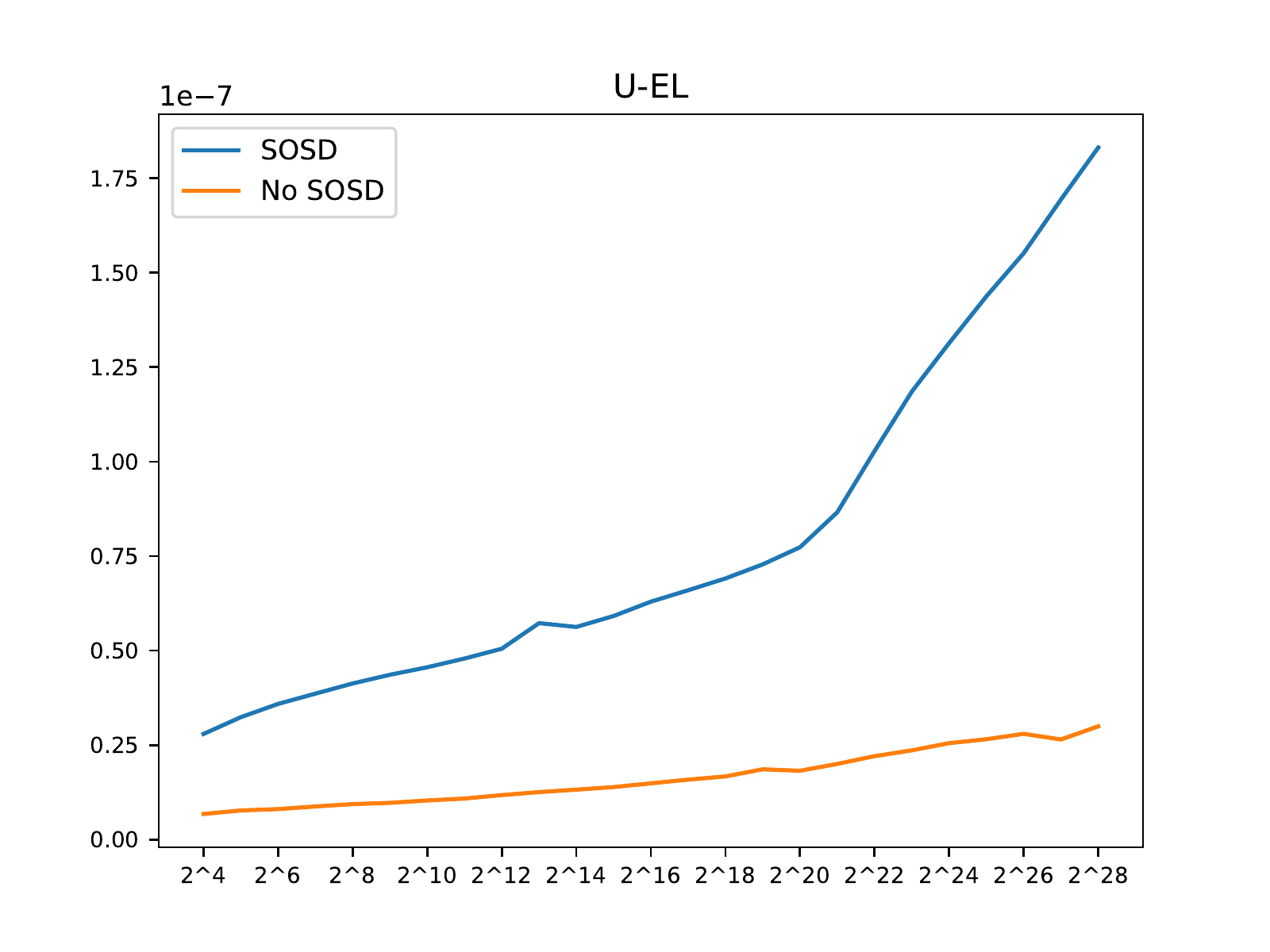}
    		\end{minipage}\hfill
    		\centering
    		(d)
    		\begin{minipage}{0.45\textwidth}
    			\includegraphics[width=\linewidth]{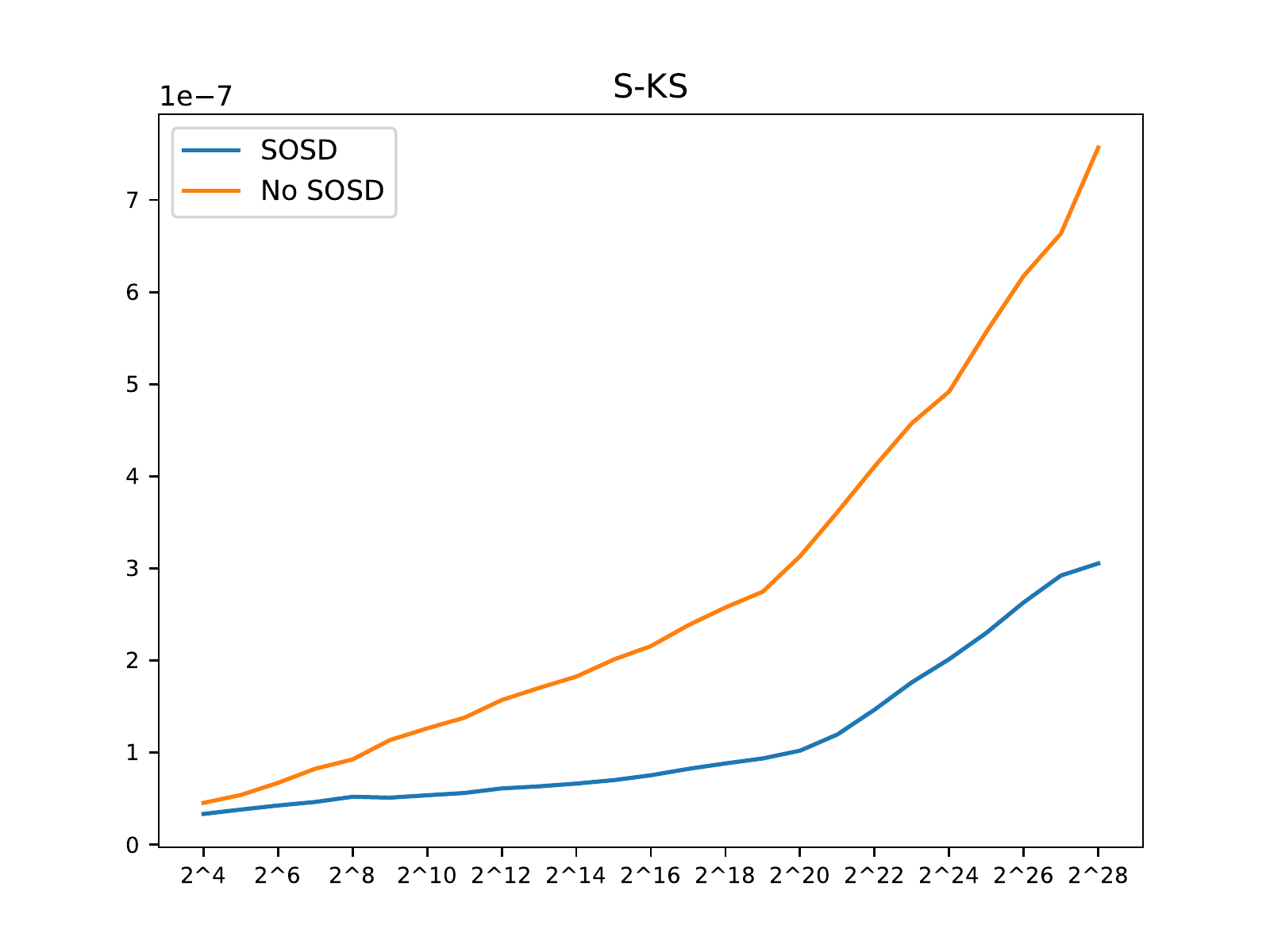}
    		\end{minipage}\hfill\\
    		\centering
    		(e)
    		\begin{minipage}{0.45\textwidth}
    			\includegraphics[width=\linewidth]{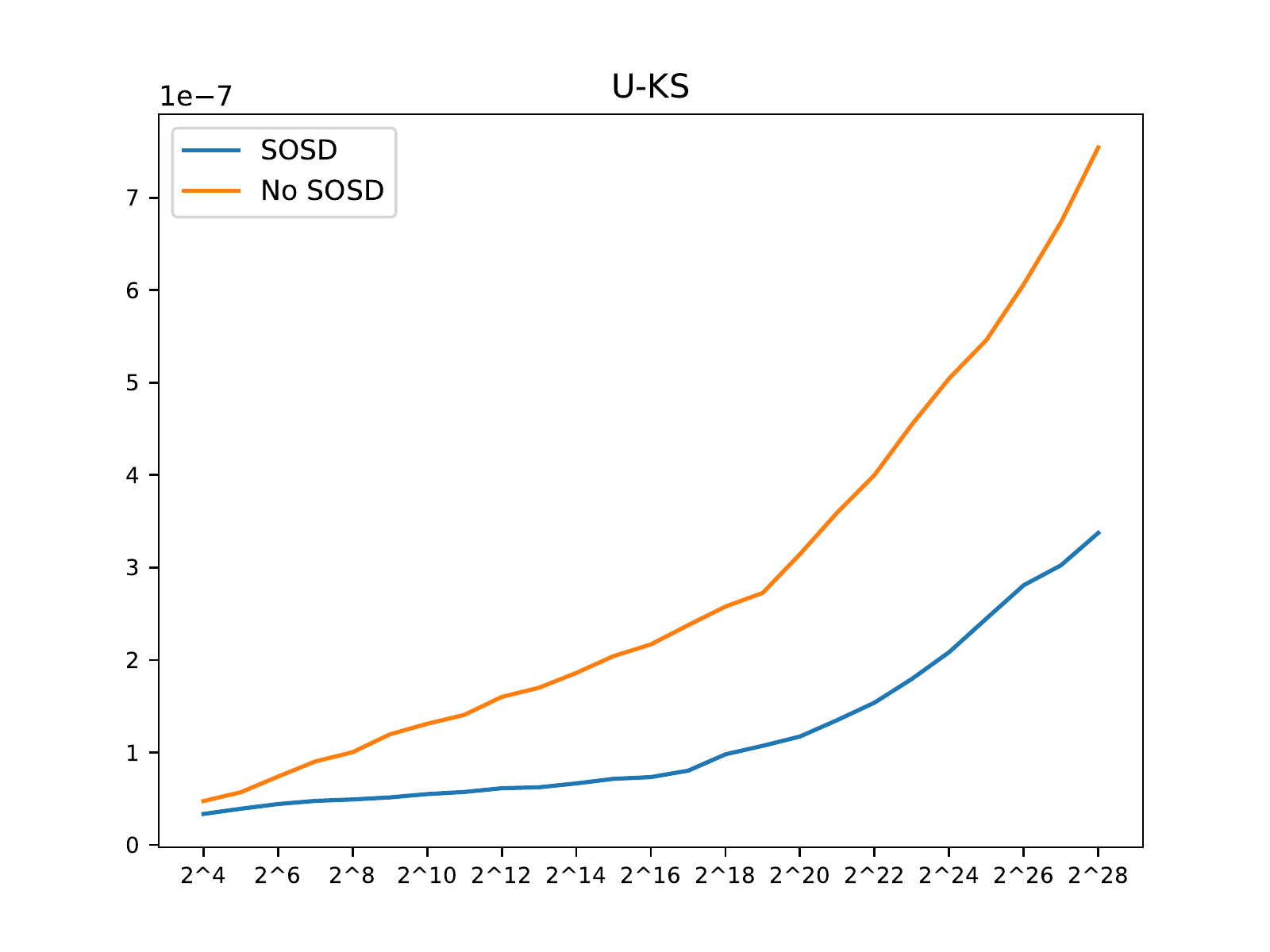}
    		\end{minipage}\hfill
    		\caption{{\bf Mean Query Times Comparison between SOSD and Stand-alone Implementations} For each method, i.e. S-BS(a), U-BS(b), U-EL(c), S-KS(d), U-KS(e), we report SOSD (blue curve) and \emph{stand-alone} (orange curve) implementations for the case of Intel I7 architectures. The abscissa reports number of elements on the Table, while ordinate the mean query time in seconds.}
    		\label{S-fig:I7sosdnososd}
    	\end{figure}
				
		\begin{figure}[tbh]
			\centering
			(a)
			\begin{minipage}{0.45\textwidth}
				\includegraphics[width=\linewidth]{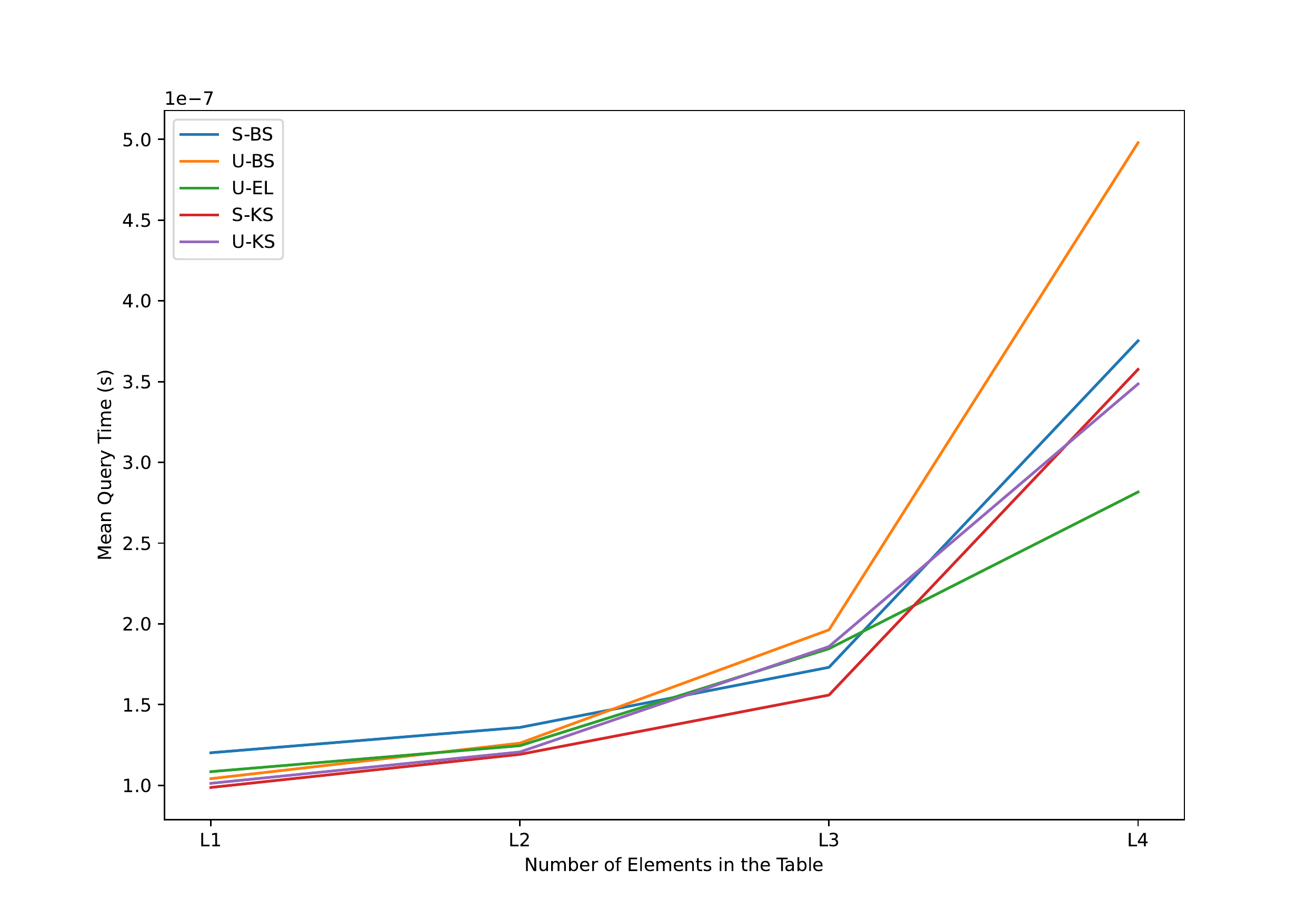}
			\end{minipage}\hfill
			\centering
			(b)
			\begin{minipage}{0.45\textwidth}
				\includegraphics[width=\linewidth]{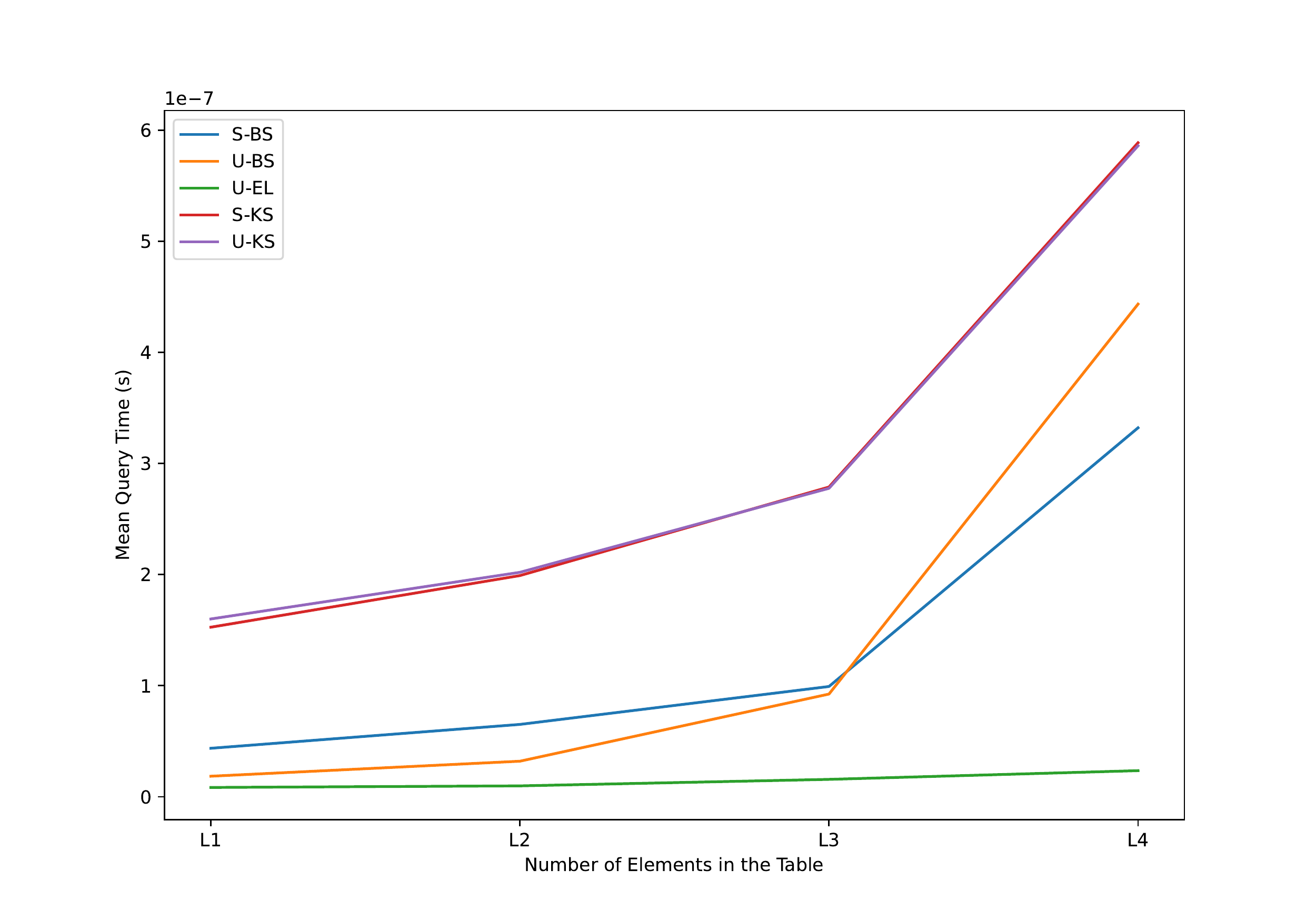}
			\end{minipage}\hfill\\
			\centering
			(c)
			\begin{minipage}{0.45\textwidth}
				\includegraphics[width=\linewidth]{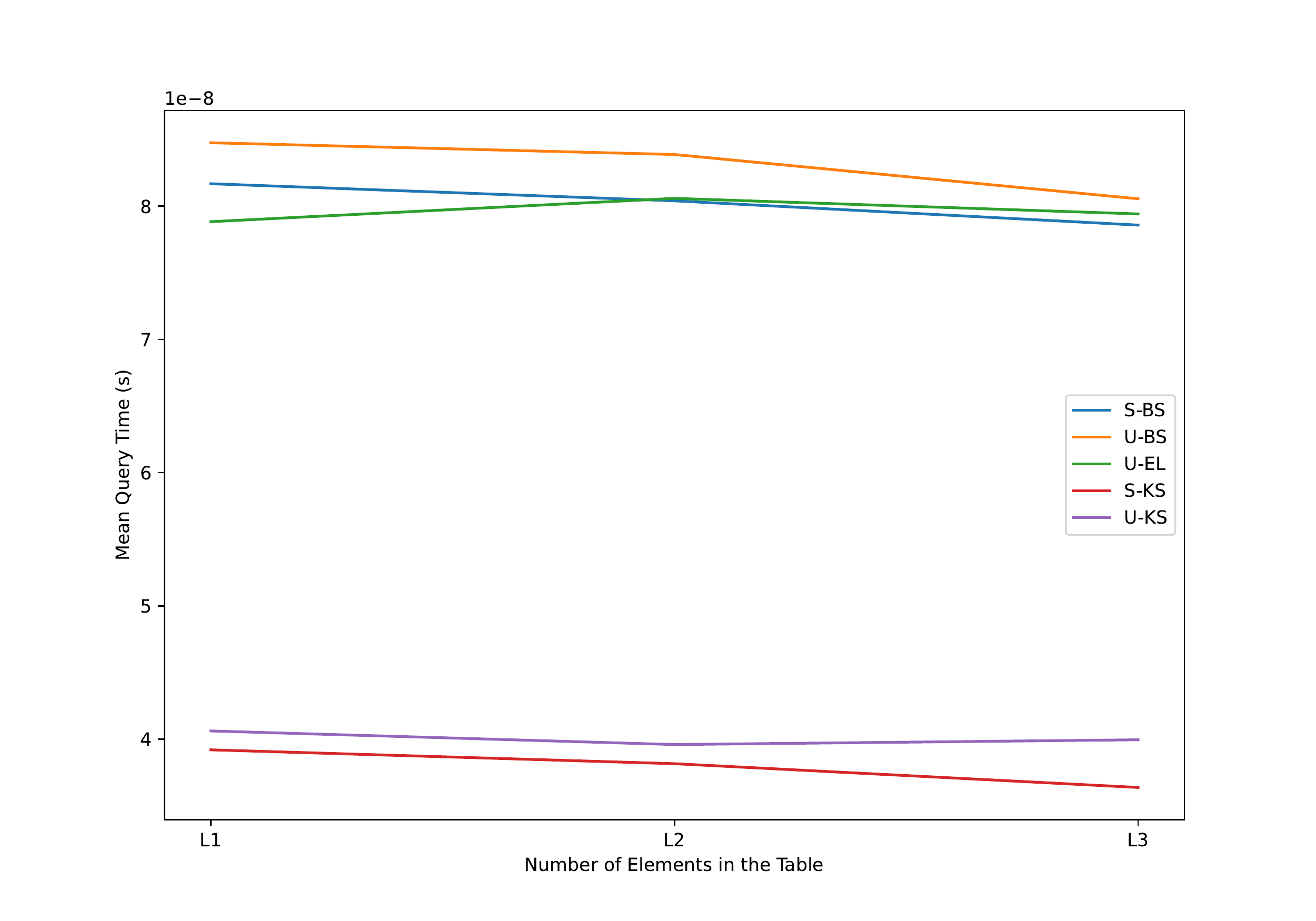}
			\end{minipage}\hfill
			(d)
			\begin{minipage}{0.45\textwidth}
				\includegraphics[width=\linewidth]{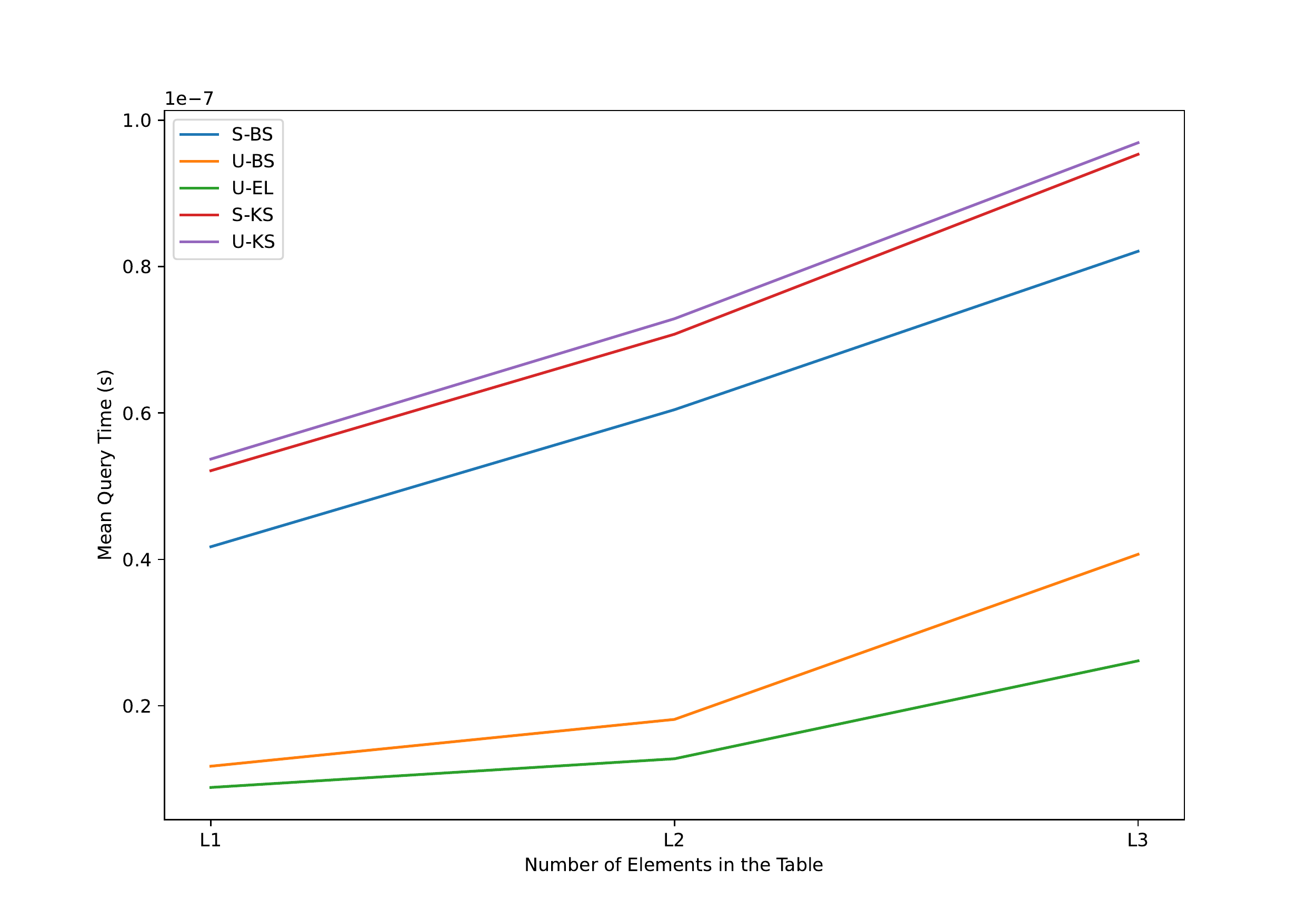}
			\end{minipage}\hfill
			\caption{{\bf Mean Query times on osm Datasets on Intel I7 (a,b) and Apple M1 (c,d) Architectures}. Figures (a,c) show only the comparison of the routines executed within the {\bf SOSD} platform, while Figures (b,d) with a Standard C++ implementation. On the abscissa, we report the hierarchical memory level in which datasets fit  and, on the ordinates, the mean query time in seconds.}
			\label{S-fig:I7M1osm}
		\end{figure}
		
		\begin{figure}[tbh]
			\centering
			(a)
			\begin{minipage}{0.45\textwidth}
				\includegraphics[width=\linewidth]{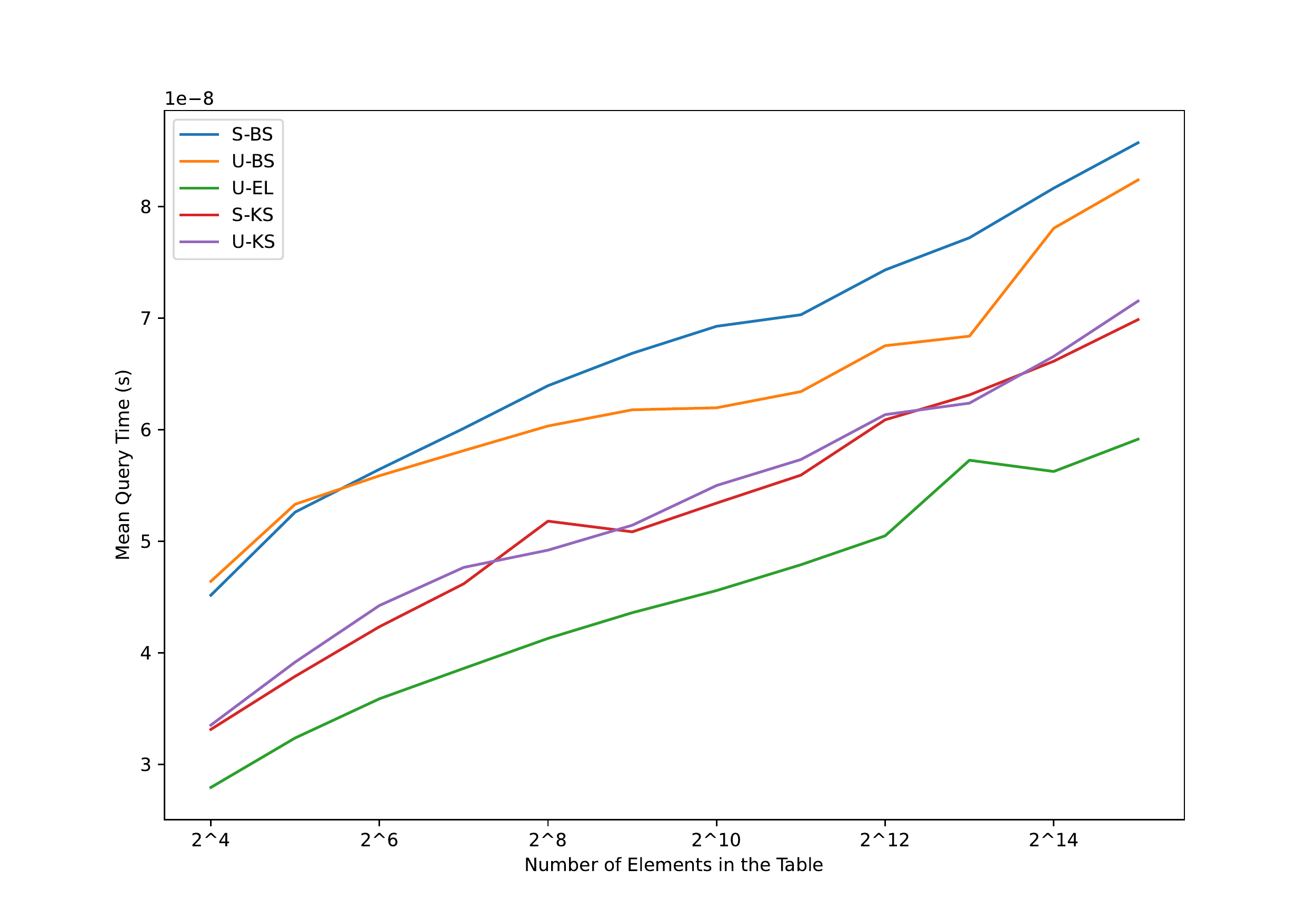}
			\end{minipage}\hfill
			\centering
			(b)
			\begin{minipage}{0.45\textwidth}
				\includegraphics[width=\linewidth]{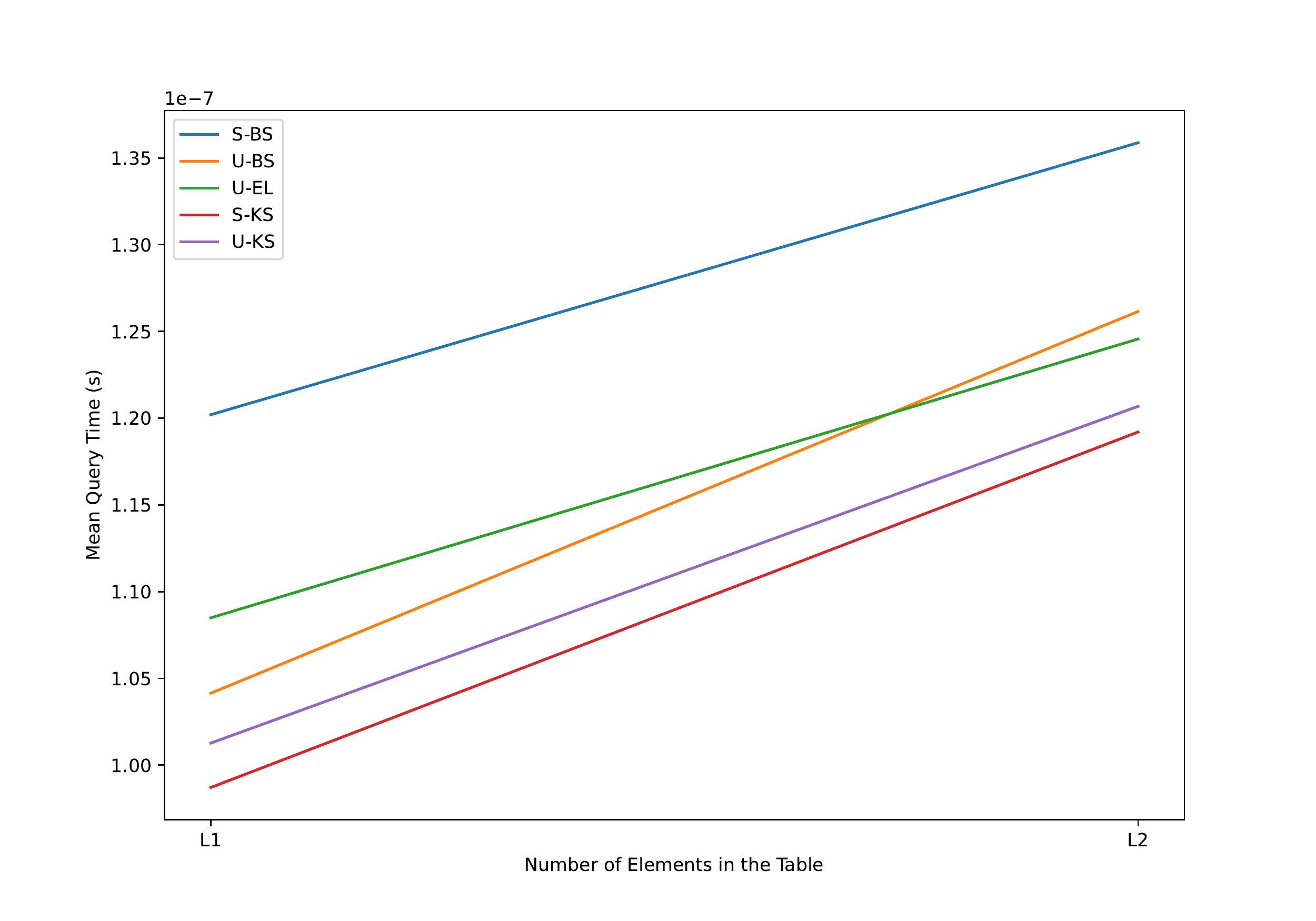}
			\end{minipage}\hfill\\
			(c)
			\begin{minipage}{0.45\textwidth}
				\includegraphics[width=\linewidth]{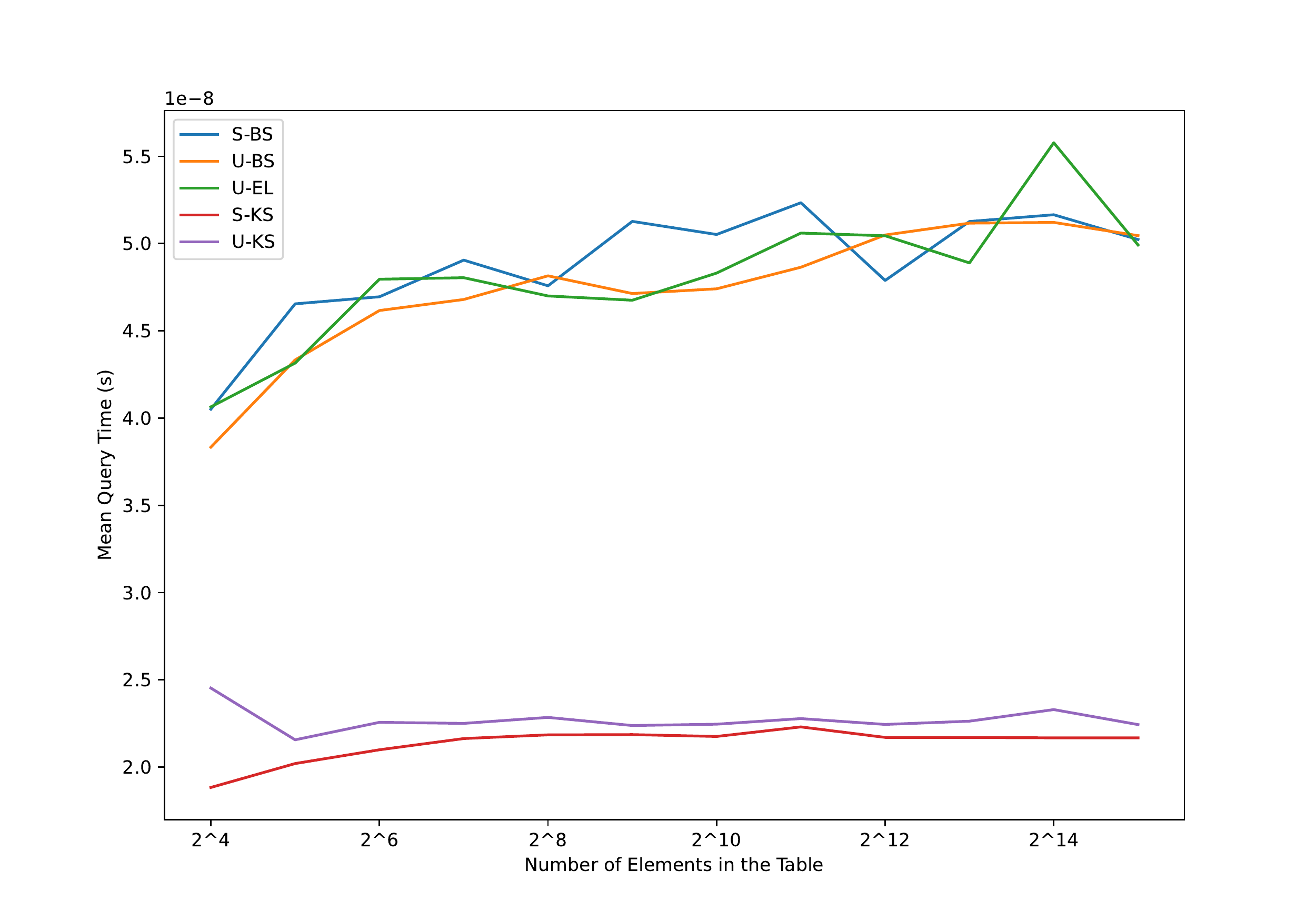}
			\end{minipage}\hfill
			\centering
			(d)
			\begin{minipage}{0.45\textwidth}
				\includegraphics[width=\linewidth]{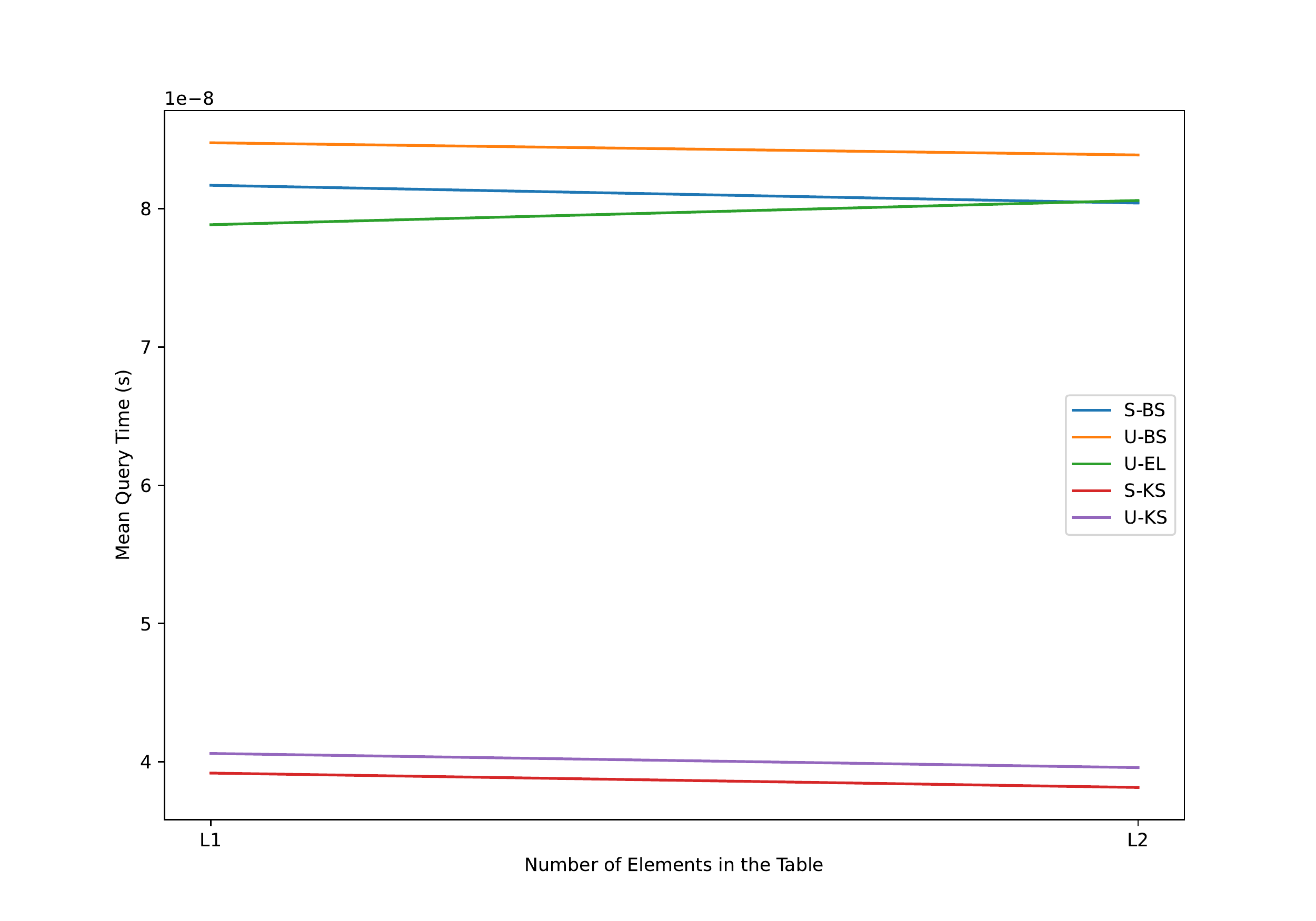}
			\end{minipage}\hfill\\
			\centering
			\caption{{\bf Zoom-in on Mean Query Time with SOSD }. The Figures (a) and (b) reports a zoom-in of \ref{M-fig:Morin}(a) and of the Figure \ref{S-fig:I7M1osm}(a), respectively, while the Figures (c) and (d) reports a zoom-in of \ref{M-fig:Morin}(c) and of the Figure \ref{S-fig:I7M1osm}(c), respectively. The legend is as in the corresponding Figures.}
			\label{S-fig:I7MorinOsmZoom}
		\end{figure}

		\paragraph{Profiler Analysis}\label{S-sec:profiler}
		
		In order to look more closely into the results in Section \ref{M-sec:morin}, we make use of the {\bf profiler} on the routines to monitor the following parameters.
				
		\begin{itemize}
			\item {\bf Percentage of Retiring operation}. That is the percentage of the pipeline slots used to perform useful work. It is divided into light operations, which are instructions that require at most one micro-operation, and heavy, which require two or more of such operations. A high percentage of this parameter implies a very effective use of the CPU.
			\item {\bf Percentage of Front-end operation}. That is the percentage of pipeline slots where the processor is not able to fetch instructions to the Back-end part. This issue can be caused, for example, by stalls due to instruction-cache misses. 
			\item {\bf Percentage of Back-end operation}. That is the percentage of pipeline slots where the Back-end part is not able to deliver micro-operation for the retiring. This issue can be caused, for example, by stalls due to a lack of resources.
			\item {\bf Percentage of Bad Speculation operation}. That is the percentage of pipeline slots where a wrong pipeline of instructions is loaded, due to a mispredicted branch.
		\end{itemize}
		
    	The results are summarized in Tables \ref{S-T:ProfilingMorin10} and \ref{S-T:ProfilingMorin24}.

    	\begin{table}[h]
    		\center
    		\small
    	
    			\begin{tabular}{|c|c|c|c|c|c|c|c|c|}
    				\hline
    				&
    				\multicolumn{2}{|c}{S-BS} & 
    				\multicolumn{2}{|c}{U-BS} & 
    				\multicolumn{2}{|c}{U-EL} & 
    				\multicolumn{2}{|c|}{S-KS} \\ \hline
    				Type & SOSD & NO SOSD & SOSD & NO SOSD & SOSD & NO SOSD & SOSD & NO SOSD \\ \hline
    				Retiring & 34.2-95.3& 18.7-95.7  & 38.1-100 &  36.8-100 & 17.0-100 & 23.8-100 & 33.5-88.1 & 25.6-32.4 \\ \hline
    				Front-end & 22.6 & 16.5 & 15.8 & 8.9 & 25.2 & 37.5 & 21.8 & 42.3 \\ \hline
    				Back-end & 28.5 & 21.9 & 41.1 & 51.3 & 29.7 & 25.5 & 28.2 & 15.1 \\ \hline
    				Bad Spec. & 14.6 & 43.0 & 5.0 & 3.0 & 28.2 & 13.2 & 16.5 & 17.1\\ \hline
    				
    			\end{tabular}
    			
    			\hfill 
    			\caption{{\bf Profiling of S-BS, U-BS, U-EL and S-KS on the Intel I7}. The results refer to a synthetic table of $2^{10}$ elements. We consider {\bf S-KS} only, since its  performance is analogous to {\bf U-KS}. The first row reports the methods and the second row, for each method, indicates the execution environment, i.e. with or without {\bf SOSD}. The next rows represent hardware operation types. All the values are reported in percentage. In particular, the second value on the Retiring row is the percentage of light operations.}\label{S-T:ProfilingMorin10}
    		
    	\end{table}
    	
    	\begin{table}[h]
    		\center
    		\small
 
    			\begin{tabular}{|c|c|c|c|c|c|c|c|c|}
    				\hline
    				&
    				\multicolumn{2}{|c}{S-BS} & 
    				\multicolumn{2}{|c}{U-BS} & 
    				\multicolumn{2}{|c}{U-EL} & 
    				\multicolumn{2}{|c|}{S-KS} \\ \hline
    				Type & SOSD & NO SOSD & SOSD & NO SOSD & SOSD & NO SOSD & SOSD & NO SOSD \\ \hline
    				Retiring & 19.1-75.4 & 5.5-100 & 16.4-76.2 & 6.8-79.4 & 7.2-62.5 & 30.8-94.8 & 9.1-71.4 & 15-46.2 \\ \hline
    				Front-end & 14.5 & 4.4 & 9.1 & 3 & 13.3 & 30.8 & 13.1 & 34.4 \\ \hline
    				Back-end & 51.3 & 61.4 & 71.9 &  89.2 & 51.9 & 21.0 & 45.9 & 32.5 \\ \hline
    				Bad Spec. & 15.2 & 28.7 & 2.7 & 1 & 27.0 &  17.0 & 31.9 & 18.1\\ \hline
    				
    			\end{tabular}
    			
    			\hfill 
    			\caption{{\bf Profiling of S-BS, U-BS and S-KS on the Intel I7}. The results refer to a synthetic table of $2^{24}$ elements. The legend is as in Table \ref{S-T:ProfilingMorin10}}\label{S-T:ProfilingMorin24}
    		
    	\end{table}
			
		\subsection{Pros and Cons of Prefetching}\label{S-sec:prefetch}
			
			This Section presents the details of the profiler analysis indicated in Section \ref{M-sec:pref}.  
			A deeper analysis with the profiler, as just done in the previous Section, is reported in Tables \ref{S-T:PipelinePrefetch}-\ref{S-T:BackendPrefetch}. Such analysis allows us to see that the utilisation of pipeline slots is relatively identical with or without explicit prefetching on a dataset that is small enough to be stored in cache memory, e.g. uniform with $2^{10}$ elements. On the other hand, it can be seen that if the dataset resides in central memory, e.g. uniform with $2^{24}$ elements, the percentage of stalls in the back-end component of the CPU is higher without the use of explicit prefetching. In particular, as shown in Table \ref{S-T:BackendPrefetch}, 70\% of these are due to a data cache miss problem that explicit prefetching manages to avoid.

			\begin{table}[h]
				\begin{center}
					\caption{{\bf U-BS with or without Explicit Prefetching on Intel I7 Pipeline Usage Percentage}  The first column reports CPU pipeline operation types. For each of them, we report the percentage of slots used.}\label{S-T:PipelinePrefetch}					\begin{tabular}{|c|r|r|r|r|}
						\hline
						&
						\multicolumn{2}{c|}{$2^{10}$} & \multicolumn{2}{c|}{$2^{24}$} \\ \hline
						Type & Prefetch & No Prefetch & Prefetch & No Prefetch \\ \hline
						Retiring & 38.4 & 39.2 & 24.4 & 19.7  \\ \hline
						Front-end  & 16.2 & 16.3 & 8.2 & 6.8 \\ \hline
						Back-end  & 40.4 & 39.4 & 64.4 & 71.2 \\ \hline
						Bad Spec.  & 5.1 & 5.1 & 2.7 & 2.3 \\ \hline
					\end{tabular}
				\end{center}
			\end{table}
			
			\begin{table}[h]
				\begin{center}
					\caption{{\bf U-BS with or without Explicit Prefetching on Intel I7 Back-end Usage Percentage}  The first column reports the types of Back-end stalls. Memory indicates stalls occurred due to a data cache miss error, while Core indicates stalls due to an overloaded execution unit. For each of them, we report the percentage of slots used.}\label{S-T:BackendPrefetch}
					\begin{tabular}{|c|r|r|r|r|}
							\hline
							&
							\multicolumn{2}{c|}{Uniform $2^{10}$} & \multicolumn{2}{c|}{Uniform $2^{24}$} \\ \hline
							Type & Prefetch & No Prefetch & Prefetch & No Prefetch \\ \hline
							Memory & 23.4 & 22.5 & 58.4 &  70.2 \\ \hline
							Core  & 76.6 & 77.5 & 41.6 & 29.8 \\ \hline
						\end{tabular}
					
				\end{center}
			\end{table}

\section{Experiments: Searching using Learned Indexes, With or Without SOSD - Additional Results}

		We report here the details of the experiments mentioned in Section \ref{M-Exp:Learned_Index}. In particular, we provide the following.

		\begin{itemize}
			
			\item We, report the result of Best Learned Indexes within {\bf SOSD} other than {\bf osm} in Figures \ref{S-fig:amzn32I7I9M1}-\ref{S-fig:wikiI7I9M1}.
			\item The full list of best Models indicated by {\bf SOSD} in Table \ref{S-T:BestModels}.
		\end{itemize}
	    
		\begin{figure}[tbh]
			\begin{center}
				(a)
				\begin{minipage}{0.45\textwidth}
					\includegraphics[width=\linewidth]{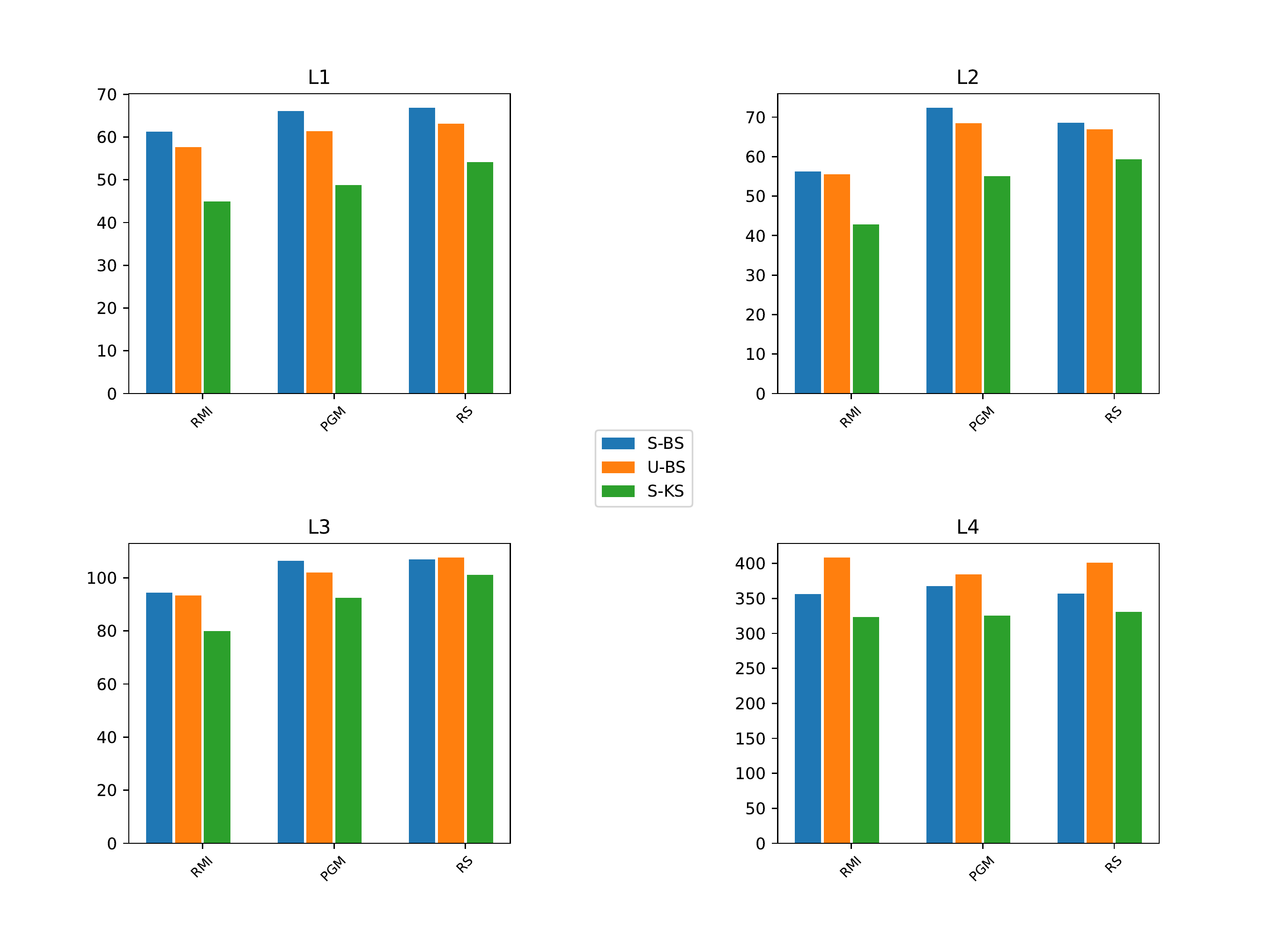}
				\end{minipage}\hfill
				(b)
				\begin{minipage}{0.45\textwidth}
					\includegraphics[width=\linewidth]{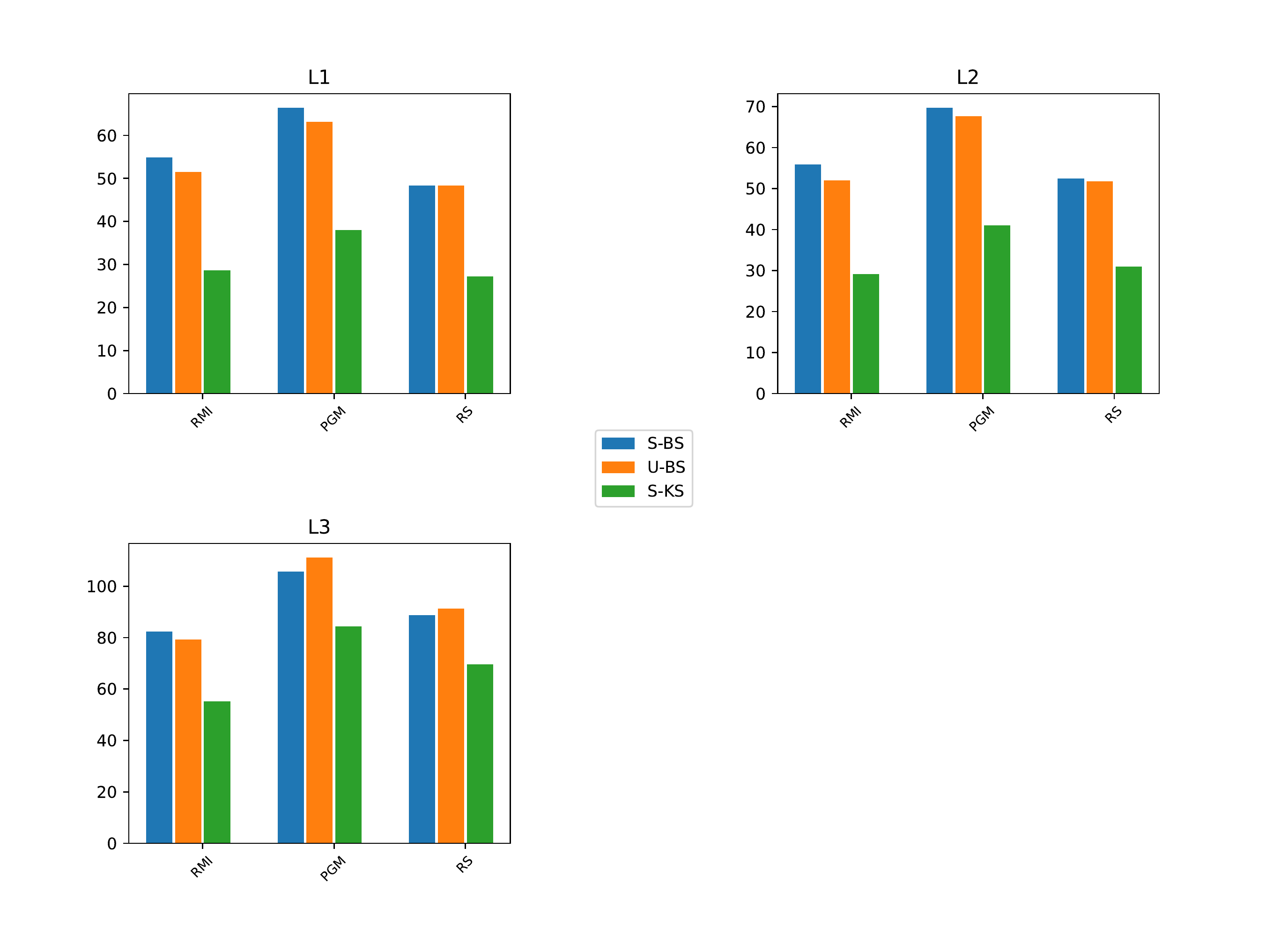}
				\end{minipage}\hfill
				\caption{{\bf Mean Query Times of Best Learned Indexes on amzn32 dataset}. Figure (a) and (b) report results using {\bf SOSD} on the Intel I7 and Apple Arm M1, respectively. For each model class, we report the mean query time of the best Learned Indexes adopting in their last stage the routines described in Section \ref{M-sec:methods}. In particular, the blue bar is {\bf S-BS}, the orange bar is {\bf U-BS} and the green bar is {\bf S-KS}.}\label{S-fig:amzn32I7I9M1}
			\end{center}
		\end{figure}
		
		\begin{figure}[tbh]
			\begin{center}
				(a)
				\begin{minipage}{0.45\textwidth}
					\includegraphics[width=\linewidth]{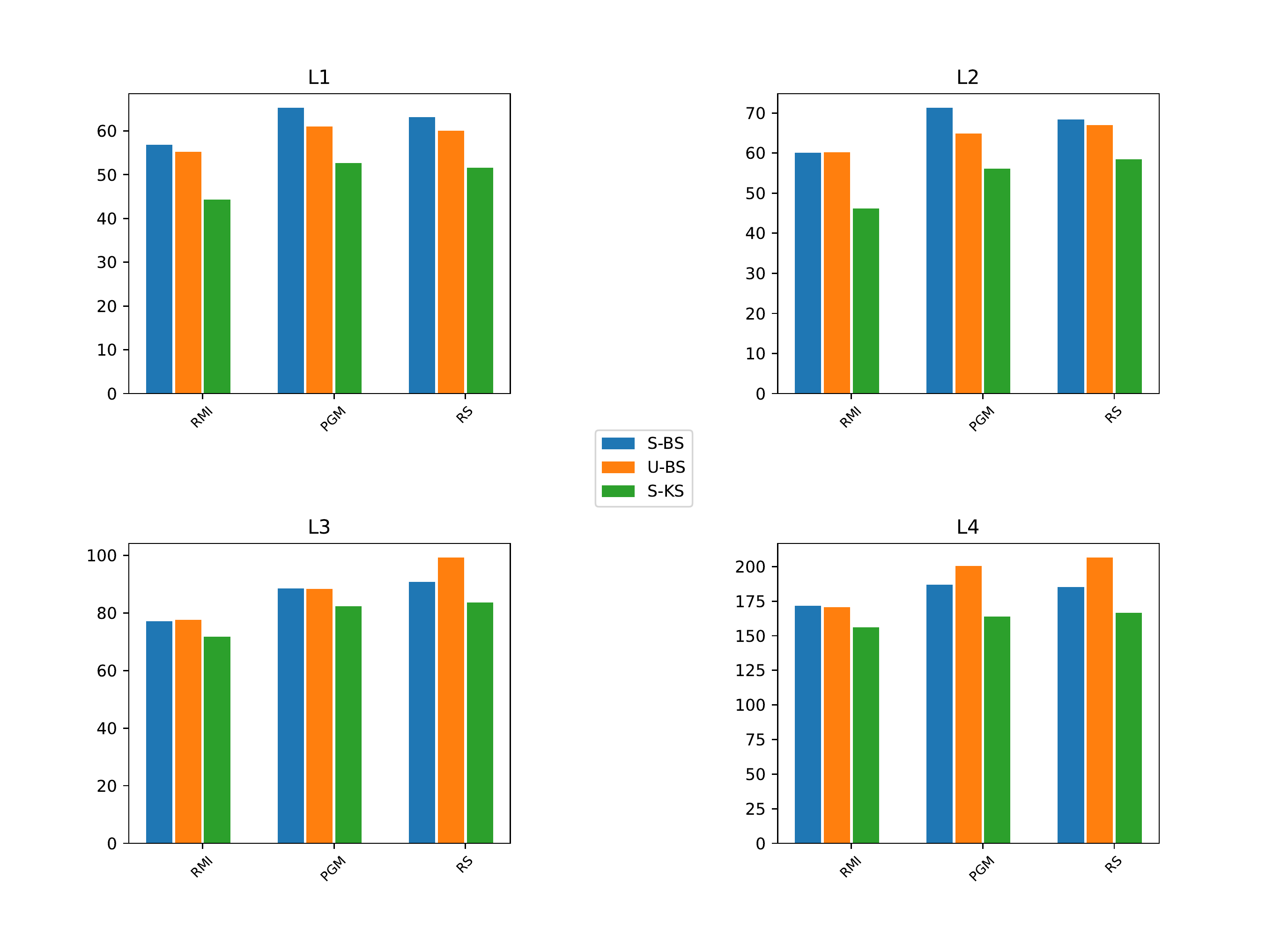}
				\end{minipage}\hfill
				(b)
				\begin{minipage}{0.45\textwidth}
					\includegraphics[width=\linewidth]{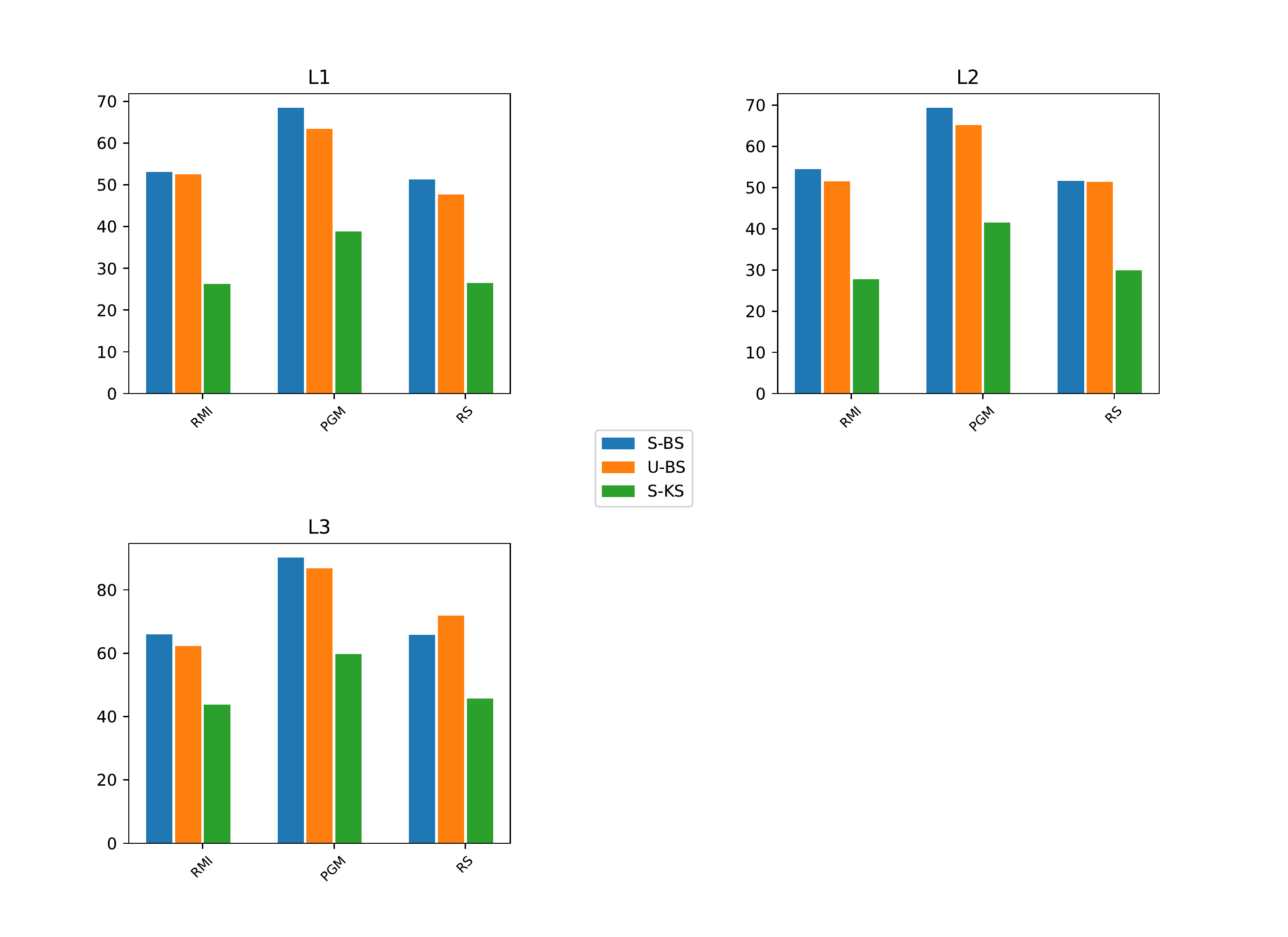}
				\end{minipage}\hfill
				\caption{{\bf Mean Query Times of Best Learned Indexes on amzn64 dataset}. The legend is as in Figure \ref{S-fig:amzn32I7I9M1}.}\label{S-fig:amzn64I7I9M1}
			\end{center}
		\end{figure}
		
		\begin{figure}[tbh]
			\begin{center}
				(a)
				\begin{minipage}{0.45\textwidth}
					\includegraphics[width=\linewidth]{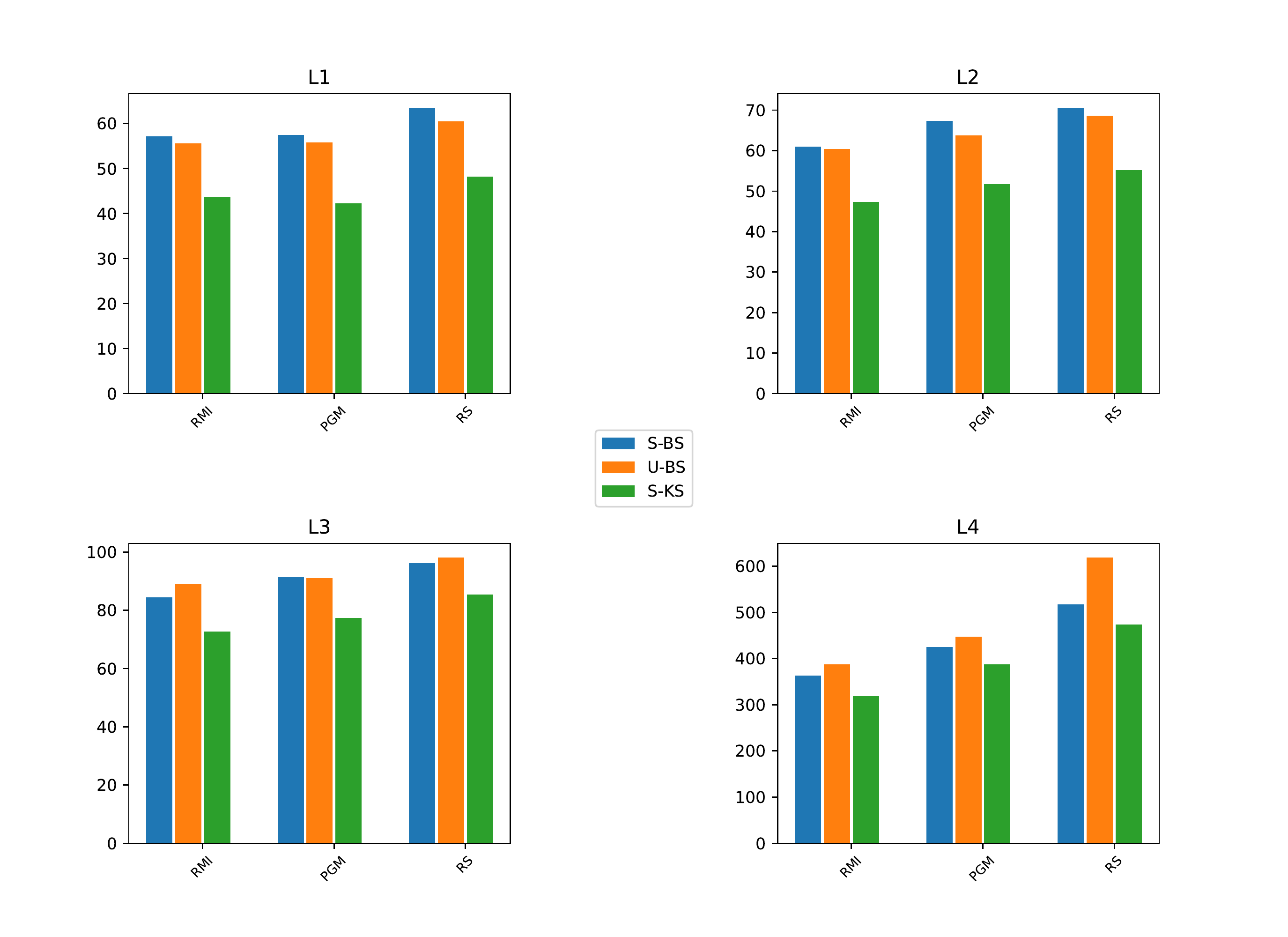}
				\end{minipage}\hfill
				(b)
				\begin{minipage}{0.45\textwidth}
					\includegraphics[width=\linewidth]{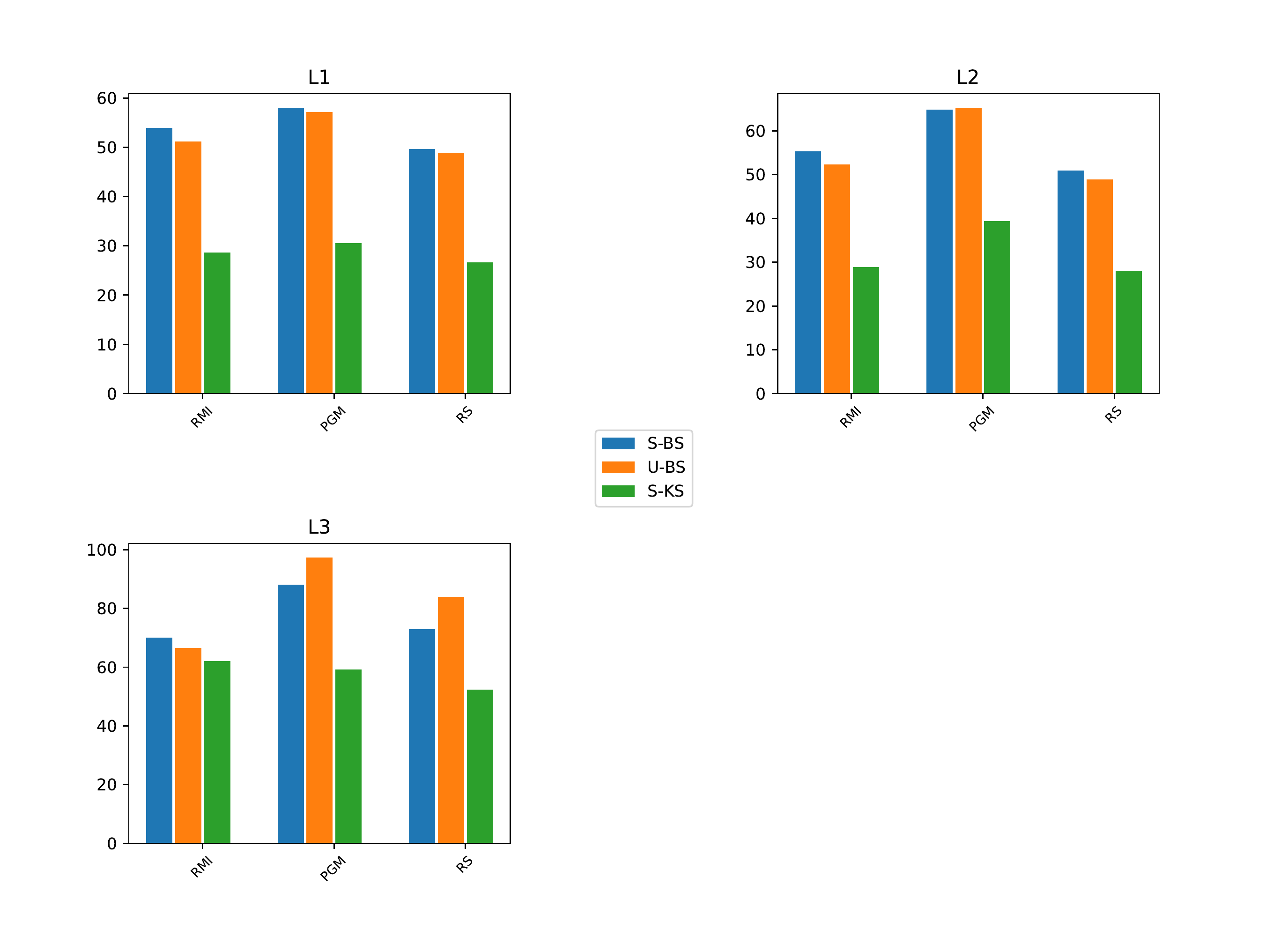}
				\end{minipage}\hfill
				\caption{{\bf Mean Query Times of Best Learned Indexes on face dataset}.The legend is as in Figure \ref{S-fig:amzn32I7I9M1}.}\label{S-fig:faceI7I9M1}
			\end{center}
		\end{figure}
		
		\begin{figure}[tbh]
			\begin{center}
				(a)
				\begin{minipage}{0.45\textwidth}
					\includegraphics[width=\linewidth]{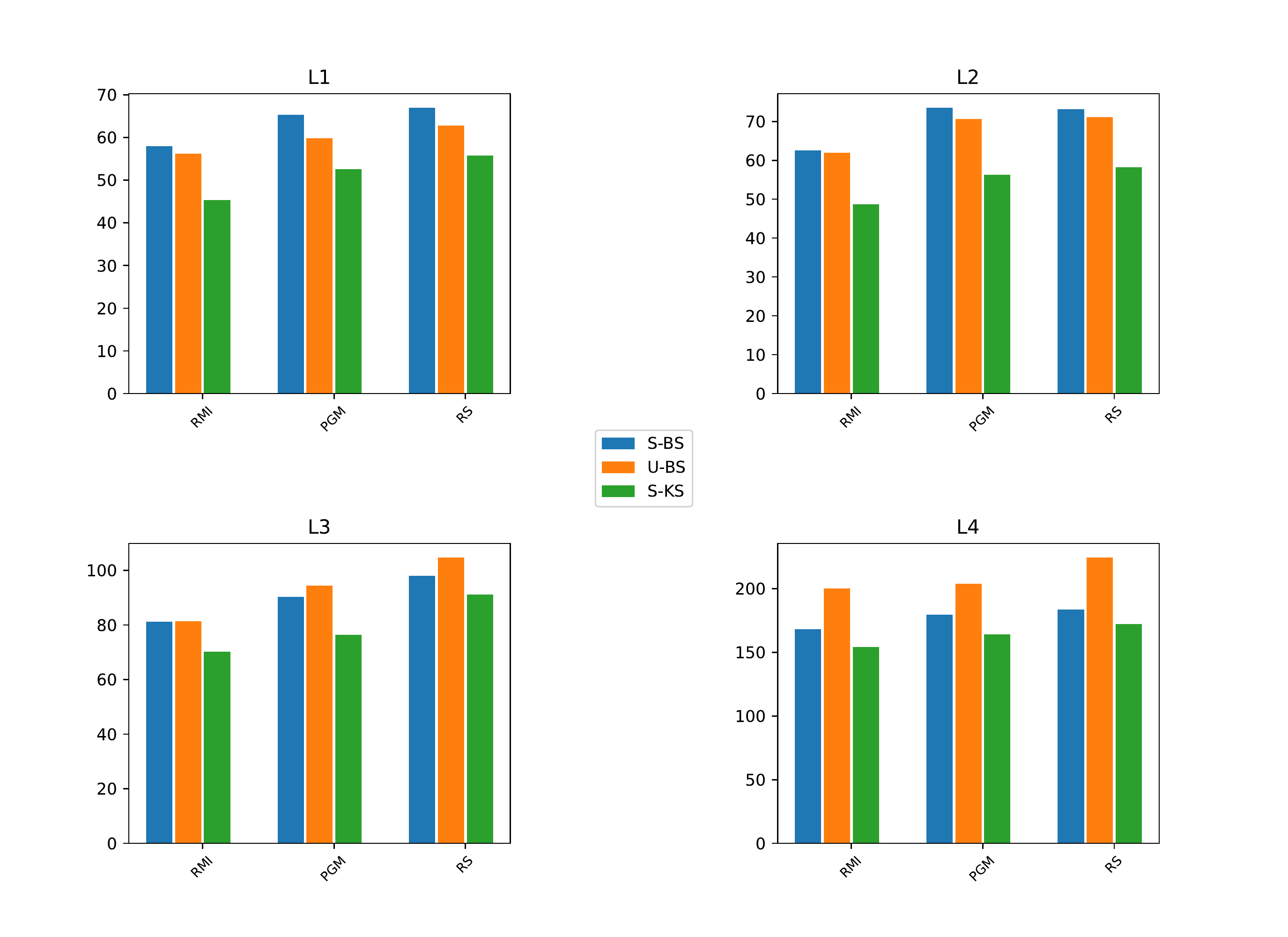}
				\end{minipage}\hfill
				(b)
				\begin{minipage}{0.45\textwidth}
					\includegraphics[width=\linewidth]{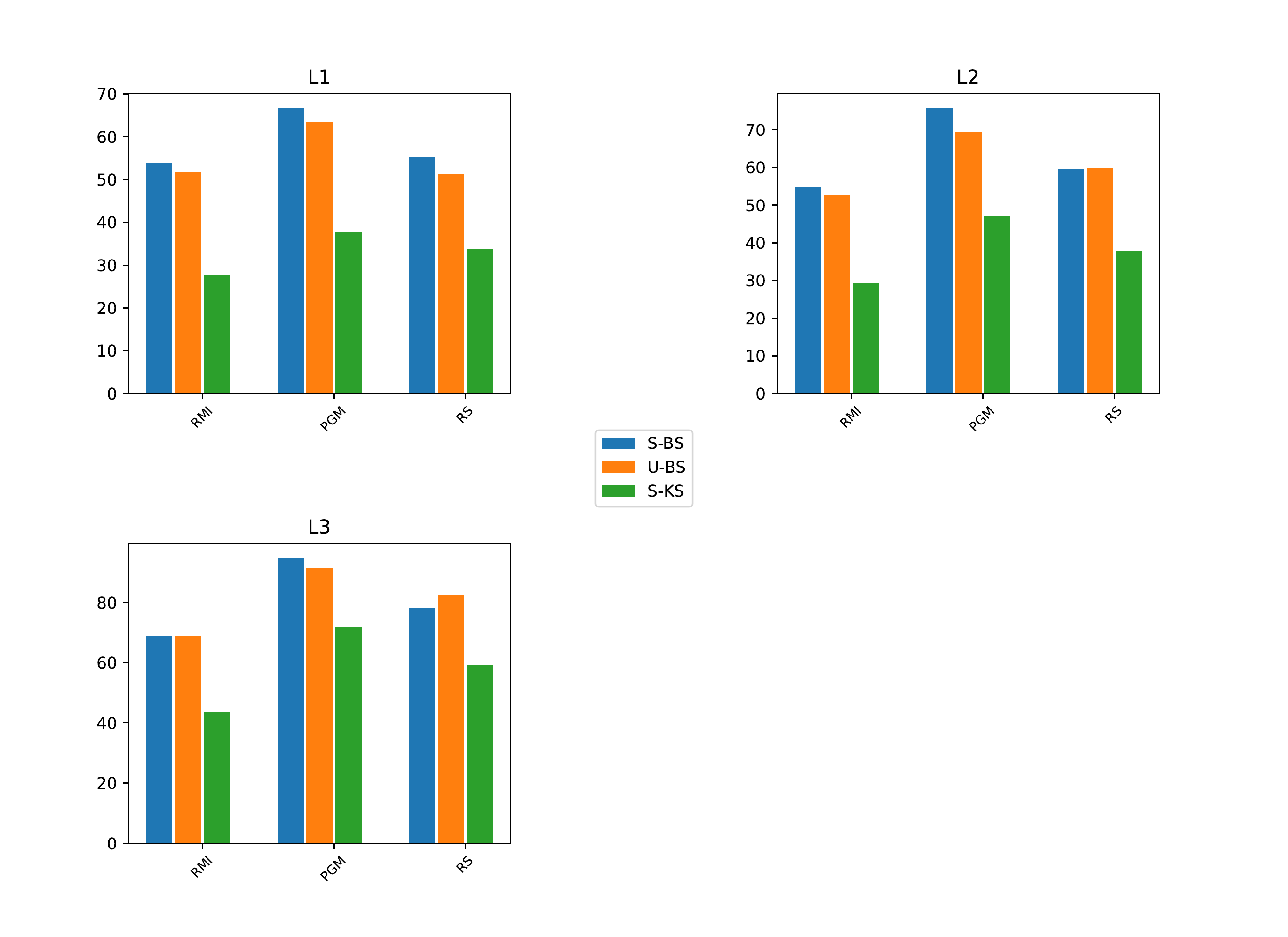}
				\end{minipage}\hfill
				\caption{{\bf Mean Query Times of Best Learned Indexes on wiki dataset}. The legend is as in Figure \ref{S-fig:amzn32I7I9M1}.}\label{S-fig:wikiI7I9M1}
			\end{center}
		\end{figure}
		
		\begin{table}[h]
			\center
	
				\begin{tabular}{|c|c|c|c|c|}
					\hline
					\multicolumn{5}{|c|}{Intel I7} \\ \hline \hline
					& L1 & L2 &L3 & L4 \\ \hline
					amzn32 & RMI & RMI & RMI & RMI \\ \hline
					amzn64 & RMI & RMI & RMI & RMI \\ \hline
					face & PGM & RMI & RMI & RMI \\ \hline
					osm & RMI & RMI & RMI & RMI \\ \hline
					wiki & RMI & RMI & RMI & RMI \\ \hline \hline
					
					\multicolumn{5}{|c|}{Intel I9} \\ \hline \hline
					
					& L1 & L2 & L3 & L4 \\ \hline
					amzn32 & RMI & RMI & RMI & RMI \\ \hline
					amzn64 & RMI & RMI & RMI & RMI \\ \hline
					face & RMI & RMI & RMI & RMI \\ \hline
					osm & RMI & RMI & RMI & RMI \\ \hline
					wiki & RMI & RMI & RMI & RMI \\ \hline \hline
					
					\multicolumn{5}{|c|}{Apple M1} \\ \hline \hline
					
					& L1 & L2 &L3 & L4 \\ \hline
					amzn32 & RMI & RMI & RMI & - \\ \hline
					amzn64 & RMI & RMI & RMI  & - \\ \hline
					face & RS & RS & RS  & - \\ \hline
					osm & RMI & RMI & RMI & - \\ \hline
					wiki & RMI & RMI & RMI & - \\ \hline 
					
				\end{tabular}
				
				\hfill 
				\caption{{\bf SOSD Best Models} . For each dataset and each memory level, the table reports the class of the best performing model indicated by {\bf SOSD}.}\label{S-T:BestModels}
	
		\end{table}

\end{document}